\documentclass[aps,prb,twocolumn,superscriptaddress]{revtex4-1}

\usepackage{graphicx,latexsym}
\usepackage{amsmath,amssymb,amsfonts}
\usepackage{bm}
\usepackage{braket}

\begin{document}


\title{Wannier representation of Floquet topological states}
\author{Masaya Nakagawa}
\email{nakagawa@cat.phys.s.u-tokyo.ac.jp}
\affiliation{Department of Physics, University of Tokyo, 7-3-1 Hongo, Bunkyo-ku, Tokyo 113-0033, Japan}
\author{Robert-Jan Slager}
\affiliation{Department of Physics, Harvard University, Cambridge MA 02138, USA}
\author{Sho Higashikawa}
\affiliation{Department of Physics, University of Tokyo, 7-3-1 Hongo, Bunkyo-ku, Tokyo 113-0033, Japan}
\author{Takashi Oka}
\affiliation{Max-Planck-Institut f\"{u}r Physik komplexer Systeme, N\"{o}thnitzer Stra\ss e 38, 01187 Dresden, Germany}
\affiliation{Max-Planck-Institut f\"{u}r Chemische Physik fester Stoffe, N\"{o}thnitzer Stra\ss e 40, 01187 Dresden, Germany}




\date{\today}

\begin{abstract}
A universal feature of topological insulators is that they cannot be adiabatically connected to an atomic limit, where individual lattice sites are completely decoupled. 
This property is intimately related to a topological obstruction to constructing a localized Wannier function from Bloch states of an insulator. Here we generalize this characterization of topological phases toward periodically driven systems. 
We show that nontrivial connectivity of hybrid Wannier centers in momentum space and time can characterize various types of topology in periodically driven systems, which include Floquet topological insulators, anomalous Floquet topological insulators with micromotion-induced boundary states, and gapless Floquet states realized with topological Floquet operators. 
In particular, nontrivial time dependence of hybrid Wannier centers indicates impossibility of continuous deformation of a driven system into an undriven insulator, and a topological Floquet operator implies an obstruction to constructing a generalized Wannier function which is localized in real and frequency spaces. 
Our results pave a way to a unified understanding of topological states in periodically driven systems as a topological obstruction in Floquet states.
\end{abstract}


\maketitle


\section{Introduction}
Topology has perpetually been playing a prominent role in physics, providing understanding of an increasing number of phenomena by relating them to rigorous mathematical insights. Following the discovery of the integer and fractional quantum Hall effects \cite{Klitzing80,Tsui82,Laughlin83}, topological invariants have been  associated with concrete observables describing phases of matter that are accessible in the laboratory. These concepts were reinvigorated with
 the prediction and experimental verification of the topological insulator \cite{Hasan10, Qi11}. In such topological insulators, the presence of symmetry, time-reversal symmetry in this case, furnishes a necessary condition for band structures to host non-trivial topology. It was shortly realized that the time-reversal symmetry is however not special in this regard, and a classification of all topological band structures due to (anti-)unitary symmetries for every spatial dimension rapidly emerged 
 			\cite{Schnyder08, Kitaev09}. In contrast to quantum Hall effects, topological band insulators nonetheless require a crystal lattice.  Recently, considerable attention has focused on the effects of these additional symmetries,   		exposing a rather rich landscape of novel topological phases \cite{Fu11, Slager12, Shiozaki14, Kruthoff17, Bradlyn17,Po17,Hoeller18, PhysRevB.91.161105}. In crude essence, these results relate symmetry concepts to compatibility relations, 		    			determining whether a Wannier description in terms of localized functions is obstructed. The possibility of cataloguing materials using these new tools has put these kinds of studies actively on the agenda \cite{Hurtubise19}. 
			
In addition to the progress in understanding equilibrium topological phases, 
the past decade has witnessed remarkable progress in extending the concept of topological phases of matter towards periodically driven systems, which are far from equilibrium\cite{OkaAoki09, Kitagawa10_walk, Kitagawa10, Kitagawa11, Lindner11, GomezLeon13, Kundu14, Jiang11, Rudner13, NathanRudner15, Carpentier15, RoyHarper17, MorimotoPoVishwanath, Yao17, Budich17, Higashikawa18, Sun18, Eckardt17, OkaKitamura19}. In a system whose Hamiltonian varies periodically in time, the discrete time-translation symmetry leads to a time analog of the Bloch theorem, which is called the Floquet theorem. The Floquet picture has enabled various intriguing possibilities of topological photo-dressed bands termed Floquet topological insulators\cite{Lindner11,OkaAoki09}. Furthermore, it has been revealed that Floquet systems possess even richer topological structures than static systems, leading to unique topological phenomena absent in equilibrium\cite{Kitagawa10, Rudner13, RoyHarper17, MorimotoPoVishwanath}. Experimental realizations of Floquet topological states have been reported with the help of developments in engineering laser-driven quantum materials\cite{Wang13, Mahmood16, McIver18}, photonic systems\cite{Kitagawa12, Rechtsman13, Maczewsky17}, acoustic systems\cite{Fleury16, Peng16}, and cold atoms\cite{Jotzu14, Flaschner16}.

All such Floquet topological states of non-interacting fermions discussed in literature fall into one of the following three distinct types of topology (precise definitions of operators in the following are given in Sec.~\ref{sec_Wannier}): (i) topology of a gapped effective Hamiltonian\cite{OkaAoki09, Kitagawa10_walk, Kitagawa10, Kitagawa11, Lindner11, GomezLeon13, Kundu14} $H_{\mathrm{eff}}(\bm{k})$, (ii) topology of a time-evolution operator\cite{Jiang11, Rudner13, NathanRudner15, Carpentier15, RoyHarper17, MorimotoPoVishwanath, Yao17} $U(\bm{k},t)$ during one period, and (iii) topology of a Floquet operator\cite{Kitagawa10, Budich17, Higashikawa18, Sun18} $U(\bm{k})$ which is a time-evolution operator over one period. 
The topology of type (i) has been discussed in the context of Floquet topological insulators \cite{OkaAoki09, Kitagawa11, Lindner11}. In this case, the topological properties of effective photo-dressed bands are defined through the effective static Hamiltonian $H_{\mathrm{eff}}$. 
The topology of type (ii) leads to anomalous Floquet topological insulators which fall outside the topological classification of static insulators\cite{Jiang11, Rudner13, NathanRudner15, Carpentier15, RoyHarper17, MorimotoPoVishwanath, Yao17}. In this case, even if the topology of the effective Hamiltonian is trivial, the topology of a time-evolution operator characterizes nontrivial micromotion during one period and leads to anomalous edge states which are absent in static cases. In contrast to these two cases, the third class of topology (iii) does not require a gapped Floquet band. 
It characterizes gapless quasienergy spectra which cannot be gapped out under continuous deformation of a Floquet operator\cite{Kitagawa10, Budich17, Higashikawa18, Sun18}. 
While the type (i) is quite similar to the topology in static systems, the types (ii) and (iii) are genuinely drive-induced topology, which has no analog in equilibrium cases.

Despite the above rich structures in Floquet topological states, the three types of topology (i)-(iii) are defined through three distinct operators, and the mathematical classification of each type of operators requires different conditions on the smooth deformation of operators\cite{Kitagawa10, NathanRudner15, RoyHarper17, Higashikawa18}. 
On the other hand, topological insulators in static systems can be characterized as a property of Bloch states rather than a property of Hamiltonians. 
In fact, Bloch states of topological insulators form a nontrivial vector bundle over the Brillouin zone, thereby leading to a topological obstruction to adiabatic deformation into an atomic limit. This property is closely related to the fact that one cannot construct a Wannier function localized in real space from Bloch states of topological insulators\cite{Thonhauser06, Coh09, Soluyanov11, Soluyanov11_2, Yu11, Read17}. However, Floquet topological states have been characterized with topology of operators, and topological characterization directly with Floquet states is missing, 
except for Anderson-localized Floquet insulators\cite{Nathan17,Nathan19} and a few limiting cases\cite{MondragonShem18}. 
If the three types of topology (i)-(iii) can be rewritten with topological obstructions in Floquet states, such description may provide a coherent understanding of topology in periodically driven systems, and the various types of Floquet topological states can be discussed from a unified viewpoint.

In this paper, we develop a state-based characterization of Floquet topological states. To this end, we utilize geometric phases, Berry connection, and Berry curvature of Floquet states, and express topological invariants of periodically driven systems with Floquet states. In addition, we introduce a notion of ``Wannier functions" in periodically driven systems, and demonstrate that the topological invariants can be extracted from nontrivial connectivity of the (hybrid) Wannier centers over momentum space and time. From these results, we clarify what kinds of topological obstructions exist in Floquet topological states of each type.

In contrast to Bloch states of static insulators, a single-particle Floquet state depends on momentum and time. Thus, it is legitimate to expect that a Floquet topological state has an obstruction to deforming a system into an undriven insulator in an atomic limit, of which a state is independent of momentum and time. Through the Fourier transformation over the momentum and time, such an undriven trivial insulator is characterized by a generalized Wannier function which is localized not only in real space, but also in the frequency domain. From the topological characterization of Floquet states, we find that each type of Floquet topological state has the following obstructions to an undriven trivial insulator. First, a Floquet topological insulator [type (i)] is characterized by Floquet states at a specific time slice located at each driving period, and thus its topological obstruction is related only to the locality of Wannier functions in real space. In contrast, a gapless Floquet topological state [type (iii)] is characterized by an obstruction to constructing a generalized Wannier function localized in both real and frequency spaces, and can thus be regarded as a counterpart of topological insulators defined on a coordinate-frequency lattice. Finally, an anomalous Floquet topological insulator [type (ii)] is characterized by nontrivial time dependence of Wannier centers, giving an obstruction to continuously switching off the driving while keeping a Floquet-band gap. Our work therefore demonstrates that all the topological information in periodically driven systems can indeed be extracted from Floquet wavefunctions over spacetime, providing a coherent framework parallel to static topological phases of matter.

The rest of the paper is organized as follows. In Sec.~\ref{sec_Wannier}, we introduce a notion of Wannier functions in periodically driven systems, and show that various geometric phases in Floquet states can be regarded as centers of the generalized Wannier functions. On the basis of the Wannier representation in Floquet systems, we characterize each type of Floquet topological states in subsequent sections. We start in Sec.~\ref{sec_FloquetTI} with Floquet topological insulators [type (i)]. In Sec.~\ref{sec_gapless}, we consider gapless Floquet topological states [type (iii)] characterized by topological Floquet operators. In Sec.~\ref{sec_anom}, we proceed to anomalous Floquet topological insulators [type (ii)]. We finish this paper by summarizing our results and discussing some outlooks in Sec.~\ref{sec_discussion}. 
In Appendix \ref{sec_pump}, we perform an explicit calculation of the geometric phases of Floquet states using a solvable model of non-adiabatic topological pumping. 
Appendix \ref{sec_sym} provides a summary of symmetries in periodically driven systems and corresponding Altland-Zirnbauer symmetry class considered in the main text. A detail of a calculation of a topological invariant of time-reversal-symmetric gapless Floquet states is presented in Appendix \ref{sec_AIIinv}.

\section{Wannier functions in periodically driven systems\label{sec_Wannier}}
We first introduce a notion of Wannier functions in periodically driven systems. 
We consider a periodically driven system of non-interacting fermions described by a time-dependent Bloch Hamiltonian $H(\bm{k},t)$. Here $\bm{k}=(k_1,\cdots,k_d)$ is a crystal momentum in $d$-dimensional space, and the Hamiltonian satisfies a time-periodicity condition $H(\bm{k},t+T)=H(\bm{k},t)$ where $T$ is the period of driving. In periodically driven systems, the Floquet theorem states that a solution of the Schr\"{o}dinger equation $i\partial_t\ket{\psi(\bm{k},t)}=H(\bm{k},t)\ket{\psi(\bm{k},t)}$ is written as\cite{Shirley65, Sambe73}
\begin{equation}
\ket{\psi_\alpha(\bm{k},t)}=e^{-i\varepsilon_\alpha(\bm{k}) t}\ket{\Phi_\alpha(\bm{k},t)}.
\label{eq_Floquetstate}
\end{equation}
Here $\varepsilon_\alpha(\bm{k})$ is quasienergy of the $\alpha$-th Floquet band, and $\ket{\Phi_\alpha(\bm{k},t)}$ is a Floquet-Bloch state, which satisfies the time-periodicity $\ket{\Phi_\alpha(\bm{k},t+T)}=\ket{\Phi_\alpha(\bm{k},t)}$. The quasienergy and the Floquet-Bloch state are obtained from an eigenvalue equation for a Floquet operator $U(\bm{k})\equiv\mathcal{T}\exp\Bigl[-i\int_0^Tdt H(\bm{k},t)\Bigr]$:
\begin{align}
U(\bm{k})\ket{\Phi_\alpha(\bm{k},0)}=e^{-i\varepsilon_\alpha(\bm{k})T}\ket{\Phi_\alpha(\bm{k},0)}.
\label{eq_eigen1}
\end{align}
The time-evolved Floquet state $\ket{\Phi_\alpha(\bm{k},t)}$ is given by
\begin{align}
\ket{\Phi_\alpha(\bm{k},t)}&=e^{i\varepsilon_\alpha(\bm{k}) t}\ket{\psi_\alpha(\bm{k},t)}\notag\\
&=e^{i\varepsilon_\alpha(\bm{k}) t}U(\bm{k},t)\ket{\Phi_\alpha(\bm{k},0)},
\label{eq_evolFloquet}
\end{align}
where $U(\bm{k},t)\equiv\mathcal{T}\exp\Bigl[-i\int_0^tdt' H(\bm{k},t')\Bigr]$ is the time-evolution operator (note that $\ket{\psi_\alpha(\bm{k},0)}=\ket{\Phi_\alpha(\bm{k},0)}$). From Eq.\ \eqref{eq_eigen1}, we define an effective Hamiltonian 
$H_{\mathrm{eff}}(\bm{k})\equiv \frac{i}{T}\log U(\bm{k})$, 
which satisfies
\begin{equation}
H_{\mathrm{eff}}(\bm{k})\ket{\Phi_\alpha(\bm{k},0)}=\varepsilon_\alpha(\bm{k})\ket{\Phi_\alpha(\bm{k},0)}.
\label{eq_Heff}
\end{equation}
Here, we set the branch of the logarithm using a condition $-\pi/T\leq \varepsilon_\alpha(\bm{k})<\pi/T$.

In static systems, a Wannier function is defined through the Fourier transformation of Bloch states\cite{Marzari12}. Analogously, we introduce a time-dependent Wannier function from a Floquet-Bloch state as
\begin{equation}
\ket{w_\alpha(\bm{R},t)}\equiv\int_{\mathrm{BZ}} \frac{d\bm{k}}{(2\pi)^d}e^{-i\bm{k}\cdot\bm{R}}\ket{\Phi_\alpha(\bm{k},t)},
\label{eq_Wannier}
\end{equation}
where BZ denotes the first Brillouin zone. 
We also introduce a hybrid Wannier function
\begin{equation}
\ket{w_\alpha(R_j;\bm{k}_\perp,t)}\equiv\int_{-\pi}^{\pi} \frac{dk_j}{2\pi}e^{-ik_jR_j}\ket{\Phi_\alpha(\bm{k},t)},
\label{eq_hybWannier}
\end{equation}
which is localized only in the $j$-th direction of the real space, as in static cases\cite{Marzari12}. Here $\bm{k}_\perp$ is the crystal momentum perpendicular to the $j$-th direction. In static topological insulators, hybrid Wannier functions play an important role in characterizing a topological obstruction to constructing a Wannier function that is localized in all directions of the real space\cite{Soluyanov11, Soluyanov11_2, Yu11, Taherinejad14, Taherinejad15, Gresch17}.

In periodically driven systems, we can perform a Fourier transformation in time direction
\begin{equation}
\ket{\Phi_\alpha^{(m)}(\bm{k})}\equiv\frac{1}{T}\int_0^T dte^{im\omega t}\ket{\Phi_\alpha(\bm{k},t)},
\label{eq_FloquetWannier}
\end{equation}
where $\omega=2\pi/T$ is the frequency of the driving, and $m\in\mathbb{Z}$. In analogy with Eq.~\eqref{eq_hybWannier}, the $m$-th harmonics \eqref{eq_FloquetWannier} in the Floquet state can be regarded as a hybrid Wannier function in the frequency domain. This interpretation naturally leads to a definition of a generalized Wannier function
\begin{equation}
\ket{w_\alpha^{(m)}(\bm{R})}\equiv\frac{1}{T}\int_0^T dt\int_{\mathrm{BZ}}\frac{d\bm{k}}{(2\pi)^d}e^{-i\bm{k}\cdot\bm{R}+im\omega t}\ket{\Phi_\alpha(\bm{k},t)},
\label{eq_genWannier}
\end{equation}
which is localized in the real and frequency spaces.

Here we note that we require the continuity and periodicity of the Floquet-Bloch state $\ket{\Phi_\alpha(\bm{k},t)}$ in the crystal momentum $\bm{k}$ to obtain localized Wannier functions \eqref{eq_Wannier} 
and \eqref{eq_genWannier}. If there exists an obstruction to taking a gauge in which the Floquet-Bloch state is continuous and periodic in $\bm{k}$, the Wannier functions are not localized in the real space\cite{Vanderbilt_book}.  
We also remark that one can take $\ket{\psi_\alpha(\bm{k},t)}$ instead of $\ket{\Phi_\alpha(\bm{k},t)}$ to define the Wannier functions, since the two states differ only by a phase factor in Eq.~\eqref{eq_Floquetstate}. 
This gauge degree of freedom is also important for understanding of the localizability of the Wannier functions of Floquet topological states. We come back to this point in Sec.~\ref{sec_discussion}.

In static insulators, displacement of averaged positions of Wannier functions from a lattice site is closely related to the Berry phase of Bloch states. This property is also important in modern formulation of electric polarization in crystals\cite{KingSmith93, Resta94, Vanderbilt_book}. In the case of periodically driven systems, we define the Berry phase\cite{Berry84, Zak89} of Floquet states as
\begin{equation}
\gamma_j^{(\alpha)}(\bm{k}_\perp,t)\equiv\int_{-\pi}^\pi dk_j\bra{\Phi_\alpha(\bm{k},t)}i\partial_{k_j}\ket{\Phi_\alpha(\bm{k},t)},
\label{eq_Berry}
\end{equation}
which correspond to a hybrid Wannier center of Eq.~\eqref{eq_hybWannier} and may also be regarded as electric polarization of the $\alpha$-th Floquet band. 
As a multiband generalization of the Berry phase, eigenvalues of a Wilson loop\cite{WilczekZee84}
\begin{equation}
W_j(\bm{k}_\perp,t)\equiv \mathcal{P}\exp\left[i\oint_{C_j}d\bm{k}\cdot\bm{A}(\bm{k},t)\right]
\label{eq_Wilson}
\end{equation}
can be used for a subset of Floquet bands $\{ \ket{\Phi_\alpha(\bm{k},t)}\}_{\alpha=1}^{N_b}$. Here, $A_\mu^{\alpha\beta}(\bm{k},t)\equiv \bra{\Phi_\alpha(\bm{k},t)}i\partial_{k_\mu}\ket{\Phi_\beta(\bm{k},t)}\ (\alpha,\beta=1,\cdots,N_b)$ is the non-Abelian Berry connection, $C_j$ is a closed path parallel to the $j$-th axis in the momentum space, and $\mathcal{P}$ denotes the path ordering.

On the other hand, a similar quantity in the time direction
\begin{equation}
\gamma_{t}^{(\alpha)}(\bm{k})\equiv\int_0^T dt\bra{\Phi_\alpha(\bm{k},t)}i\partial_t\ket{\Phi_\alpha(\bm{k},t)}
\label{eq_AA}
\end{equation}
is not an adiabatic Berry phase, but a \textit{non-adiabatic} geometric phase which was considered by Aharonov and Anandan\cite{Aharonov87}. Notably, an averaged position of the Floquet state in the frequency domain is given by the Aharonov-Anandan phase as follows \cite{MondragonShem18}:
\begin{align}
\sum_{m=-\infty}^\infty\bra{\Phi_\alpha^{(m)}(\bm{k})}m\omega\ket{\Phi_\alpha^{(m)}(\bm{k})}=-\frac{1}{T}\gamma_{t}^{(\alpha)}(\bm{k}).
\label{eq_AApol}
\end{align}
This is in a complete analogy with the modern theory of polarization in insulators\cite{KingSmith93, Resta94, Vanderbilt_book}. Also, as a non-Abelian generalization of the non-adiabatic geometric phase\cite{Anandan88}, a non-adiabatic Wilson loop for a subset of Floquet bands $\{ \ket{\Phi_\alpha(\bm{k},t)}\}_{\alpha=1}^{N_b}$ is defined by
\begin{equation}
W_t(\bm{k})\equiv\mathcal{T}\exp\left[i\int_0^TdtA_t(\bm{k},t)\right],
\label{eq_timeWilson}
\end{equation}
where $A_t^{\alpha\beta}(\bm{k},t)\equiv\bra{\Phi_\alpha(\bm{k},t)}i\partial_t\ket{\Phi_\beta(\bm{k},t)}\ (\alpha,\beta=1,\cdots,N_b)$.

In general, we can use a Wilson loop along an arbitrary closed path in a momentum-time torus $T^{d+1}=\mathrm{BZ}\times [0,T]$ instead of straight paths in Eqs.~\eqref{eq_Wilson} and \eqref{eq_timeWilson}. Such Wilson loops along general paths may be useful in characterizing Floquet topological states with crystalline symmetries\cite{MorimotoPoVishwanath, Ladovrechis18, Franca18, Bomantara19, RodriguezVega18, Huang18, Peng18}. In this paper, we use only simple Wilson loops defined by Eqs.~\eqref{eq_Wilson} and \eqref{eq_timeWilson}.

In practical numerical implementation, one can only know Floquet states on discrete points in the momentum-time space. In this case, one may first discretize a path $C$ into points $(\bm{k}^{(1)},t^{(1)}), (\bm{k}^{(2)},t^{(2)}), \cdots, (\bm{k}^{(M)},t^{(M)}),$ and then approximate a Wilson loop along $C$ by a product\cite{Vanderbilt_book}
\begin{equation}
D(C)\equiv\prod_{j=1}^M \mathcal{M}(\bm{k}^{(j)},t^{(j)};\bm{k}^{(j+1)},t^{(j+1)})
\end{equation}
with an overlap matrix $\mathcal{M}_{\alpha\beta}(\bm{k}^{(j)},t^{(j)};\bm{k}^{(j+1)},t^{(j+1)})\equiv\braket{\Phi_\alpha(\bm{k}^{(j+1)},t^{(j+1)})|\Phi_\beta(\bm{k}^{(j)},t^{(j)})}$, where we set $(\bm{k}^{(M+1)},t^{(M+1)})\equiv(\bm{k}^{(1)},t^{(1)})$. We note that one should take an argument of eigenvalues of $D(C)$ to obtain a numerical approximation of the Wilson-loop eigenvalues, since $D(C)$ is not a unitary matrix in general.

\section{Floquet topological insulators: topology of $H_\mathrm{eff}(\bm{k})$\label{sec_FloquetTI}}
To characterize Floquet topological states with Wannier representation, let us begin with Floquet topological insulators, which are defined through the effective Hamiltonian $H_\mathrm{eff}(\bm{k})$. Suppose that a Floquet eigenspectrum of $H_\mathrm{eff}(\bm{k})$ has a finite gap between Floquet bands. Then, applying topological band theory of static insulators, we can classify Floquet bands into certain topological equivalence classes of insulators. When the Floquet eigenspectrum possesses a topologically nontrivial band in this sense, we say that this system is a Floquet topological insulator\cite{OkaAoki09, Kitagawa10, Kitagawa11, Lindner11}.

From the above definition, we see that Floquet topological insulators can be characterized by a property of Floquet states, since the Floquet states $\ket{\Phi_\alpha(\bm{k},0)}$ at $t=0$ play the role exactly same as Bloch states in static insulators due to Eq.~\eqref{eq_Heff}. In particular, Floquet states of a topological Floquet band cannot be continuously deformed into momentum-independent states as long as the Floquet band gap and symmetries of the system are kept. This gives a state-based characterization of Floquet topological insulators.

Furthermore, the correspondence between Floquet states at $t=0$ and Bloch states in static insulators enables us to characterize Floquet topological insulators with Wannier functions defined by Eqs.~\eqref{eq_Wannier} and \eqref{eq_hybWannier}. For example, let us consider a Floquet Chern insulator, which possesses a Floquet band with a nonzero Chern number\cite{OkaAoki09, Kitagawa11, Inoue10, PerezPiskunow14, Usaj14, Dehghani14, Dehghani15_1, Dehghani15_2, DAlessio15, Mikami16, Yates16}. It is known that in static Chern insulators, one cannot construct a localized Wannier function due to a topological obstruction\cite{Thonhauser06, Coh09}. The topological obstruction can be characterized by nontrivial connectivity of hybrid Wannier centers in momentum space\cite{Coh09}. By the same token, in a Floquet Chern insulator, a localized Wannier function \eqref{eq_Wannier} cannot be constructed since one cannot take a gauge of $\ket{\Phi_\alpha(\bm{k},0)}$ that is smooth over the whole Brillouin zone. Also, nontrivial connectivity of hybrid Floquet-Wannier centers \eqref{eq_Berry} at $t=0$ as a function of $k_2$ can characterize a Floquet Chern insulator. Similarly, by applying known Wilson-loop characterization of static topological insulators\cite{Soluyanov11, Soluyanov11_2, Yu11, Taherinejad14, Taherinejad15, Alexandradinata14, Alexandradinata16, Bouhon18}, we can characterize various Floquet topological insulators, which are protected by time-reversal symmetry\cite{Lindner11, Lindner13, Nakagawa14, Takasan17} or crystalline symmetries\cite{Ladovrechis18}, with the $t=0$ Floquet-Wilson loop \eqref{eq_Wilson} or its generalization to appropriate closed paths in the Brillouin zone\cite{Alexandradinata14, Alexandradinata16, Bouhon18}.

\section{Gapless Floquet topological states: topology of $U(\bm{k})$\label{sec_gapless}}
\subsection{Preliminaries}
Next, we consider gapless Floquet topological states characterized by topology of Floquet operators, which has no analog in static systems\cite{Kitagawa10,Higashikawa18}. A Floquet operator $U(\bm{k})$ defines a map from a $d$-dimensional Brillouin zone to the space of $N\times N$ unitary matrices U($N$) ($N$ is the number of Floquet bands within $-\pi/T\leq \varepsilon_\alpha<\pi/T$). If this map is homotopically inequivalent to a trivial map given by the $N\times N$ identity matrix $U(\bm{k})=1_{N\times N}$, the Floquet operator leads to a gapless quasienergy spectrum which cannot be gapped out by a smooth deformation of $U(\bm{k})$, since any topologically trivial Floquet operator can be deformed into $U(\bm{k})=1_{N\times N}$, which possesses a gapped (flat) quasienergy spectrum $\varepsilon_\alpha(\bm{k})=0$. A topological classification of Floquet operators was performed in Ref.~\onlinecite{Higashikawa18}. It was shown that the classification of Floquet operators in $d$ spatial dimensions coincides with that of gapless surface states of static topological insulators/superconductors in $d$ dimensions. This implies that topologically nontrivial Floquet operators lead to gapless quasienergy spectra akin to surface states of static topological insulators/superconductors. 
For instance, a chiral (helical) fermion dispersion emerges from a topological Floquet operator in one-dimensional class A (AII) systems \cite{Kitagawa10, Budich17}. Recently, it was shown that a single Weyl fermion, which appears as a surface state of a four-dimensional topological insulator, can be realized with a topological Floquet operator in three spatial dimensions \cite{Higashikawa18, Sun18}.

However, to achieve a topologically nontrivial Floquet operator, one generally needs some additional condition on Floquet driving. To see this, let us define $U_\lambda(\bm{k})\equiv\mathcal{T}\exp[-i\int_0^T dt \lambda H(\bm{k},t)]$ with $0\leq\lambda\leq 1$. This one-parameter family of Floquet operators clearly connects a Floquet operator $U(\bm{k})=U_{\lambda=1}(\bm{k})$ with a trivial unitary $U_{\lambda=0}(\bm{k})=1_{N\times N}$ without changing symmetries of the unitary operators. This situation is similar to that of static Bloch bands, for which an arbitrary Bloch Hamiltonian $H(\bm{k})$ can be continuously deformed into a trivial Hamiltonian $H(\bm{k})=0$ by using $H_\lambda(\bm{k})\equiv \lambda H(\bm{k})$, if one does not impose an assumption that an occupied band is separated from other bands by an energy gap. 
In the case of Floquet operators, we assume that a Floquet operator has a block-diagonal structure
\begin{equation}
U(\bm{k})=
\begin{pmatrix}
U_1(\bm{k}) & 0\\
0 & U_2(\bm{k})
\end{pmatrix},
\label{eq_block}
\end{equation}
where $U_1(\bm{k})$ and $U_2(\bm{k})$ are $N_1\times N_1$ and $N_2\times N_2$ unitary matrices, respectively. 
Physically, this condition means that any initial state taken from the $N_1$-dimensional Hilbert subspace returns to the same Hilbert subspace after one driving period, while a time-dependent state may make a detour from the subspace in intermediate time. 
The above condition is achieved by several manners. For example, one may assume generalized adiabaticity\cite{Kitagawa10, Higashikawa18, Sun18}, which confine a time-dependent quantum state to the lower $N_1$ bands due to large separation of energies between the lower and higher bands. 
Note that this condition allows non-adiabatic dynamics within the lower bands, which makes Floquet states different from instantaneous eigenstates of a time-dependent Hamiltonian\cite{Kitagawa10}. 
Also, one can fine-tune a driving protocol to achieve a block-diagonalized Floquet unitary\cite{Budich17, Higashikawa18} (for an example, see Appendix \ref{sec_pump}). 
As another realization, a topological unitary operator emerges as an edge unitary of an anomalous Floquet topological insulator\cite{Rudner13, Higashikawa18, Roy17}, where $U_1(\bm{k})$ and $U_2(\bm{k})$ correspond to Floquet operators for states localized at one boundary and the other boundary, respectively. 
The restriction to $U_1(\bm{k})$ plays a role similar to the restriction to occupied bands in static topological insulators, which enables us to discuss consequences arising from nontrivial topology. 
In the rest part of this section, we assume this condition and consider the topology of $U_1(\bm{k})$.

\subsection{Class A in $d=1$\label{sec_gaplessAd1}}
\begin{figure}
\includegraphics[width=8.5cm]{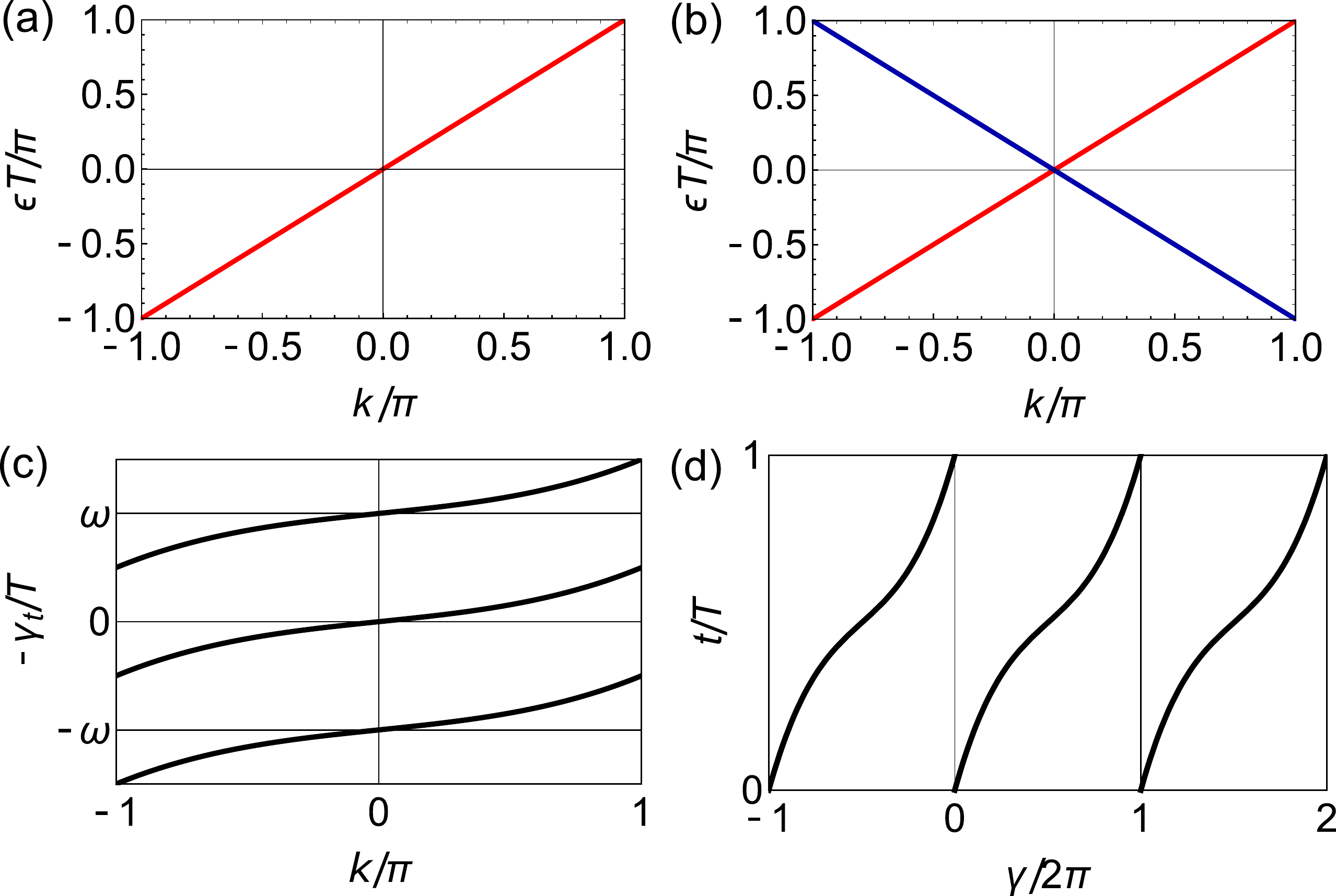}
\caption{(a) An example of quasienergy spectrum that gives a unit winding number $W_1=1$. (b) An example of quasienergy spectrum that gives a nontrivial $\mathbb{Z}_2$ invariant $\nu=1$. The colors of the lines correspond to different Floquet bands. (c) Schematic illustration of change in the Aharonov-Anandan phase $\gamma_{t}(k)$ (mod $2\pi$) in the Thouless pumping. (d) Schematic illustration of change in real-space Wannier centers (the Berry phase $\gamma(t)$ mod $2\pi$) during the Thouless pumping.}
\label{fig_WannierThouless}
\end{figure}

As the simplest example of gapless Floquet topological states, we consider a one-dimensional system without any symmetry except for charge conservation. The topological invariant of a Floquet operator is given by a winding number\cite{Kitagawa10}
\begin{align}
W_1&\equiv\frac{1}{2\pi}\int_{-\pi}^\pi dk \mathrm{Tr}[U_1^\dag(k)i\partial_kU_1(k)]\notag\\
&=\frac{T}{2\pi}\int_{-\pi}^\pi dk\sum_{\alpha=1}^{N_1}\frac{\partial\varepsilon_\alpha(k)}{\partial k},
\label{eq_1Dwind}
\end{align}
which takes an integer value. In Eq.~\eqref{eq_1Dwind}, $\mathrm{Tr}$ denotes the trace over the $N_1$-dimensional Hilbert subspace. As inferred from the second line of Eq.~\eqref{eq_1Dwind}, this topological invariant counts a winding number of quasienergy spectra over the Brillouin zone. If the winding number is nonzero, the quasienergy dispersion is topologically equivalent to $W_1$ chiral fermions\cite{Kitagawa10} [see Fig.~\ref{fig_WannierThouless} (a)]. An example of topological Floquet operators in this class is given by the Thouless pumping\cite{Thouless83} and its non-adiabatic generalization\cite{Budich17, Mizuta18, Lindner17, Privitera18}, in which the quantized pumped charge is equal to the winding number\cite{Kitagawa10}. In Appendix \ref{sec_pump}, we present an explicit calculation using a model of the non-adiabatic Thouless pumping.

The topological invariant \eqref{eq_1Dwind} is defined through the Floquet operator. However, we can express the same invariant using the Floquet state itself. To show this, let us substitute Eq.\ \eqref{eq_Floquetstate} into the Schr\"{o}dinger equation. We obtain
\begin{equation}
(H(k,t)-i\partial_t)\ket{\Phi_\alpha(k,t)}=\varepsilon_\alpha(k)\ket{\Phi_\alpha(k,t)}.
\end{equation}
Integrating the both sides in this equation over one period, we have\cite{OkaAoki09}
\begin{equation}
\varepsilon_\alpha(k)=\frac{1}{T}\int_0^Tdt\bra{\Phi_\alpha(k,t)}H(k,t)\ket{\Phi_\alpha(k,t)}-\frac{1}{T}\gamma_t^{(\alpha)}(k),
\label{eq_quasienergyAA}
\end{equation}
where $\gamma_t^{(\alpha)}(k)$ is the Aharonov-Anandan phase defined in Eq.~\eqref{eq_AA}. Thus, we obtain
\begin{equation}
W_1=-\frac{1}{2\pi}\int_{-\pi}^\pi dk\sum_{\alpha=1}^{N_1}\frac{\partial\gamma_t^{(\alpha)}(k)}{\partial k},
\label{eq_windAA}
\end{equation}
since the first term of the right hand side in Eq.~\eqref{eq_quasienergyAA}  (the dynamical phase) is a periodic function of $k$. Since the Aharonov-Anandan phase is defined with the Floquet state, the winding number \eqref{eq_windAA} is calculated directly from the Floquet state. If the winding number is nonzero, it leads to nontrivial connectivity of centers of hybrid Wanner functions in the frequency domain defined in Eq.~\eqref{eq_FloquetWannier}. Namely, since the frequency-domain polarization is ``pumped" by $W_1$ (in the unit of $\omega$) in total when $k$ is swept over the Brillouin zone, the hybrid Wannier centers are switched to their neighbors at the edge of the Brillouin zone, as illustrated in Fig.~\ref{fig_WannierThouless} (c) (see also Ref.~\onlinecite{MondragonShem18}). 

On the other hand, the same invariant \eqref{eq_1Dwind} can be rewritten as\cite{Kitagawa10}
\begin{align}
W_1=&\frac{1}{2\pi}\int_{-\pi}^\pi dk\sum_{\alpha=1}^{N_1}\bra{\Phi_\alpha(k,0)}U_1^\dag(k)i\partial_kU_1(k)\ket{\Phi_\alpha(k,0)}\notag\\
=&\frac{1}{2\pi}\sum_{\alpha=1}^{N_1}(\gamma^{(\alpha)}(T)-\gamma^{(\alpha)}(0)),
\label{eq_1DwindBerry}
\end{align}
where $\gamma^{(\alpha)}(t)\equiv \int_{-\pi}^\pi dk\bra{\psi_\alpha(k,t)}i\partial_k\ket{\psi_\alpha(k,t)}$ is the Berry phase. Since the Berry phase $\gamma^{(\alpha)}(t)$ describes a polarization of a time-dependent Wannier function \eqref{eq_hybWannier}, Eq.~\eqref{eq_1DwindBerry} can be interpreted as quantized pumping of Wannier centers over one period, which is nothing but the Thouless pumping\cite{Thouless83}. 
Consequently, the hybrid Wannier center \eqref{eq_hybWannier} in real space also exhibits nontrivial connectivity in the time domain [see Fig.~\ref{fig_WannierThouless} (d)]. 
This situation is completely analogous to that in static Chern insulators\cite{Coh09}. In fact, the expression \eqref{eq_1DwindBerry} leads to
\begin{align}
W_1=C_2\equiv\frac{1}{2\pi}\int_0^Tdt\int_{-\pi}^\pi dk\sum_\alpha F_\alpha(k,t),
\label{eq_1stChern}
\end{align}
where $F_\alpha(k,t)\equiv\braket{\partial_t\psi_\alpha(k,t)|\partial_k\psi_\alpha(k,t)}-\braket{\partial_k\psi_\alpha(k,t)|\partial_t\psi_\alpha(k,t)}$ is the Berry curvature, and thus $C_2$ is the first Chern number of the Floquet states\cite{Kitagawa10}. 
A nonzero Chern number signifies that one cannot take a gauge of Floquet states which is continuous and periodic in the momentum-time torus, since such a gauge makes a Chern number \eqref{eq_1stChern} vanishing by the Stokes' theorem. The absence of a global gauge of Floquet states implies that one cannot construct a generalized Wannier function \eqref{eq_genWannier} which is localized in the real space and the frequency space. 
Hence, we arrive at a conclusion that the winding number \eqref{eq_1Dwind} characterizes a topological obstruction of Floquet states to constructing a generalized Wannier function \eqref{eq_genWannier}. 
We note that a model for the Thouless pumping is actually regarded as a Chern insulator defined on a coordinate-frequency lattice \cite{OkaKitamura19}.

\subsection{Class AII in $d=1,2$}
As the second example, we consider class AII Floquet systems which have the time-reversal symmetry\cite{Kitagawa10, NathanRudner15, Carpentier15, RoyHarper17}
\begin{equation}
\Theta H(\bm{k},t)\Theta^{-1}=H(-\bm{k},-t)
\label{eq_TRS}
\end{equation}
with $\Theta^2=-1$ (see Appendix \ref{sec_sym}). In spatial dimension $d=1$, Floquet operators in class AII are classified\cite{Higashikawa18} by $\mathbb{Z}_2$. This coincides with the classification of $d=2$ class AII topological insulators and also with the classification of $d=1$ adiabatic spin pumping\cite{FuKane06}. 
The $\mathbb{Z}_2$ topological invariant of a class AII Floquet operator in $d=1$ is given by\footnote{This formula \eqref{eq_AIIgapless} of the topological invariant has not been derived in earlier literature. Equation \eqref{eq_AIIgapless} is derived by combining the formula of a topological invariant of static $d = 1$ class DIII topological superconductors\cite{Qi10,Schnyder11} and the fact that the classification of class AII Floquet operators is performed by mapping a Floquet unitary to an artificial Hamiltonian of a static class DIII superconductor\cite{Higashikawa18}.}
\begin{equation}
(-1)^\nu=\frac{\mathrm{Pf}[V_\Theta^\dag U_1(0)]}{\sqrt{\det[V_\Theta^\dag U_1(0)]}}\frac{\mathrm{Pf}[V_\Theta^\dag U_1(\pi)]}{\sqrt{\det[V_\Theta^\dag U_1(\pi)]}},
\label{eq_AIIgapless}
\end{equation}
where $V_\Theta$ denotes the unitary part of the time-reversal operator $\Theta=V_\Theta K$ with $V_\Theta V_\Theta^*=-1_{N\times N}$ ($K$ is complex conjugation). Here $\mathrm{Pf}[A]$ is the Pfaffian of an antisymmetric matrix $A$. 
In fact, the time-reversal symmetry of the Floquet operator \eqref{eq_TRSFloq} leads to
\begin{equation}
V_\Theta U_1^*(k)=U_1^\dag(-k)V_\Theta,
\end{equation}
which indicates that $V_\Theta^\dag U_1(k=0\ \mathrm{or}\ \pi)$ is an antisymmetric matrix. If the topological invariant takes a nontrivial value $\nu=1$, the quasienergy spectrum hosts gapless helical dispersion which cannot be gapped out under the time-reversal symmetry due to the Kramers degeneracy\cite{Higashikawa18} [see Fig.~\ref{fig_WannierThouless} (b)].

We can rewrite the $\mathbb{Z}_2$ invariant using Floquet states as (see Appendix \ref{sec_AIIinv})
\begin{align}
(-1)^{\nu}=&\frac{\mathrm{Pf}[w(0,0)]}{\sqrt{\det[w(0,0)]}}\frac{\mathrm{Pf}[w(\pi,0)]}{\sqrt{\det[w(\pi,0)]}}\notag\\
&\times\frac{\mathrm{Pf}[w(0,T/2)]}{\sqrt{\det[w(0,T/2)]}}\frac{\mathrm{Pf}[w(\pi,T/2)]}{\sqrt{\det[w(\pi,T/2)]}}
\label{eq_AIITI}
\end{align}
with $w_{\alpha\beta}(k,t)\equiv \bra{\psi_\alpha(-k,-t)}\Theta\ket{\psi_\beta(k,t)}$ ($\alpha,\beta=1,\cdots,N_1$). 
Notably, Eq.~\eqref{eq_AIITI} has the form same as the $\mathbb{Z}_2$ invariant of adiabatic spin pumping\cite{FuKane06} (see also Ref.~\onlinecite{ShindouSpinPump}), which characterizes difference of time-reversal polarization between $t=0$ and $t=T/2$. The same invariant also characterizes a static class AII topological insulators in two spatial dimensions, by regarding the time as a momentum in the second dimension. 
However, here $\ket{\psi_\alpha(k,t)}$ is a Floquet state, which is not necessarily an eigenstate of an instantaneous Hamiltonian.

Building on this observation, we can compute the $\mathbb{Z}_2$ topological invariant \eqref{eq_AIITI} from the connectivity of hybrid Wannier centers. As in adiabatic spin pumping or $\mathbb{Z}_2$ topological insulators\cite{Soluyanov11, Soluyanov11_2, Yu11}, time evolution of time-reversal polarization can be tracked by computing eigenvalues of a Wilson loop \eqref{eq_Wilson} along the spatial direction. Since the time-reversal polarization corresponds to difference of charge polarizations between time-reversal pairs, the hybrid Wannier centers switch their time-reversal partners during the time evolution from $t=0$ to $t=T/2$ if the $\mathbb{Z}_2$ invariant takes the nontrivial value ($\nu=1$). This also gives a clear physical interpretation of a topological Floquet operator characterized by the $\mathbb{Z}_2$ invariant \eqref{eq_AIIgapless}; under a Floquet driving with a $\mathbb{Z}_2$ topological Floquet operator, a time-reversal partner of Floquet-Wannier centers is pumped in opposite directions during the half of the period. Thus, the class AII topological Floquet driving provides a non-adiabatic generalization of the Fu-Kane $\mathbb{Z}_2$ spin pumping.

Here we note that the same invariant \eqref{eq_AIITI} can also be calculated from the non-adiabatic Wilson loop \eqref{eq_timeWilson}, whose eigenvalues give non-adiabatic geometric phases, since the role of the momentum and time can be interchanged in Eq.~\eqref{eq_AIITI}. In a Floquet driving with a topological Floquet operator with $\nu=1$, the non-adiabatic geometric phases show nontrivial evolution between $k=0$ and $k=\pi$.

The characterization of topological invariant with Floquet states [Eq.~\eqref{eq_AIITI}] also indicates a topological obstruction in the Floquet states. Let us consider a time-dependent gauge transformation of Floquet states given by
\begin{equation}
\ket{\bar{\psi}_\alpha(k,t)}=\sum_\beta V_{\beta\alpha}(k,t)\ket{\psi_\beta(k,t)},
\label{eq_AIIgauge}
\end{equation}
where $V(k,t)$ is a unitary matrix. Since the first Chern number \eqref{eq_1stChern} vanishes in time-reversal-symmetric systems, 
we can take a gauge in which the transformed Floquet state $\ket{\bar{\psi}_\alpha(k,t)}$ is continuous and periodic in $(k,t)$. Although $\ket{\bar{\psi}_\alpha(k,t)}$ is no longer a solution of the Schr\"{o}dinger equation, the generalized Wannier function \eqref{eq_genWannier} constructed from $\ket{\bar{\psi}_\alpha(k,t)}$ is well localized in the real and frequency spaces. 
However, in a class AII system, the Floquet states form time-reversal pairs $\ket{\psi_{1,a}(k,t)}, \ket{\psi_{2,a}(k,t)} (a=1,\cdots,N_1/2)$, where we switch the label of Floquet states from $\alpha$ to $(m,a) (m=1,2)$. 
Then, let us additionally impose a time-reversal condition on the Floquet states\cite{FuKane06, Soluyanov11, Yu11}:
\begin{subequations}
\begin{align}
\ket{\bar{\psi}_{1,a}(-k,-t)}&=\Theta\ket{\bar{\psi}_{2,a}(k,t)},\label{eq_AIITRgauge1}\\
\ket{\bar{\psi}_{2,a}(-k,-t)}&=-\Theta\ket{\bar{\psi}_{1,a}(k,t)}\label{eq_AIITRgauge2}.
\end{align}
\end{subequations}
Since the $\mathbb{Z}_2$ invariant \eqref{eq_AIITI} is invariant under the gauge transformation \eqref{eq_AIIgauge}, the time-reversal polarization computed in this gauge exhibits nonzero pumping between $t=0$ and $t=T/2$. However, since the time-reversal condition \eqref{eq_AIITRgauge1}, \eqref{eq_AIITRgauge2} forces the time-reversal polarization to vanish, a nontrivial value of the $\mathbb{Z}_2$ invariant ($\nu=1$) implies that the time-reversal condition cannot be satisfied in the whole $(k,t)$ space. Conversely, if we impose the time-reversal condition \eqref{eq_AIITRgauge1}, \eqref{eq_AIITRgauge2} in the whole $(k,t)$ space, 
the Floquet states $\ket{\bar{\psi}_{1,a}(k,t)}, \ket{\bar{\psi}_{2,a}(k,t)}$ with $\nu=1$ must have discontinuity at some $(k,t)$, since otherwise the continuous gauge leads to $\nu=0$ (see Refs.~\onlinecite{FuKane06, Soluyanov11, Yu11}). This discontinuity means that there is no generalized Wannier functions localized in real and frequency spaces under the time-reversal condition. 
This topological obstruction is completely analogous to that of $\mathbb{Z}_2$ topological insulators\cite{FuKane06, Soluyanov11, Yu11}.

The above argument for $d=1$ systems can easily be generalized to $d=2$ Floquet systems. In $d=2$, the topological classification of Floquet operators in class AII\cite{Higashikawa18} is also $\mathbb{Z}_2$, and the invariant is given by
\begin{equation}
(-1)^{\nu'}=\prod_{\bm{k}_*}\frac{\mathrm{Pf}[V_\Theta^\dag U_1(\bm{k}_*)]}{\sqrt{\det[V_\Theta^\dag U_1(\bm{k}_*)]}},
\end{equation}
where $\bm{k}_*$ denotes the four time-reversal-invariant momenta $(0,0),(\pi,0),(0,\pi)$, and $(\pi,\pi)$. As in $d=1$, we can rewrite this invariant as
\begin{equation}
(-1)^{\nu'}=\prod_{t=0,T/2}\prod_{\bm{k}_*}\frac{\mathrm{Pf}[w(\bm{k}_*,t)]}{\sqrt{\det[w(\bm{k}_*,t)]}}.
\end{equation}
This invariant can be computed from $(\bm{k}_\perp,t)$-dependence of eigenvalues of a Wilson loop \eqref{eq_Wilson} or $\bm{k}$-dependence of eigenvalues of a non-adiabatic Wilson loop \eqref{eq_timeWilson}, by using the method in Ref.~\onlinecite{Taherinejad14}.

\subsection{Class A or AII in $d=3$}
In three spatial dimensions, a Floquet operator is characterized by a winding number\cite{Kitagawa10}
\begin{align}
W_3\equiv&\int \frac{d^3\bm{k}}{24\pi^2}
\varepsilon^{\mu\nu\lambda}\mathrm{Tr}[(U_1^\dag\partial_{k_\mu} U_1)(U_1^\dag\partial_{k_\nu}U_1)(U_1^\dag\partial_{k_\lambda}U_1)],
\label{eq_wind3D}
\end{align}
which takes an integer value. Here $\varepsilon^{\mu\nu\lambda}$ is the antisymmetric tensor with $\mu,\nu,\lambda=1,2,3$. The winding number $W_3$ characterizes a topological gapless quasienergy spectra in three dimensions. 
In fact, when $W_3\neq 0$, the quasienergy spectrum possesses gapless Weyl fermions of nonvanishing total chirality, which cannot be realized in static systems due to the Nielsen-Ninomiya theorem\cite{NielsenNinomiya1, NielsenNinomiya2, Higashikawa18, Sun18}.

The winding number \eqref{eq_wind3D} can be also expressed with Floquet states. In Ref.~\onlinecite{Kitagawa10}, it was shown that the winding number \eqref{eq_wind3D} is written as the second Chern number of Floquet states
\begin{equation}
W_3=C_4\equiv\frac{1}{32\pi^2}\int_0^Tdt\int d^3\bm{k}
\varepsilon^{\nu\nu\lambda\rho}\mathrm{Tr}[F_{\mu\nu}F_{\lambda\rho}],
\label{eq_2ndChern}
\end{equation}
where $\varepsilon^{\mu\nu\lambda\rho}$ is the antisymmetric tensor, $F_{\mu\nu}^{\alpha\beta}(\bm{k},t)=\partial_\mu A_\nu^{\alpha\beta}(\bm{k},t)-\partial_\nu A_\mu^{\alpha\beta}(\bm{k},t)+i[A_\mu(\bm{k},t),A_\nu(\bm{k},t)]^{\alpha\beta}$ is the non-Abelian Berry curvature, and $A_\mu^{\alpha\beta}(\bm{k},t)=\bra{\psi_\alpha(\bm{k},t)}i\partial_\mu\ket{\psi_\beta(\bm{k},t)}$ is the non-Abelian Berry connection, with $\alpha,\beta=1,\cdots,N_1$ and $\mu,\nu,\lambda,\rho=1,2,3,4$ (here $\partial_{1,2,3}\equiv\partial_{k_{1,2,3}}$ and $\partial_4\equiv\partial_t$).

From Eq.~\eqref{eq_2ndChern}, it is clear that the winding number \eqref{eq_wind3D} detects a topological obstruction of Floquet states on the four-dimensional momentum-time torus. As in static four-dimensional topological insulators characterized by the second Chern number\cite{Taherinejad15, Qi08}, the topological invariant \eqref{eq_2ndChern} can be extracted by tracking a trajectory of hybrid Wannier centers (eigenvalues of the Wilson loop \eqref{eq_Wilson}) over the $(\bm{k}_\perp,t)$ space, or that of non-adiabatic geometric phases (eigenvalues of Eq.~\eqref{eq_timeWilson}) over the momentum space. In addition, a nonzero second Chern number \eqref{eq_2ndChern} implies that there cannot be a generalized Wannier function \eqref{eq_genWannier} localized in real and frequency spaces, since a continuous and periodic gauge in the momentum-time torus gives a vanishing Chern number.

\section{Anomalous Floquet topological insulators: topology of $U(\bm{k},t)$\label{sec_anom}}
\subsection{Preliminaries}
Finally, as the third class of topology in periodically driven systems, we consider characterization of anomalous Floquet topological insulators. Anomalous Floquet topological insulators possess gapped Floquet bands in bulk, and exhibit gapless boundary states under the open boundary condition\cite{Rudner13}. These anomalous boundary states cannot be predicted by an effective Hamiltonian $H_{\mathrm{eff}}(\bm{k})$ for one driving period, and thus are induced by nontrivial micromotion between each driving period. 
To characterize anomalous Floquet topological phases, here we assume that a quasienergy spectrum of a system has a finite gap at $\varepsilon=\pi/T$. In this case, topological invariants of a Floquet operator of the system discussed in Sec.~\ref{sec_gapless} should be trivial, since they characterize gapless quasienergy spectra. Although this may indicate that there is no gauge obstruction of the Floquet states over the momentum-time torus, we will see that there is yet another topological obstruction which prevents deformation of the Floquet states into those of undriven insulators. 

Topological invariants for anomalous Floquet topological insulators have been discussed in several literatures\cite{Rudner13,NathanRudner15,Yao17}. When a Floquet operator satisfies $U(\bm{k})=1_{N\times N}$, the time-evolution operator $U(\bm{k},t)$ defines a map from a $(d+1)$-dimensional momentum-time torus $T^{d+1}=\mathrm{BZ} \times [0,T]$ to U($N$) since $U(\bm{k},0)=U(\bm{k},T)=1_{N\times N}$. The topological invariants are obtained from topological numbers of $U(\bm{k},t)$. 
When a Floquet operator is not an identity, we can instead use a deformed evolution operator\cite{RoyHarper17} (which is also called a micromotion operator\cite{Eckardt17, Harper19}; see Eq.~\eqref{eq_evolFloquet2} below)
\begin{equation}
\tilde{U}(\bm{k},t)\equiv U(\bm{k},t)e^{iH_{\mathrm{eff}}(\bm{k})t},
\label{eq_deformevol}
\end{equation}
which satisfies the periodicity condition $\tilde{U}(\bm{k},0)=\tilde{U}(\bm{k},T)=1_{N\times N}$. The anomalous Floquet topological insulators are characterized by topology of $\tilde{U}(\bm{k},t)$.

\subsection{Class AIII in $d=1$}
We first consider one-dimensional cases. In $d=1$, anomalous Floquet topological insulators exist in class AIII, BDI, D, DIII, and CII\cite{RoyHarper17}. To consider insulators (i.e. systems without a particle-hole symmetry), here we focus on class AIII. A class AIII Floquet system have a chiral symmetry\cite{RoyHarper17, Fruchart16, Liu18} (see Appendix~\ref{sec_sym})
\begin{equation}
\Gamma H(k,t)\Gamma^{-1}=-H(k,-t),
\label{eq_AIIIchiralsym}
\end{equation}
where $\Gamma$ is a unitary operator. This leads to a symmetry of the time-evolution operator as
\begin{equation}
\Gamma U(k,t)\Gamma^{-1}=U(k,T-t)U^\dag(k,T).
\end{equation}
When $U(k,T)=1_{N\times N}$, this symmetry leads to
\begin{equation}
[\Gamma, U(k,T/2)]=0.
\end{equation}
Therefore, the time-evolution operator at $t=T/2$ can be block-diagonalized as
\begin{equation}
U(k,T/2)=
\begin{pmatrix}
U_+(k) & 0 \\
0 & U_-(k)
\end{pmatrix},
\end{equation}
according to eigenvalues $\pm1$ of the chiral operator $\Gamma$. 
The topological invariant is given by a difference of winding numbers of $U_\pm(k)$ as\cite{Fruchart16, Liu18}
\begin{align}
W&=\int_{-\pi}^\pi \frac{dk}{4\pi}\Bigl(\mathrm{Tr}[U_+^\dag(k)i\partial_k U_+(k)]-\mathrm{Tr}[U_-^\dag(k)i\partial_k U_-(k)]\Bigr)\notag\\
&=\int_{-\pi}^\pi \frac{dk}{4\pi}\mathrm{Tr}[\Gamma U^\dag(k,T/2)i\partial_k U(k,T/2)],
\label{eq_AFTIinvAIII}
\end{align}
where $\mathrm{Tr}$ implies the trace over all the Floquet bands. 

Here we rewrite the invariant \eqref{eq_AFTIinvAIII} using Floquet states. We note that, when $U(k,T)=1_{N\times N}$, we can take an arbitrary state as an initial Floquet state:
\begin{equation}
U(k,t)\ket{\alpha}=\ket{\Phi_\alpha(k,t)},
\end{equation}
where $\ket{\alpha}$ is an arbitrary state, since any state is an eigenstate of $U(k,T)$. In this case, the quasienergy spectrum is completely degenerate: $\varepsilon_\alpha(k)=0$. Let $\{\ket{\alpha}\}$ be a basis set of the Hilbert space. We obtain
\begin{align}
W&=\int_{-\pi}^\pi \frac{dk}{4\pi}\sum_\alpha\bra{\alpha}\Gamma U^\dag(k,T/2)i\partial_k U(k,T/2)\ket{\alpha}\notag\\
&=\int_{-\pi}^\pi \frac{dk}{4\pi}\sum_{\alpha,\beta}\bra{\alpha}\Gamma\ket{\beta}\bra{\beta}U^\dag(k,T/2)i\partial_k U(k,T/2)\ket{\alpha}\notag\\
&=\sum_{\alpha,\beta}\bra{\alpha}\Gamma\ket{\beta}\int_{-\pi}^\pi\frac{dk}{4\pi}\bra{\Phi_\beta(k,T/2)}i\partial_k\ket{\Phi_\alpha(k,T/2)}.
\label{eq_AIII_loop_before}
\end{align}
Let us consider a two-band system as an example and take the basis that diagonalizes the chiral operator as $\Gamma\ket{\pm}=\pm\ket{\pm}$. Then, we obtain
\begin{align}
W=&\frac{1}{4\pi}\int_{-\pi}^\pi dk\Bigl[\bra{\Phi_+(k,T/2)}i\partial_k\ket{\Phi_+(k,T/2)}\notag\\
&-\bra{\Phi_-(k,T/2)}i\partial_k\ket{\Phi_-(k,T/2)}\Bigr],
\label{eq_AIII_loop}
\end{align}
which is expressed with the Berry phases of the Floquet states at $t=T/2$. The physical meaning of Eq.~\eqref{eq_AIII_loop} is clear. Since $[\Gamma,U(k,T/2)]=0$, the Floquet states at $t=T/2$ are eigenstates of $\Gamma$: $\Gamma\ket{\Phi_\pm(k,T/2)}=\pm\ket{\Phi_\pm(k,T/2)}$. Eq.~\eqref{eq_AIII_loop} measures the difference of polarization (i.e. Wannier centers) of the Floquet states at $t=T/2$. Note that the polarization difference is zero at $t=0$ since $\ket{\Phi_\pm(k,0)}=\ket{\pm}$ is $k$-independent and thus has zero polarization. The nonzero winding number \eqref{eq_AIII_loop} indicates that the two chiral components are displaced in opposite directions during the half period. 

So far we have assumed that $U(k,T)=1_{N\times N}$. When $U(k,T)\neq 1_{N\times N}$, we use the deformed evolution operator \eqref{eq_deformevol}. Note that
\begin{align}
\tilde{U}(k,t)\ket{\Phi_\alpha(k,0)}&=U(k,t)e^{iH_{\mathrm{eff}}(k)t}\ket{\Phi_\alpha(k,0)}\notag\\
&=e^{i\varepsilon_\alpha(k)t}U(k,t)\ket{\Phi_\alpha(k,0)}\notag\\
&=\ket{\Phi_\alpha(k,t)},
\label{eq_evolFloquet2}
\end{align}
because of Eq.~\eqref{eq_evolFloquet}. Using this relation, we have
\begin{align}
W=&\int_{-\pi}^\pi \frac{dk}{4\pi}\mathrm{Tr}[\Gamma \tilde{U}^\dag(k,T/2)i\partial_k \tilde{U}(k,T/2)]\notag\\
=&\int_{-\pi}^\pi \frac{dk}{4\pi}\sum_{\alpha,\beta}\bra{\Phi_\alpha(k,0)}\Gamma\ket{\Phi_\beta(k,0)}\notag\\
&\times\bra{\Phi_\beta(k,0)}\tilde{U}^\dag(k,T/2)(i\partial_k\tilde{U}(k,T/2))\ket{\Phi_\alpha(k,0)}\notag\\
=&\int_{-\pi}^\pi \frac{dk}{4\pi}\sum_{\alpha,\beta}\bra{\Phi_\alpha(k,0)}\Gamma\ket{\Phi_\beta(k,0)}\notag\\
&\times\Bigl[\bra{\Phi_\beta(k,T/2)}i\partial_k\ket{\Phi_\alpha(k,T/2)}\notag\\
&-\bra{\Phi_\beta(k,0)}i\partial_k\ket{\Phi_\alpha(k,0)}\Bigr]\notag\\
=&\frac{1}{4\pi}\int_{-\pi}^\pi dk\int_0^{T/2}dt\partial_t\mathrm{Tr}'[\Gamma_0(k)A_x(k,t)],
\label{eq_AIIIwind_result}
\end{align}
where $\Gamma_0^{\alpha\beta}(k)\equiv\bra{\Phi_\alpha(k,0)}\Gamma\ket{\Phi_\beta(k,0)}$, $A_x^{\alpha\beta}(k,t)\equiv\bra{\Phi_\alpha(k,t)}i\partial_k\ket{\Phi_\beta(k,t)}$, and $\mathrm{Tr}'$ denotes the trace over the Floquet band indices $\alpha, \beta$. 

To elucidate the physical meaning of the formula \eqref{eq_AIIIwind_result}, here we note that Eq.~\eqref{eq_AIIIwind_result} is invariant under a gauge transformation
\begin{gather}
\ket{\bar{\Phi}_\beta(k,0)}=\sum_\alpha V_{\beta\alpha}^*(k)\ket{\Phi_\alpha(k,0)},
\end{gather}
where $V(k)$ is a $\textit{time-independent}$ unitary matrix. Note that the state $\ket{\bar{\Phi}_\beta(k,0)}$ is no longer an eigenstate of the Floquet operator. The time-evolved state is given by
\begin{equation}
\ket{\bar{\Phi}_\beta(k,t)}=\tilde{U}(k,t)\ket{\bar{\Phi}_\beta(k,0)}=\sum_\alpha V_{\beta\alpha}^*(k)\ket{\Phi_\alpha(k,t)}.
\label{eq_AIIIgauge}
\end{equation}
Under the gauge transformation \eqref{eq_AIIIgauge}, we have
\begin{gather}
\Gamma_0(k)=V^\dag(k)\bar{\Gamma}_0(k)V(k),\\
A_x(k,t)=V^\dag(k)\bar{A}_x(k,t)V(k)-V^\dag(k)i\partial_k V(k),
\end{gather}
where $\bar{\Gamma}_0^{\alpha\beta}(k)\equiv\bra{\bar{\Phi}_\alpha(k,0)}\Gamma\ket{\bar{\Phi}_\beta(k,0)}$ and $\bar{A}_x^{\alpha\beta}(k,t)\equiv\bra{\bar{\Phi}_\alpha(k,t)}i\partial_k\ket{\bar{\Phi}_\beta(k,t)}$. 
Thus, we obtain
\begin{align}
W=&\frac{1}{4\pi}\int_{-\pi}^\pi dk\int_0^{T/2}dt\partial_t\mathrm{Tr}'[V^\dag(k)\bar{\Gamma}_0(k)V(k)\notag\\
&\cdot(V^\dag(k)\bar{A}_x(k,t)V(k)-V^\dag(k)i\partial_k V(k))]\notag\\
=&\frac{1}{4\pi}\int_{-\pi}^\pi dk\int_0^{T/2}dt\partial_t\Bigl\{\mathrm{Tr}'[\bar{\Gamma}_0(k)\bar{A}_x(k,t)]\notag\\
&-\mathrm{Tr}'[V^\dag(k)\bar{\Gamma}_0(k)i\partial_k V(k)]\Bigr\}\notag\\
=&\frac{1}{4\pi}\int_{-\pi}^\pi dk\int_0^{T/2}dt\partial_t\mathrm{Tr}'[\bar{\Gamma}_0(k)\bar{A}_x(k,t)],
\end{align}
and the expression \eqref{eq_AIIIwind_result} is gauge invariant (note that $V(k)$ does not depend on $t$). Thanks to this gauge invariance, we can choose a gauge that diagonalizes the chirality matrix $\Gamma_0(k)$ as
\begin{equation}
\bar{\Gamma}_0^{\alpha\beta}(k)=\gamma_\alpha\delta_{\alpha\beta}\ \ (\gamma_\alpha=\pm1),
\label{eq_Gamma0}
\end{equation}
and then the winding number is given by
\begin{align}
W=&\frac{1}{4\pi}\int_{-\pi}^\pi dk\int_0^{T/2}dt\sum_\alpha\gamma_\alpha\partial_t\bar{A}_x^{\alpha\alpha}(k,t)\notag\\
=&\frac{1}{2}\int_0^{T/2}dt\sum_\alpha\partial_tP_\alpha^\Gamma(t),
\label{eq_AIIIwind_result2}
\end{align}
where
\begin{equation}
P_\alpha^\Gamma(t)\equiv\frac{\gamma_\alpha}{2\pi}\int_{-\pi}^\pi dk\bra{\bar{\Phi}_\alpha(k,t)}i\partial_k\ket{\bar{\Phi}_\alpha(k,t)}
\label{eq_chiralpol}
\end{equation}
is a ``chirality polarization" of the $\alpha$-th band\footnote{A similar quantity has been discussed in a context of static chiral topological insulators in Refs.~\onlinecite{Shiozaki13, Daido19}.}. 
Equation \eqref{eq_AIIIwind_result2} is the main result of this section. 
As a result, the winding number is given by a change of chirality polarizations during the half of the period. If the Floquet operator is an identity, this result is reduced to Eq.~\eqref{eq_AIII_loop}. Since the chirality polarization is a gauge-invariant Berry phase, its calculation can easily be implemented in numerical calculations. Namely, given a set of Floquet states $\ket{\Phi_\alpha(k,0)}$, we first diagonalize the chirality matrix as Eq.~\eqref{eq_Gamma0} using a unitary matrix $V(k)$. Then, after performing a gauge transformation of the Floquet states by $V(k)$, we calculate the chirality polarization \eqref{eq_chiralpol} from $\ket{\bar{\Phi}_\alpha(k,t)}$. The topological invariant of the anomalous class AIII Floquet insulator is obtained from evolution of the chirality polarizations during the half of the period [Eq.~\eqref{eq_AIIIwind_result2}].

\begin{figure}
\includegraphics[width=8.5cm]{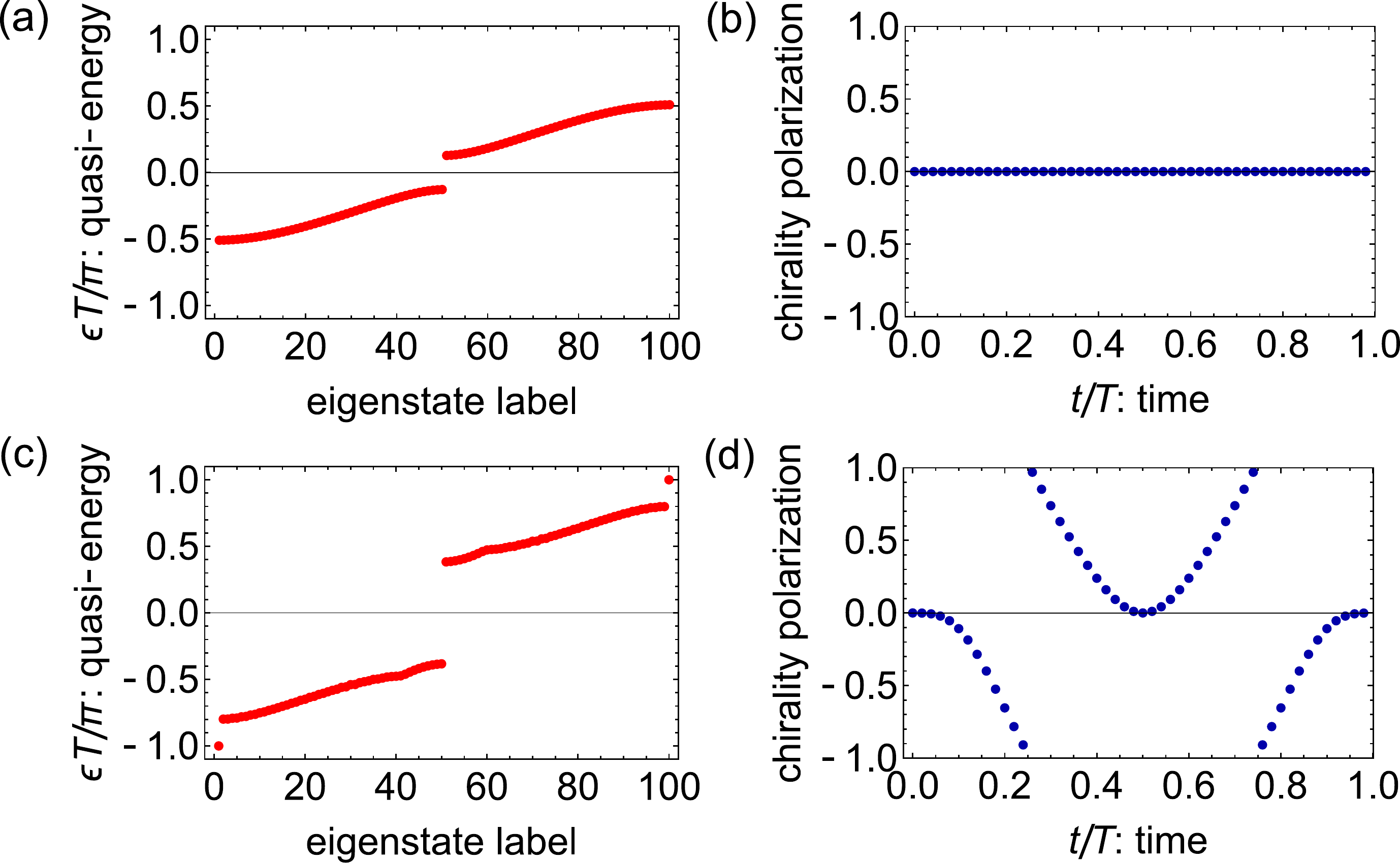}
\caption{(a) Quasienergy spectrum under the open boundary condition and (b) time evolution of total chiral polarization $\sum_\alpha P_\alpha^\Gamma(t)$ in a class AIII trivial Floquet insulator ($J_1=1, J_2=0.6, A=3, \omega=2\pi$). (c) Quasienergy spectrum under the open boundary condition and (d) time evolution of total chirality polarization $\sum_\alpha P_\alpha^\Gamma(t)$ in a class AIII anomalous Floquet topological insulator ($J_1=1, J_2=0.6, A=3, \omega=2\pi/3$). The number of lattice sites in (a) and (c) is set to $50$.}
\label{fig_AIII}
\end{figure}

To exemplify time evolution of chirality polarizations in an anomalous Floquet topological insulator, we numerically calculate it for a model proposed in Ref.~\onlinecite{Fruchart16}. The Hamiltonian is given by
\begin{equation}
H(k,t)=(J_1(t)+J_2\cos(k))\sigma_1+J_2\sin(k)\sigma_2,
\end{equation}
where $\sigma_j\ (j=1,2,3)$ are the Pauli matrices, and $J_1(t)=J_1+A\cos(\omega t)$. Here the chiral symmetry \eqref{eq_AIIIchiralsym} is satisfied for $\Gamma=\sigma_3$. 
In a trivial Floquet insulating phase with $W=0$, the quasienergy spectrum hosts no in-gap edge states under the open boundary condition [Fig.~\ref{fig_AIII} (a)], and the chirality polarization shows featureless time evolution [Fig.~\ref{fig_AIII} (b)]. On the other hand, in an anomalous Floquet topological insulator, a $\pi/T$-quasienergy edge state appears in the spectrum [Fig.~\ref{fig_AIII} (c)], and the chirality polarization is pumped by two during the half of the period, giving $W=-1$ [Fig.~\ref{fig_AIII} (d)]. We note that the chirality polarization is a bulk quantity and thus calculated under the periodic boundary condition. These numerical results clearly demonstrate that time evolution of chirality polarization captures the anomalous Floquet topological phase.

The nontrivial time evolution of chirality polarizations in class AIII anomalous Floquet topological insulators indicates a topological obstruction in Floquet states. Suppose that the time dependence of a Hamiltonian of a class AIII Floquet system can continuously be switched off without closing the quasienergy gap at $\pi/T$. Then Floquet states \eqref{eq_AIIIgauge} of the undriven system becomes time-independent, and thus the winding number \eqref{eq_AIIIwind_result2} vanishes. Therefore, a nonzero value of the winding number \eqref{eq_AIIIwind_result2} indicates an obstruction to removing the time dependence of a driven system under the chiral symmetry and a quasienergy gap at $\pi/T$.

\subsection{Class A in $d=2$}
Next, we consider two-dimensional systems. Here we focus on the simplest case without any symmetry, i.e., class A. 
An anomalous Floquet topological insulator of this class is characterized by a winding number\cite{Rudner13}
\begin{align}
W_3=&\int_0^T dt\int \frac{d^2\bm{k}}{24\pi^2}\varepsilon^{\mu\nu\lambda}
\mathrm{Tr}[(\tilde{U}^\dag\partial_{\mu}\tilde{U})(\tilde{U}^\dag\partial_{\nu}\tilde{U})(\tilde{U}^\dag\partial_{\lambda}\tilde{U})]
\label{eq_Awind3D}
\end{align}
where $\tilde{U}(\bm{k},t)$ is a deformed time-evolution operator \eqref{eq_deformevol}, $\mu,\nu,\lambda=0,1,2$, $\partial_{1,2}\equiv\partial_{k_{1,2}}$, and $\partial_0\equiv\partial_t$. 
If $W_3\neq 0$, a Floquet system hosts chiral edge states under the open boundary condition, even if the Chern numbers of Floquet bands vanish\cite{Rudner13}.

To find a topological obstruction in Floquet states of an anomalous Floquet insulator, we use a Hermitian matrix given by
\begin{equation}
H_U(\bm{k},t)\equiv
\begin{pmatrix}
0 & \tilde{U}(\bm{k},t)\\
\tilde{U}^\dag(\bm{k},t) & 0
\end{pmatrix},
\label{eq_AHU}
\end{equation}
which is used for the classification of anomalous Floquet topological insulators\cite{RoyHarper17}. A complete orthonormal set of eigenstates of the matrix \eqref{eq_AHU} is given by\cite{Ryu10}
\begin{equation}
\ket{\Xi^{\pm}_{\alpha}(\bm{k},t)}=\frac{1}{\sqrt{2}}
\begin{pmatrix}
\pm \tilde{U}(\bm{k},t)\ket{\varphi_\alpha}\\
\ket{\varphi_\alpha}
\end{pmatrix}
\ \ (\alpha=1,\cdots, N),
\end{equation}
where $\{\ket{\varphi_\alpha}\}_{\alpha=1}^N$ is a complete orthonormal set of the $N$-dimensional Hilbert space (here we assume that $\tilde{U}(\bm{k},t)$ is an $N\times N$ matrix). In particular, when we take Floquet states as a basis set, i.e. $\ket{\varphi_\alpha}=\ket{\Phi_\alpha(\bm{k},0)}$, we have
\begin{equation}
\ket{\Xi^{\pm}_{\alpha}(\bm{k},t)}=\frac{1}{\sqrt{2}}
\begin{pmatrix}
\pm \ket{\Phi_\alpha(\bm{k},t)}\\
\ket{\Phi_\alpha(\bm{k},0)}
\end{pmatrix},
\label{eq_extendedFloquet}
\end{equation}
since $\tilde{U}(\bm{k},t)\ket{\Phi_\alpha(\bm{k},0)}=\ket{\Phi_\alpha(\bm{k},t)}$ [see Eq.~\eqref{eq_evolFloquet2}].

Equation \eqref{eq_AHU} can be regarded as a Hamiltonian of a static class AIII insulator in three dimensions, where $t$ plays a role of a momentum in the third dimension. In fact, it has a chiral symmetry $\{ \Gamma_1, H_U(\bm{k},t)\}=0$ which is given by
\begin{equation}
\Gamma_1\equiv
\begin{pmatrix}
1_{N\times N} & 0 \\
0 & -1_{N\times N}
\end{pmatrix}.
\end{equation}
Since $H_U(\bm{k},t)\ket{\Xi^{\pm}_{\alpha}(\bm{k},t)}=\pm\ket{\Xi^{\pm}_{\alpha}(\bm{k},t)}$, the state $\ket{\Xi^+_{\alpha}(\bm{k},t)}$ ($\ket{\Xi^{-}_{\alpha}(\bm{k},t)}$) corresponds to a higher (lower) band. The topological invariant of the static class AIII insulator is given by the winding number \eqref{eq_Awind3D}. 
Furthermore, from the three-dimensional class AIII topological insulator \eqref{eq_AHU}, we can construct a class A topological insulator in four dimensions by\cite{TeoKane}
\begin{subequations}
\begin{align}
\tilde{H}(\bm{k},t,\theta)\equiv& \cos(\theta)H_U(\bm{k},t)+\sin(\theta)\Gamma_1\notag\\
=&
\begin{pmatrix}
\sin(\theta)1_{N\times N} & \cos(\theta)\tilde{U}(\bm{k},t)\\
\cos(\theta)\tilde{U}^\dag(\bm{k},t) & -\sin(\theta)1_{N\times N}
\end{pmatrix},
\label{eq_Hnonchiral}
\end{align}
for $\theta\in [-\pi/2,\pi/2]$, and 
\begin{equation}
\tilde{H}(\bm{k},t,\theta)\equiv
\begin{pmatrix}
\sin(\pi-\theta)1_{N\times N} & \cos(\pi-\theta)1_{N\times N}\\
\cos(\pi-\theta)1_{N\times N} & -\sin(\pi-\theta)1_{N\times N}
\end{pmatrix}
\label{eq_Hnonchiral2}
\end{equation}
\end{subequations}
for $\theta\in[-\pi,-\pi/2]\cup[\pi/2,\pi]$. 
In fact, the Hamiltonian $\tilde{H}(\bm{k},t,\theta)$ is periodic in $\theta\in[-\pi,\pi]$ and does not have the chiral symmetry associated with $\Gamma_1$. 
A four-dimensional insulator is characterized by a second Chern number $C_4$. 
Since the construction by Eqs.~\eqref{eq_Hnonchiral} and \eqref{eq_Hnonchiral2} keeps the topological equivalence of insulators, we have $C_4=W_3$ and the Hamiltonian $\tilde{H}(\bm{k},t,\theta)$ gives a class A topological insulator if $W_3\neq 0$.  

A four-dimensional topological insulator in class A can be characterized by nontrivial connectivity of a Wilson-loop spectrum of the occupied bands\cite{Taherinejad15}. 
Here a Wilson loop is defined as
\begin{equation}
\tilde{\mathcal{W}}_j(k_\perp, t,\theta)=\mathcal{P}\exp\left[i\int_{-\pi}^{\pi}dk_j\tilde{\mathcal{A}}_j(\bm{k},t,\theta)\right],
\end{equation}
where $\tilde{\mathcal{A}}_j^{\alpha\beta}(\bm{k},t,\theta)=\bra{\tilde{\Xi}_\alpha^-(\bm{k},t,\theta)}i\partial_{k_j}\ket{\tilde{\Xi}_\beta^-(\bm{k},t,\theta)}\ (\alpha,\beta=1,\cdots,N)$ is the non-Abelian Berry connection of Bloch states $\ket{\tilde{\Xi}_\alpha^-(\bm{k},t,\theta)}$ of the occupied bands of the four-dimensional insulator $\tilde{H}(\bm{k},t,\theta)$. Let $\exp[2\pi i\tilde{X}_\alpha(k_2,t,\theta)]\ (\alpha=1,\cdots, N)$ be eigenvalues of the Wilson loop $\tilde{\mathcal{W}}_1(k_2,t,\theta)$. Since $\tilde{X}_\alpha(k_2,t,\theta)$ is defined modulo 1, we may restrict it to $-1/2\leq \tilde{X}_\alpha(k_2,t,\theta)< 1/2$ without loss of generality and illustrate it with its copies $\tilde{X}_\alpha(k_2,t,\theta)+n\ (n\in\mathbb{Z})$. As a function of $k_2,t$, and $\theta$,  the Wilson-loop spectrum $\{ \tilde{X}_\alpha(k_2,t,\theta)\}_{\alpha=1}^N$ exhibits nontrivial connectivity with its neighbors $\{ \tilde{X}_\alpha(k_2,t,\theta)\pm 1\}_{\alpha=1}^N$ if $C_4\neq 0$, mimicking an energy spectrum of gapless surface states which emerge under the open boundary condition along the first direction\cite{Taherinejad14, Taherinejad15}. However, since $H_U(\bm{k},t)=\tilde{H}(\bm{k},t,\theta=0)$ gives a class AIII topological insulator, the gapless points of the surface states should be located at $\theta=0$, and thus the Wilson-loop spectrum must show the nontrivial connectivity at $\theta=0$. Therefore, we may restrict the calculation to $\theta=0$ and use
\begin{align}
\mathcal{W}_j(k_\perp, t)&\equiv\tilde{\mathcal{W}}_j(k_\perp, t,\theta=0)\notag\\
&=\mathcal{P}\exp\left[i\int_{-\pi}^{\pi}dk_j\mathcal{A}_j(\bm{k},t)\right],
\end{align}
where
\begin{align}
\mathcal{A}_j^{\alpha\beta}(\bm{k},t)\equiv&\tilde{\mathcal{A}}_j^{\alpha\beta}(\bm{k},t,\theta=0)\notag\\
=&\bra{\Xi_\alpha^-(\bm{k},t)}i\partial_{k_j}\ket{\Xi_\beta^-(\bm{k},t)}\notag\\
=&\frac{1}{2}\Bigl[\bra{\Phi_\alpha(\bm{k},t)}i\partial_{k_j}\ket{\Phi_\beta(\bm{k},t)}\notag\\
&+\bra{\Phi_\alpha(\bm{k},0)}i\partial_{k_j}\ket{\Phi_\beta(\bm{k},0)}\Bigr]
\end{align}
$(\alpha,\beta=1,\cdots,N)$ is the non-Abelian Berry connection which can be calculated from Floquet states via Eq.~\eqref{eq_extendedFloquet}. Eigenvalues of the Wilson loop $\mathcal{W}_1(k_2,t)$ is given by $\exp[2\pi iX_\alpha(k_2,t)]=\exp[2\pi i\tilde{X}_\alpha(k_2,t,\theta=0)]\ (\alpha=1,\cdots, N)$. 
Here $X_\alpha(k_2,t)$ can be regarded as a time-dependent hybrid Wannier center constructed from a generalized Floquet state $\ket{\Xi_\alpha^-(\bm{k},t)}$. 
For an anomalous Floquet topological insulator with $W_3\neq 0$, the Wilson-loop spectrum $\{ X_\alpha(k_2,t)\}_{\alpha=1}^N$ exhibits nontrivial connectivity with its neighbors $\{ X_\alpha(k_2,t)\pm 1\}_{\alpha=1}^N$ when $k_2$ and $t$ are swept over $[-\pi,\pi]\times[0,T]$.

\begin{figure}
\includegraphics[width=8.5cm]{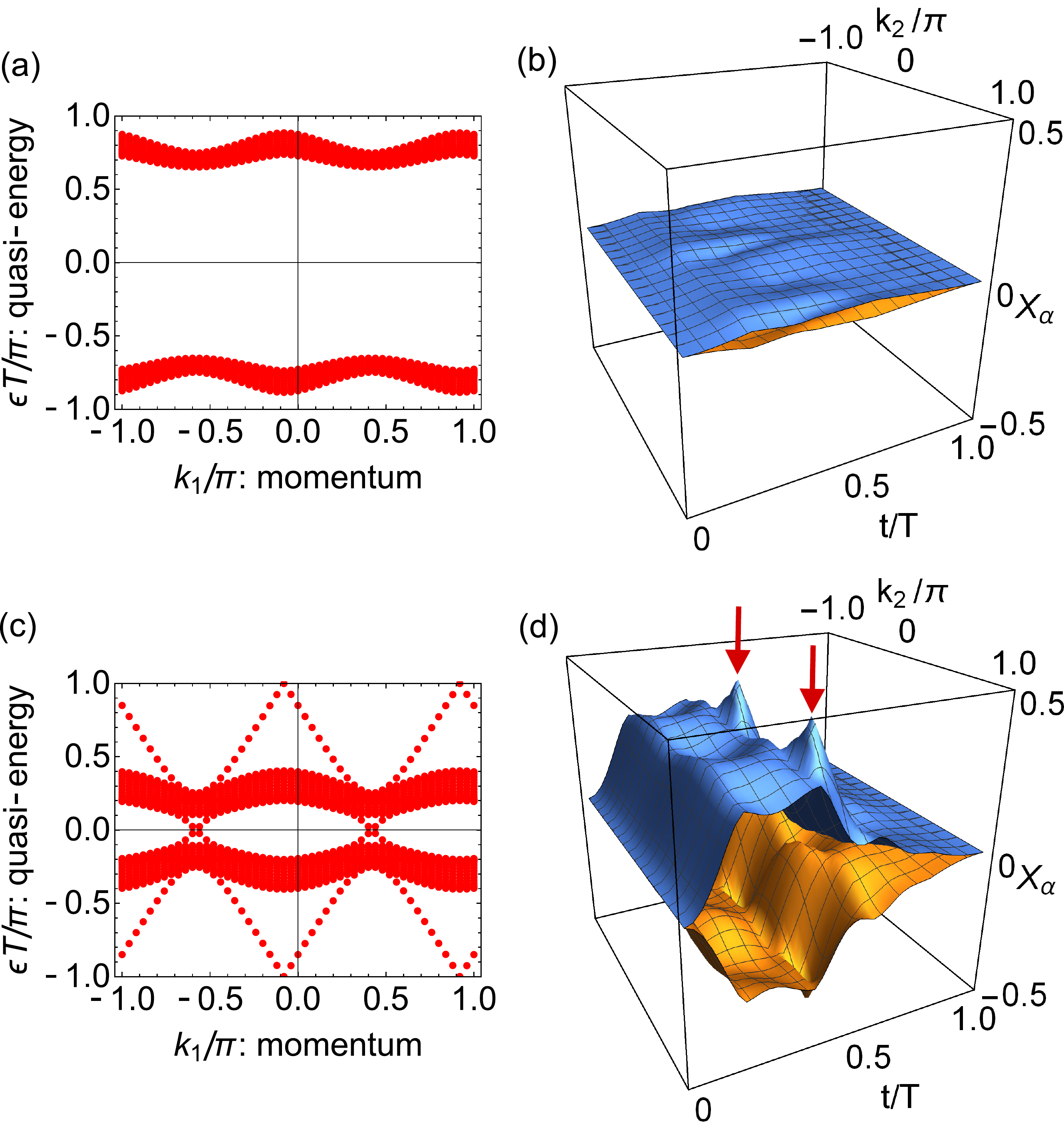}
\caption{(a) Quasienergy spectrum under the open boundary condition and (b) trajectory of Wannier centers $X(k_2,t)$ for a class A trivial Floquet insulator with $W_3=0$ ($J=0.5\pi/T, \delta=2.5\pi/T$). (c) Quasienergy spectrum under the open boundary condition and (d) trajectory of Wannier centers $X(k_2,t)$ for a class A anomalous Floquet topological insulator with $W_3=2$ ($J=2.5\pi/T, \delta=2.5\pi/T$). In (d), the arrows indicate points where the spectrum touches $1/2$.}
\label{fig_anomA}
\end{figure}

To check the validity of the above Wilson-loop characterization of an anomalous Floquet topological insulator, we perform a numerical calculation of a Wilson-loop spectrum. The model Hamiltonian is given by a five-step model in Ref.~\onlinecite{Rudner13}:
\begin{align}
H(\bm{k},t)=-\sum_{n=1}^4 J_n(t)(e^{i\bm{b}_n\cdot\bm{k}}\sigma^++e^{-i\bm{b}_n\cdot\bm{k}}\sigma^-)+\delta\sigma_3,
\end{align}
where $\bm{b}_1=-\bm{b}_3=(1,0), \bm{b}_2=-\bm{b}_4=(0,1)$, $\sigma^\pm=(\sigma_1\pm i\sigma_2)/2$, and
\begin{align}
J_n(t)&=J \ \ ((n-1)T/5\leq t < nT/5),\notag\\
J_n(t)&=0 \ \ (\mathrm{otherwise}).\notag
\end{align}
Here we consider a case where all the Floquet bands have vanishing Chern numbers. 
For a trivial Floquet topological insulator with $W_3=0$, the quasienergy spectrum under the open boundary condition hosts no edge states [Fig.~\ref{fig_anomA} (a)], and the Wilson-loop spectrum $\{ X_\alpha(k_2,t)\}_{\alpha=1,2}$ does not reach $\pm 1/2$ [Fig.~\ref{fig_anomA} (b)]. On the other hand, when $W_3\neq0$, an anomalous Floquet topological insulator exhibits chiral edge states under the open boundary even if the Chern numbers of the Floquet bands are zero [Fig.~\ref{fig_anomA} (c)]. In this case, the Wilson-loop spectrum touches at $X_\alpha=\pm 1/2$ and the number of the connecting points reflects the winding number $W_3$ [Fig.~\ref{fig_anomA} (d)]. Here we note that the Wilson loops are calculated under the periodic boundary condition, and thus creating boundaries is not necessary for diagnosing the topological phase. While the same phase has been detected by using the Aharonov-Anandan phase of \textit{edge} states in Ref.~\onlinecite{MondragonShem18}, our calculation shows that the topological information of the anomalous Floquet insulator is indeed encoded in the \textit{bulk} Floquet-Bloch states, resulting in the nontrivial connectivity in the Wilson-loop spectrum. 
We also note that an anomalous Floquet insulator in the same class is characterized by quantized orbital magnetization of Floquet states when the Floquet states are localized due to strong disorder\cite{Nathan17,Nathan19}. Since orbital magnetization for spatially extended Floquet-Bloch states still remains elusive, our result provides another complementary characterization of the anomalous Floquet insulator in the translationally invariant case.

The nontrivial connectivity of the Wilson-loop spectrum in Fig.~\ref{fig_anomA} (d) can be  interpreted as a topological obstruction in the Floquet states. To see this, we note that the Wilson-loop eigenvalues at $t=0$ are trivial [$X_\alpha(k_2,0)=0$] since $\mathcal{A}_j^{\alpha\beta}(\bm{k},0)=\bra{\Phi_\alpha(\bm{k},0)}i\partial_{k_j}\ket{\Phi_\beta(\bm{k},0)}$ and $\sum_\alpha \ket{\Phi_\alpha(\bm{k},0)}\bra{\Phi_\alpha(\bm{k},0)}=1$. This implies that an undriven insulator gives $X_\alpha(k_2,t)=0$. Thus, the nontrivial connectivity of the Wilson-loop spectrum indicates that one cannot remove the time dependence of the Floquet states by a continuous deformation. From these observations, we can conclude that an anomalous Floquet topological insulator is characterized by a topological obstruction to continuously switching off the driving without closing a quasienergy gap at $\pi/T$. 
Such an obstruction is a manifestation of the fact that an anomalous Floquet topological insulator originates from a nontrivial micromotion of Floquet states. Indeed, all the characterization of anomalous Floquet insulators proposed so far express such obstructions \cite{Rudner13, NathanRudner15, RoyHarper17, MondragonShem18}. Compared to the previous work, our message here is that the obstruction can be directly extracted from time dependence of the Wilson-loop spectra of Floquet-Bloch states.

\begin{figure}
\includegraphics[width=8.5cm]{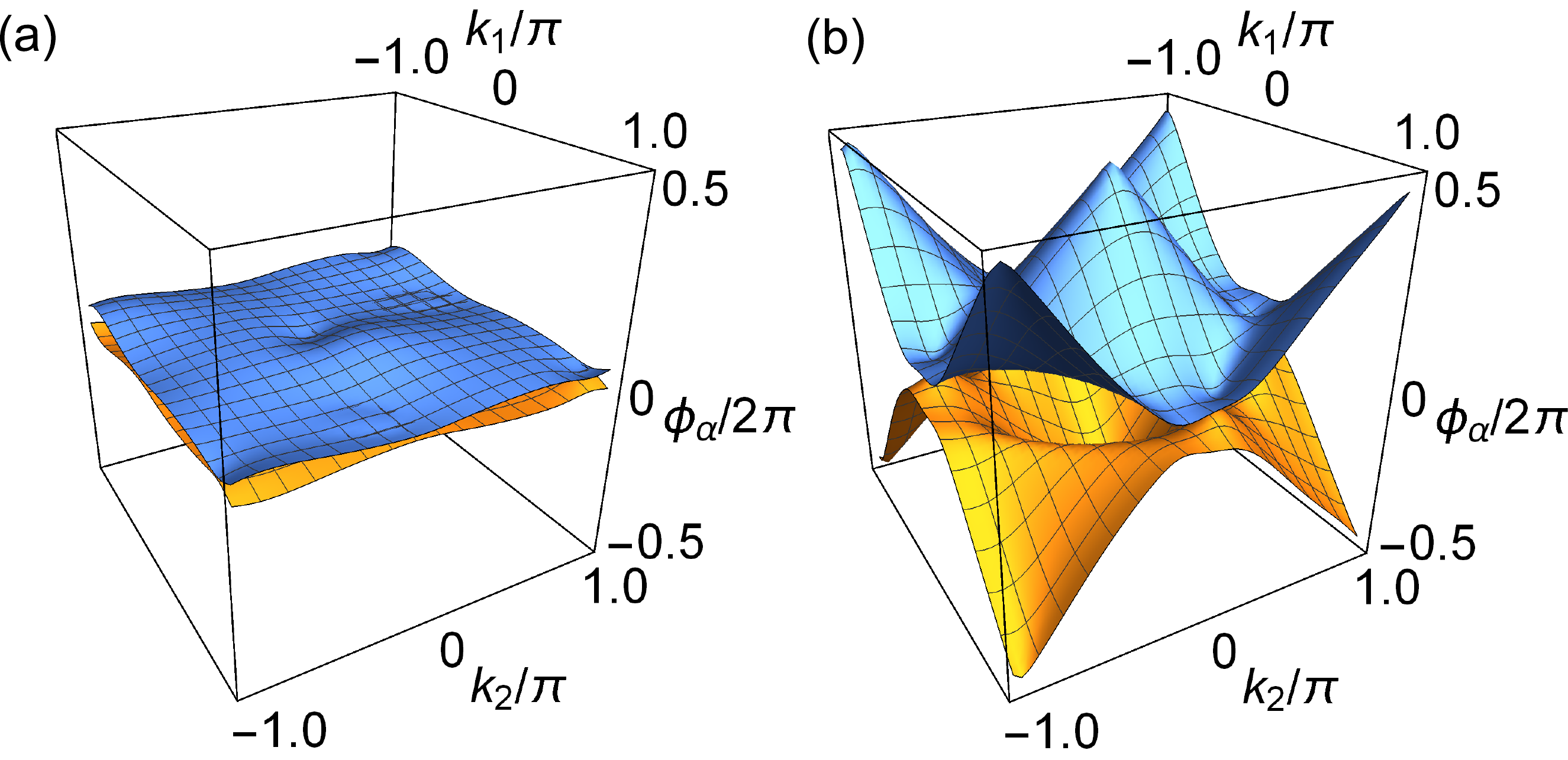}
\caption{Eigenvalues of the Wilson loop along the time direction for (a) a trivial Floquet insulator and (b) an anomalous Floquet topological insulator. The model and parameters are the same as in Fig.~\ref{fig_anomA}.}
\label{fig_anomA_Wiloop_t}
\end{figure}

Finally, we note that the same topological information can be also extracted from a Wilson loop along the time direction defined by
\begin{equation}
\mathcal{W}_t(\bm{k})\equiv\mathcal{T}\exp\left[i\int_0^T dt\mathcal{A}_t(\bm{k},t)\right],
\end{equation}
where
\begin{align}
\mathcal{A}_t^{\alpha\beta}(\bm{k},t)&\equiv\bra{\Xi_\alpha^-(\bm{k},t)}i\partial_t\ket{\Xi_\beta^-(\bm{k},t)}\notag\\
&=\frac{1}{2}\bra{\Phi_\alpha(\bm{k},t)}i\partial_t\ket{\Phi_\beta(\bm{k},t)}
\end{align}
$(\alpha,\beta=1,\cdots,N)$ is the non-Abelian Berry connection which is slightly different from that of the Floquet states $A_t^{\alpha\beta}(\bm{k},t)$ due to the factor $1/2$. In Fig.~\ref{fig_anomA_Wiloop_t}, we show the numerical results of the eigenvalues $\{\exp[i\phi_\alpha(\bm{k})]\}_{\alpha=1,2}$ of the Wilson loop $\mathcal{W}_t(\bm{k})$ for the model same as in Fig.~\ref{fig_anomA}. Whereas the Wilson-loop spectrum is featureless for the trivial Floquet insulator [Fig.~\ref{fig_anomA_Wiloop_t}(a)], the phase of the eigenvalues $\phi_\alpha(\bm{k})$ touches at $\pm\pi$ for the anomalous Floquet topological insulator  [Fig.~\ref{fig_anomA_Wiloop_t}(b)], manifesting the topological structure. The nontrivial connectivity of the Wilson-loop spectrum is consistent with the topological obstruction in the anomalous Floquet insulator, since an undriven insulator leads to $\phi_\alpha(\bm{k})=0$.

\section{Conclusion and Discussion\label{sec_discussion}}
In this paper, we have developed a state-based characterization of topological phases in periodically driven systems. Floquet topological states have been diagnosed by their surface states under the open boundary condition or topological invariants defined through time-evolution operators. However, here we have shown that topological information of periodically driven systems is indeed encoded in bulk Floquet-Bloch states, and it can be extracted from nontrivial connectivity of hybrid Wannier centers in momentum-time space. On the basis of the topological characterization of Floquet-Bloch states, we have clarified what kinds of topological obstructions exist in Floquet-Bloch states given by topological Floquet driving.

Let us summarize the results with highlighting the role of Wannier functions in all types of topological Floquet states. In a Floquet topological insulator characterized by an effective Hamiltonian $H_{\mathrm{eff}}(\bm{k})$, Floquet states at each period $\ket{\Phi_\alpha(\bm{k},0)}$ form a nontrivial vector bundle over the Brillouin zone. If there exists a topological obstruction to taking a continuous and periodic gauge of $\ket{\Phi_\alpha(\bm{k},0)}$, the Wannier function \eqref{eq_Wannier} at $t=0$ cannot be localized in real space. If the topology of $H_{\mathrm{eff}}(\bm{k})$ is trivial, one may make the initial Floquet state $\ket{\Phi_\alpha(\bm{k},0)}$ continuous and periodic in $\bm{k}$. In that case, $\ket{\psi_\alpha(\bm{k},t)}=U(\bm{k},t)\ket{\Phi_\alpha(\bm{k},0)}$ should be periodic in $\bm{k}$, but not in $t$. On the other hand, $\ket{\Phi_\alpha(\bm{k},t)}=e^{i\varepsilon_\alpha(\bm{k})t}\ket{\psi_\alpha(\bm{k},t)}$ is manifestly periodic in $t$. Whether or not $\ket{\Phi_\alpha(\bm{k},t)}$ can be periodic in $\bm{k}$ is determined from the topology of a Floquet operator. If the topology of a Floquet operator is nontrivial, Floquet states form a nontrivial vector bundle over the \textit{momentum-time} space. Consequently, one cannot take a momentum-periodic gauge of $\ket{\Phi_\alpha(\bm{k},t)}$, implying that the generalized Wannier function \eqref{eq_genWannier} cannot be localized in \textit{real and frequency} space. In fact, for the gapless quasienergy spectra given by topological Floquet operators in Figs.~\ref{fig_WannierThouless} (a) and (b), the quasienergy loses its periodicity as $\varepsilon_\alpha(k+2\pi)\neq\varepsilon_\alpha(k)$, indicating that $\ket{\Phi_\alpha(k,t)}$ is not periodic in $k$. In contrast to the above two types of Floquet topological states, anomalous Floquet topological insulators are characterized by a topological obstruction to continuously remove the time dependence of Wannier functions. 
The existence of such an obstruction is consistent with the previous work\cite{Rudner13, NathanRudner15, RoyHarper17, MondragonShem18}. The obstruction encoded in Floquet-Bloch states is  detected from nontrivial connectivity of properly defined Wilson-loop spectra as shown in Sec.~\ref{sec_anom}.

Furthermore, we emphasize that our approach based on gauge-invariant Wilson-loop spectra will be useful in numerical calculations of Floquet topological phases, since one can diagnose the topology of periodically driven systems from a data of Floquet states on discretized momentum and time. 
The efficient numerical calculation of a topological invariant using a Wilson loop has been extensively used for finding of (static) topological insulators in real materials\cite{Soluyanov11, Soluyanov11_2, Yu11}. 
Hence, we hope that our results not only serve to deepen the understanding of topological phases out of equilibrium, but also facilitate a future discovery of new Floquet topological phases.

An appealing, concrete, future pursuit entails the incorporation of crystalline symmetries. As shown by our formalism, the topology, as indexed by winding numbers, roughly is determined by symmetric values of time during the driving period $T$. Indeed, in class AIII for example, one focuses on $t=0$ and $t=T/2$ contributions. This is rather reminiscent of the symmetry classification in equilibrium, where one classifies phases according to quantities, such as irreducible representations, at high-symmetric momenta in the Brillouin zone \cite{Slager12, Kruthoff17, Po17}, which can similarly be evaluated using Wilson loop operators\cite{Bouhon18}. We are therefore optimistic that this construction can be lifted to comprise space-time points, e.g. high-symmetric momenta for both $t=0,T/2$ planes in the AIII (and AII) cases. Similarly, reformulating these notions into an elementary band representation analysis \cite{Bradlyn17}, is also anticipated to give access to additional non-equilibrium fragile topological phases \cite{Bouhon18, Po18}. 
As for the equilibrium phases we suspect that these crystalline topological quasi-energy band structures can similarly be probed by bulk-boundary correspondence \cite{ Slager15,Hatsugai93,Slager16,Essin11,Rhim18}, whereas the generalization of defect responses \cite{Ran09, Juricic12, Slager14, SlagerJPC, bbcweyl, TeoKane, Mesaros13, Imura11} of such crystalline phases to the Floquet setting provides for another interesting avenue. 

We also remark that characterization of Floquet states of Floquet topological superconductors is not covered in this paper. While the characterization in terms of Wannier functions may not be useful in superconductors, the geometric phase and the Wilson loops are still applicable to Floquet topological superconductors\cite{MondragonShem18}. It remains as a future issue to provide a complete topological characterization of Floquet states in all the Altland-Zirnbauer classes. 
Interestingly, a recently discovered Floquet topological phase beyond the K-theoretic classification\cite{MondragonShem18} is characterized by a topological invariant defined from Floquet-Bloch states.

A final possible research direction would be to generalize our formalism to the context of higher-order Floquet topological insulators \cite{Huang18, RodriguezVega18, Peng18}. In the equilibrium case
 such higher order topologies can be addressed by considering nested Wilson loops \cite{Benalcazar17}. Furthermore, the quantities can then again be related to symmetry indicators \cite{Khalaf18}.  Given the connection of our construction to the static case, it is reasonable to expect that this procedure can be similarly generalized. Given the these concrete future perspectives, we are convinced that our work will serve as an inspiration for future work.

\begin{acknowledgments} 
We are grateful to A.~Bouhon, S.~Kitamura, and I.~Maruyama for valuable discussions. M.N. was supported by RIKEN Special Postdoctoral Researcher Program. R.-J.~S appreciatively acknowledges funding via A. Vishwanath from the Center for Advancement of Topological Semimetals, an Energy Frontier Research Center funded by the U.S. Department of Energy Office of Science, Office of Basic Energy Sciences, through 
Ames Laboratory under Contract No. DE-AC02-07CH11358. S.H. was supported by the Japan Society for the Promotion of Science (JSPS) through Program for Leading Graduate Schools (ALPS) and by JSPS Grant No.~JP16J03619.
\end{acknowledgments}

\appendix

\section{Non-adiabatic topological pumping\label{sec_pump}}
\begin{figure}
\includegraphics[width=8.5cm]{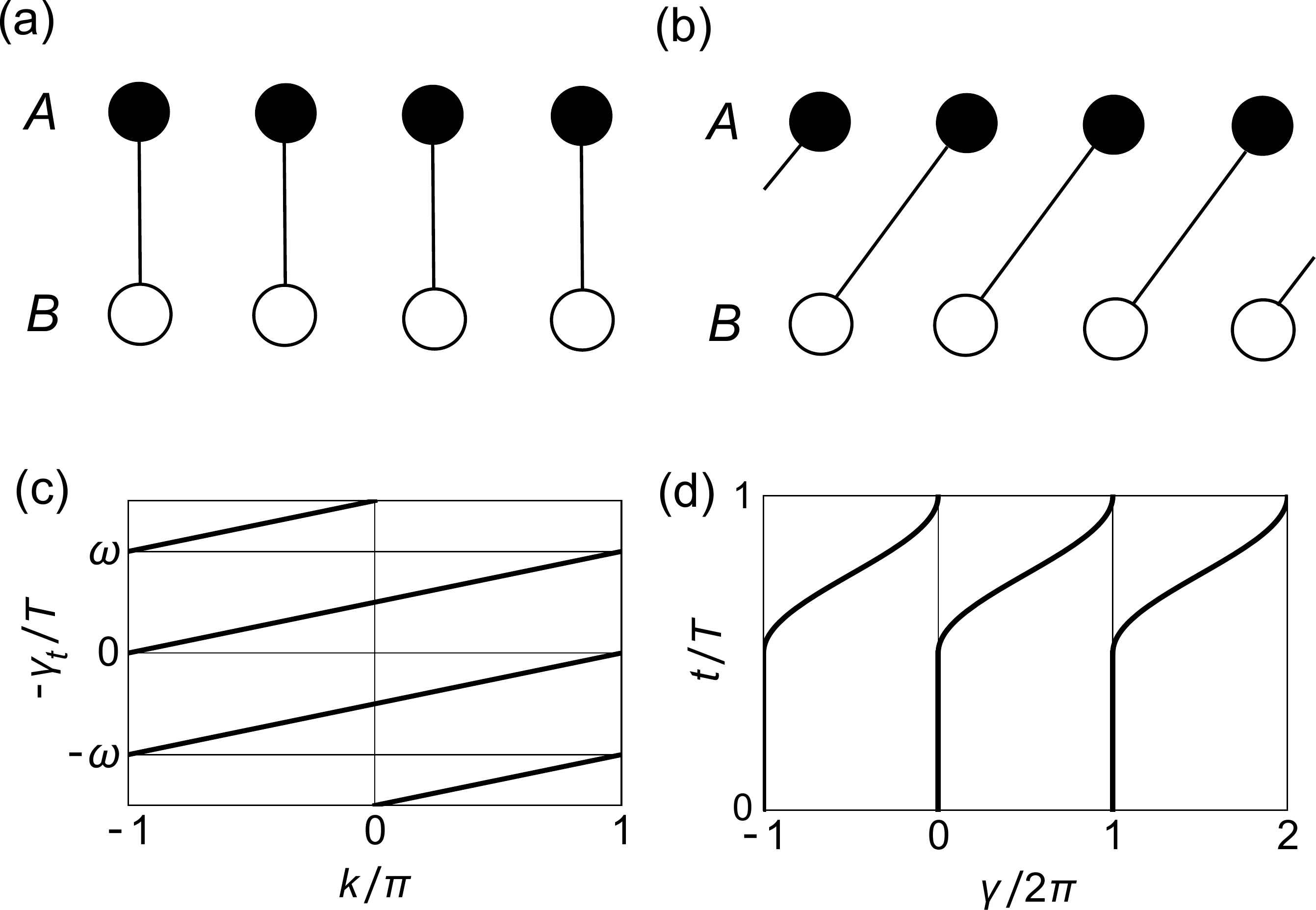}
\caption{(a) (b) Schematic illustration of the Hamiltonian $H_1$ [(a)] and $H_2$ [(b)] [Eq.~\eqref{eq_nonad_model}]. The black (white) dots denote lattice sites of sublattice A (B), and the lines represent the hopping between sites. (c) The Aharonov-Anandan phase $\gamma_{t}^A(k)$ (mod $2\pi$) in the non-adiabatic Thouless pumping. (d) The Berry phase $\gamma^A(t)$ (mod $2\pi$) in the non-adiabatic Thouless pumping.}
\label{fig_nonadThouless}
\end{figure}

In this appendix, for a demonstration of the result presented in Sec.~\ref{sec_gaplessAd1}, we calculate the Berry phase and the Aharonov-Anandan phase of Floquet states using an exactly solvable model of non-adiabatic Thouless pumping\cite{Budich17,Mizuta18}. The model Hamiltonian is given by
\begin{align}
H(t)=
\begin{cases}
H_1 & (0\leq t< T/2)\\
H_2 & (T/2\leq t< T)
\end{cases}
\label{eq_nonad_model}
\end{align}
where
\begin{subequations}
\begin{align}
H_1&=-\frac{\pi}{T}\sum_j(c_{j,A}^\dag c_{j,B}+\mathrm{h.c.}),\\
H_2&=-\frac{\pi}{T}\sum_j(c_{j+1,A}^\dag c_{j,B}+\mathrm{h.c.}).
\end{align}
\end{subequations}
Here we consider a one-dimensional ladder system, and $c_{j,\alpha}$ denotes an annihilation operator of a fermion at site $j$ in sublattice $\alpha$ [see Figs.~\ref{fig_nonadThouless} (a) and \ref{fig_nonadThouless} (b)]. After the Fourier transformation, the Hamiltonian reads
\begin{subequations}
\begin{align}
H_1&=\sum_k (c_{k,A}^\dag, c_{k,B}^\dag)H_1(k)
\begin{pmatrix}
c_{k,A}\\
c_{k,B}
\end{pmatrix},\\
H_2&=\sum_k (c_{k,A}^\dag, c_{k,B}^\dag)H_2(k)
\begin{pmatrix}
c_{k,A}\\
c_{k,B}
\end{pmatrix},
\end{align}
and
\begin{align}
H_1(k)&=-\frac{\pi}{T}
\begin{pmatrix}
0 & 1 \\
1 & 0
\end{pmatrix},\\
H_2(k)&=-\frac{\pi}{T}
\begin{pmatrix}
0 & e^{-ik} \\
e^{ik} & 0
\end{pmatrix},
\end{align}
\end{subequations}
where $c_{k,\alpha}$ is an annihilation operator of a fermion in sublattice $\alpha$ with momentum $k$. The single-particle time-evolution operator of this model is then easily calculated as
\begin{align}
U(k,t)=
\begin{cases}
e^{-iH_1(k)t} & (0\leq t \leq T/2)\\
e^{-iH_2(k)(t-T/2)}e^{-iH_1(k)T/2} & (T/2\leq t \leq T)
\end{cases}
\end{align}
where
\begin{subequations}
\begin{align}
e^{-iH_1(k)t}=&
\begin{pmatrix}
\cos\left(\frac{\pi t}{T}\right) & i\sin\left(\frac{\pi t}{T}\right)\\
i\sin\left(\frac{\pi t}{T}\right) & \cos\left(\frac{\pi t}{T}\right)
\end{pmatrix},\\
e^{-iH_2(k)t}=&
\begin{pmatrix}
\cos\left(\frac{\pi t}{T}\right) & i\sin\left(\frac{\pi t}{T}\right)e^{-ik}\\
i\sin\left(\frac{\pi t}{T}\right)e^{ik} & \cos\left(\frac{\pi t}{T}\right)
\end{pmatrix}.
\end{align}
\end{subequations}
Thus, the Floquet operator is given by
\begin{align}
U(k)&=U(k,T)\notag\\
&=
\begin{pmatrix}
-e^{-ik} & 0\\
0 & -e^{ik}
\end{pmatrix}.
\end{align}
Note that the Floquet operator has a block-diagonal structure \eqref{eq_block} thanks to the fine-tuning of the driving protocol, whereas the time-evolution operator $U(k,t)$ mixes the two sublattice degrees of freedom in intermediate time. The Floquet states are given by
\begin{align}
\ket{\Phi_A(k,t)}=
\begin{cases}
e^{i\varepsilon_A(k)t}\begin{pmatrix}\cos\left(\frac{\pi t}{T}\right)\\ i\sin\left(\frac{\pi t}{T}\right) \end{pmatrix} \\
\hspace{75pt} (0\leq t \leq T/2)\\
e^{i\varepsilon_A(k)t}\begin{pmatrix}-\sin\left(\frac{\pi t}{T}-\frac{\pi}{2}\right)e^{-ik}\\ i\cos\left(\frac{\pi t}{T}-\frac{\pi}{2}\right) \end{pmatrix}\\
\hspace{75pt}  (T/2\leq t\leq T)
\end{cases}
\label{eq_FloquetstateA}
\end{align}
with quasienergy $\varepsilon_A(k)=(k+\pi)/T$, and
\begin{align}
\ket{\Phi_B(k,t)}=
\begin{cases}
e^{i\varepsilon_B(k)t}\begin{pmatrix}i\sin\left(\frac{\pi t}{T}\right) \\ \cos\left(\frac{\pi t}{T}\right)\end{pmatrix} \\
\hspace{75pt} (0\leq t \leq T/2)\\
e^{i\varepsilon_B(k)t}\begin{pmatrix}i\cos\left(\frac{\pi t}{T}-\frac{\pi}{2}\right)\\ -\sin\left(\frac{\pi t}{T}-\frac{\pi}{2}\right)e^{ik} \end{pmatrix}\\
\hspace{75pt}  (T/2\leq t\leq T)
\end{cases}
\end{align}
with quasienergy $\varepsilon_B(k)=(-k+\pi)/T$. This result indicates that the fermions initially in sublattice A (B) shows the Thouless pumping and the chiral-fermion quasienergy spectrum with the winding number [Eq.~\eqref{eq_1Dwind}] $+1$ ($-1$).

Using Eq.~\eqref{eq_FloquetstateA}, the Aharonov-Anandan phase of the Floquet state $\ket{\Phi_A(k,t)}$ is obtained as
\begin{align}
\gamma_t^A(k)=&\int_0^T dt\bra{\Phi_A(k,t)}i\partial_t\ket{\Phi_A(k,t)}\notag\\
=&-k-\pi,
\end{align}
which means that the quasienergy contains only the geometric-phase contribution in Eq.~\eqref{eq_quasienergyAA} (i.e. the dynamical phase vanishes). Similarly, the Berry phase of the Floquet state is calculated as
\begin{align}
\gamma^A(t)=&\int_{-\pi}^\pi dk\bra{\psi_A(k,t)}i\partial_k\ket{\psi_A(k.t)}\notag\\
=&
\begin{cases}
0 & (0\leq t\leq T/2)\\
2\pi\sin^2\left(\frac{\pi t}{T}-\frac{\pi}{2}\right) & (T/2 \leq t \leq T)
\end{cases}
\end{align}
where $\ket{\psi_A(k,t)}=e^{-i\varepsilon_A(k)t}\ket{\Phi_A(k,t)}$. The trajectories of hybrid Wannier centers are illustrated in Figs.~\ref{fig_nonadThouless} (c) and \ref{fig_nonadThouless} (d), which are consistent with the result in Sec.~\ref{sec_gaplessAd1}.

\section{Symmetries in periodically driven systems\label{sec_sym}}
In this appendix, we summarize symmetries in periodically driven systems\cite{Kitagawa10, NathanRudner15, Carpentier15, RoyHarper17} according to the Altland-Zirnbauer symmetry class \cite{Schnyder08, Kitaev09}. The time-reversal symmetry is written as
\begin{equation}
\Theta H(\bm{k},t)\Theta^{-1}=H(-\bm{k},-t)\label{eq_defTRS}
\end{equation}
with an antiunitary operator $\Theta$ which satisfies $\Theta^2=\pm1$. In periodically driven superconductors, the particle-hole symmetry
\begin{equation}
C H(\bm{k},t)C^{-1}=-H(-\bm{k},t)\label{eq_defPHS}
\end{equation}
is satisfied, where $C$ is another antiunitary operator with $C^2=\pm1$. As a third symmetry, the chiral symmetry is given by
\begin{equation}
\Gamma H(\bm{k},t)\Gamma^{-1}=-H(\bm{k},-t)\label{eq_defCS}
\end{equation}
with a unitary operator $\Gamma$. Each symmetry can be translated into that of the time-evolution operator as\cite{Kitagawa10, RoyHarper17}
\begin{align}
\Theta U(\bm{k},t)\Theta^{-1} &= U(-\bm{k},T-t)U^\dag(-\bm{k},T),\label{eq_TRSevol}\\
C U(\bm{k},t)C^{-1}&=U(-\bm{k},t),\\
\Gamma U(\bm{k},t)\Gamma^{-1} &=U(\bm{k},T-t)U^\dag(\bm{k},T).
\end{align}
By setting $t=T$ in these equations, we obtain symmetries of the Floquet operator as
\begin{align}
\Theta U(\bm{k})\Theta^{-1} &= U^\dag(-\bm{k}),\label{eq_TRSFloq}\\
C U(\bm{k})C^{-1}&=U(-\bm{k}),\\
\Gamma U(\bm{k})\Gamma^{-1} &=U^\dag(\bm{k}).
\end{align}
The above symmetries of the Floquet operator lead to symmetries of the effective Hamiltonian
\begin{align}
\Theta H_\mathrm{eff}(\bm{k})\Theta^{-1}&=H_\mathrm{eff}(-\bm{k}),\\
CH_\mathrm{eff}(\bm{k})C^{-1}&=-H_\mathrm{eff}(-\bm{k}),\\
\Gamma H_\mathrm{eff}(\bm{k})\Gamma^{-1}&=-H_\mathrm{eff}(\bm{k}),
\end{align}
which are consistent with the Altland-Zirnbauer symmetry class of static Hamiltonians. In the main text, we consider three symmetry classes; class A means absence of any symmetries of Eqs.~\eqref{eq_defTRS}-\eqref{eq_defCS}, class AII has the time-reversal symmetry \eqref{eq_defTRS} with $\Theta^2=-1$, and class AIII has the chiral symmetry \eqref{eq_defCS}.

\section{Derivation of Eq.~\eqref{eq_AIITI}\label{sec_AIIinv}}
Here we show the equality between Eq.~\eqref{eq_AIIgapless} and the right hand side of Eq.~\eqref{eq_AIITI}. For simplicity, we consider a two-band case in which the time-reversal operator is given by $\Theta=i\sigma_2K$. Here $\sigma_2$ is the Pauli matrix. In this case, we have $V_\Theta = i\sigma_2$ and
\begin{equation}
V_\Theta^\dag U_1(k_*)=
\begin{pmatrix}
0 & -\exp[-i\varepsilon(k_*)T]\\
\exp[-i\varepsilon(k_*)T] & 0
\end{pmatrix},
\label{eq_twobandmat}
\end{equation}
where $k_*=0$ or $\pi$. Equation \eqref{eq_twobandmat} is an antisymmetric matrix due to the Kramers degeneracy $\varepsilon_1(k_*)=\varepsilon_2(k_*)\equiv\varepsilon(k_*)$. In a general $N$-band case, one can take a gauge in which $V_\Theta^\dag U_1(k_*)$ is decomposed into blocks for each Kramars pair like the right hand side of Eq.~\eqref{eq_twobandmat}. 
The $\mathbb{Z}_2$ topological invariant \eqref{eq_AIIgapless} for the Floquet operator reads
\begin{equation}
(-1)^{\nu}=\frac{\exp[-i\varepsilon(0)T]}{\sqrt{(\exp[-i\varepsilon(0)T])^2}}\frac{\exp[-i\varepsilon(\pi)T]}{\sqrt{(\exp[-i\varepsilon(\pi)T])^2}},
\end{equation}
since the Pfaffian is given by $\mathrm{Pf}[V_\Theta^\dag U_1(k)]=-\exp[-i\varepsilon(k)T]$. On the other hand, the $w$ matrix in the two-band case is
\begin{equation}
w(k,t)=
\begin{pmatrix}
0 & w_{12}(k,t)\\
-w_{12}(k,t) & 0
\end{pmatrix}
\end{equation}
at $(k,t)=(0,0),(0,T/2),(\pi,0)$, and $(\pi,T/2)$. The right hand side of Eq.~\eqref{eq_AIITI} is therefore
\begin{align}
\frac{w_{12}(0,0)}{\sqrt{w_{12}(0,0)^2}}\frac{w_{12}(\pi,0)}{\sqrt{w_{12}(\pi,0)^2}}
\frac{w_{12}(0,T/2)}{\sqrt{w_{12}(0,T/2)^2}}\frac{w_{12}(\pi,T/2)}{\sqrt{w_{12}(\pi,T/2)^2}}.
\label{eq_Pfw}
\end{align}
From the time-reversal symmetry \eqref{eq_TRSevol}, we have
\begin{equation}
\Theta U_1(k,T/2)\Theta^{-1}=U_1(-k,T/2)U_1^\dag(-k).
\end{equation}
Using this relation, we obtain
\begin{align}
&w_{12}(0,T/2)=\bra{\psi_1(0,T/2)}\Theta\ket{\psi_2(0,T/2)}\notag\\
=&\bra{\psi_1(0,0)}U_1^\dag(0,T/2)\Theta U_1(0,T/2)\ket{\psi_2(0,0)}\notag\\
=&\bra{\psi_1(0,0)}U_1^\dag(0,T/2)U_1(0,T/2)U_1^\dag(0)\Theta\ket{\psi_2(0,0)}\notag\\
=&\bra{\psi_1(0,0)}U_1^\dag(0)\Theta\ket{\psi_2(0,0)}\notag\\
=&\exp[i\varepsilon(0)T]w_{12}(0,0),
\end{align}
and similarly $w_{12}(\pi,T/2)=\exp[i\varepsilon(\pi)T]w_{12}(\pi,0)$. Therefore, we arrive at
\begin{equation}
\mathrm{Eq}.~\eqref{eq_Pfw}~=\frac{\exp[i\varepsilon(0)T]}{\sqrt{(\exp[i\varepsilon(0)T])^2}}\frac{\exp[i\varepsilon(\pi)T]}{\sqrt{(\exp[i\varepsilon(\pi)T])^2}},
\end{equation}
which completes the proof of Eq.~\eqref{eq_AIITI}.

\bibliography{FloquetWilson_ref.bib}

\begin{thebibliography}{125}%
\makeatletter
\providecommand \@ifxundefined [1]{%
 \@ifx{#1\undefined}
}%
\providecommand \@ifnum [1]{%
 \ifnum #1\expandafter \@firstoftwo
 \else \expandafter \@secondoftwo
 \fi
}%
\providecommand \@ifx [1]{%
 \ifx #1\expandafter \@firstoftwo
 \else \expandafter \@secondoftwo
 \fi
}%
\providecommand \natexlab [1]{#1}%
\providecommand \enquote  [1]{``#1''}%
\providecommand \bibnamefont  [1]{#1}%
\providecommand \bibfnamefont [1]{#1}%
\providecommand \citenamefont [1]{#1}%
\providecommand \href@noop [0]{\@secondoftwo}%
\providecommand \href [0]{\begingroup \@sanitize@url \@href}%
\providecommand \@href[1]{\@@startlink{#1}\@@href}%
\providecommand \@@href[1]{\endgroup#1\@@endlink}%
\providecommand \@sanitize@url [0]{\catcode `\\12\catcode `\$12\catcode
  `\&12\catcode `\#12\catcode `\^12\catcode `\_12\catcode `\%12\relax}%
\providecommand \@@startlink[1]{}%
\providecommand \@@endlink[0]{}%
\providecommand \url  [0]{\begingroup\@sanitize@url \@url }%
\providecommand \@url [1]{\endgroup\@href {#1}{\urlprefix }}%
\providecommand \urlprefix  [0]{URL }%
\providecommand \Eprint [0]{\href }%
\providecommand \doibase [0]{http://dx.doi.org/}%
\providecommand \selectlanguage [0]{\@gobble}%
\providecommand \bibinfo  [0]{\@secondoftwo}%
\providecommand \bibfield  [0]{\@secondoftwo}%
\providecommand \translation [1]{[#1]}%
\providecommand \BibitemOpen [0]{}%
\providecommand \bibitemStop [0]{}%
\providecommand \bibitemNoStop [0]{.\EOS\space}%
\providecommand \EOS [0]{\spacefactor3000\relax}%
\providecommand \BibitemShut  [1]{\csname bibitem#1\endcsname}%
\let\auto@bib@innerbib\@empty
\bibitem [{\citenamefont {Klitzing}\ \emph {et~al.}(1980)\citenamefont
  {Klitzing}, \citenamefont {Dorda},\ and\ \citenamefont
  {Pepper}}]{Klitzing80}%
  \BibitemOpen
  \bibfield  {author} {\bibinfo {author} {\bibfnamefont {K.~v.}\ \bibnamefont
  {Klitzing}}, \bibinfo {author} {\bibfnamefont {G.}~\bibnamefont {Dorda}}, \
  and\ \bibinfo {author} {\bibfnamefont {M.}~\bibnamefont {Pepper}},\ }\href
  {\doibase 10.1103/PhysRevLett.45.494} {\bibfield  {journal} {\bibinfo
  {journal} {Phys. Rev. Lett.}\ }\textbf {\bibinfo {volume} {45}},\ \bibinfo
  {pages} {494} (\bibinfo {year} {1980})}\BibitemShut {NoStop}%
\bibitem [{\citenamefont {Tsui}\ \emph {et~al.}(1982)\citenamefont {Tsui},
  \citenamefont {Stormer},\ and\ \citenamefont {Gossard}}]{Tsui82}%
  \BibitemOpen
  \bibfield  {author} {\bibinfo {author} {\bibfnamefont {D.~C.}\ \bibnamefont
  {Tsui}}, \bibinfo {author} {\bibfnamefont {H.~L.}\ \bibnamefont {Stormer}}, \
  and\ \bibinfo {author} {\bibfnamefont {A.~C.}\ \bibnamefont {Gossard}},\
  }\href {\doibase 10.1103/PhysRevLett.48.1559} {\bibfield  {journal} {\bibinfo
   {journal} {Phys. Rev. Lett.}\ }\textbf {\bibinfo {volume} {48}},\ \bibinfo
  {pages} {1559} (\bibinfo {year} {1982})}\BibitemShut {NoStop}%
\bibitem [{\citenamefont {Laughlin}(1983)}]{Laughlin83}%
  \BibitemOpen
  \bibfield  {author} {\bibinfo {author} {\bibfnamefont {R.~B.}\ \bibnamefont
  {Laughlin}},\ }\href {\doibase 10.1103/PhysRevLett.50.1395} {\bibfield
  {journal} {\bibinfo  {journal} {Phys. Rev. Lett.}\ }\textbf {\bibinfo
  {volume} {50}},\ \bibinfo {pages} {1395} (\bibinfo {year}
  {1983})}\BibitemShut {NoStop}%
\bibitem [{\citenamefont {Hasan}\ and\ \citenamefont {Kane}(2010)}]{Hasan10}%
  \BibitemOpen
  \bibfield  {author} {\bibinfo {author} {\bibfnamefont {M.~Z.}\ \bibnamefont
  {Hasan}}\ and\ \bibinfo {author} {\bibfnamefont {C.~L.}\ \bibnamefont
  {Kane}},\ }\href {\doibase 10.1103/RevModPhys.82.3045} {\bibfield  {journal}
  {\bibinfo  {journal} {Rev. Mod. Phys.}\ }\textbf {\bibinfo {volume} {82}},\
  \bibinfo {pages} {3045} (\bibinfo {year} {2010})}\BibitemShut {NoStop}%
\bibitem [{\citenamefont {Qi}\ and\ \citenamefont {Zhang}(2011)}]{Qi11}%
  \BibitemOpen
  \bibfield  {author} {\bibinfo {author} {\bibfnamefont {X.-L.}\ \bibnamefont
  {Qi}}\ and\ \bibinfo {author} {\bibfnamefont {S.-C.}\ \bibnamefont {Zhang}},\
  }\href {\doibase 10.1103/RevModPhys.83.1057} {\bibfield  {journal} {\bibinfo
  {journal} {Rev. Mod. Phys.}\ }\textbf {\bibinfo {volume} {83}},\ \bibinfo
  {pages} {1057} (\bibinfo {year} {2011})}\BibitemShut {NoStop}%
\bibitem [{\citenamefont {Schnyder}\ \emph {et~al.}(2008)\citenamefont
  {Schnyder}, \citenamefont {Ryu}, \citenamefont {Furusaki},\ and\
  \citenamefont {Ludwig}}]{Schnyder08}%
  \BibitemOpen
  \bibfield  {author} {\bibinfo {author} {\bibfnamefont {A.~P.}\ \bibnamefont
  {Schnyder}}, \bibinfo {author} {\bibfnamefont {S.}~\bibnamefont {Ryu}},
  \bibinfo {author} {\bibfnamefont {A.}~\bibnamefont {Furusaki}}, \ and\
  \bibinfo {author} {\bibfnamefont {A.~W.~W.}\ \bibnamefont {Ludwig}},\ }\href
  {\doibase 10.1103/PhysRevB.78.195125} {\bibfield  {journal} {\bibinfo
  {journal} {Phys. Rev. B}\ }\textbf {\bibinfo {volume} {78}},\ \bibinfo
  {pages} {195125} (\bibinfo {year} {2008})}\BibitemShut {NoStop}%
\bibitem [{\citenamefont {Kitaev}(2009)}]{Kitaev09}%
  \BibitemOpen
  \bibfield  {author} {\bibinfo {author} {\bibfnamefont {A.}~\bibnamefont
  {Kitaev}},\ }\href {\doibase 10.1063/1.3149495} {\bibfield  {journal}
  {\bibinfo  {journal} {AIP Conference Proceedings}\ }\textbf {\bibinfo
  {volume} {1134}},\ \bibinfo {pages} {22} (\bibinfo {year}
  {2009})}\BibitemShut {NoStop}%
\bibitem [{\citenamefont {Fu}(2011)}]{Fu11}%
  \BibitemOpen
  \bibfield  {author} {\bibinfo {author} {\bibfnamefont {L.}~\bibnamefont
  {Fu}},\ }\href {\doibase 10.1103/PhysRevLett.106.106802} {\bibfield
  {journal} {\bibinfo  {journal} {Phys. Rev. Lett.}\ }\textbf {\bibinfo
  {volume} {106}},\ \bibinfo {pages} {106802} (\bibinfo {year}
  {2011})}\BibitemShut {NoStop}%
\bibitem [{\citenamefont {Slager}\ \emph {et~al.}(2012)\citenamefont {Slager},
  \citenamefont {Mesaros}, \citenamefont {Juri{\v c}i{\'c}},\ and\
  \citenamefont {Zaanen}}]{Slager12}%
  \BibitemOpen
  \bibfield  {author} {\bibinfo {author} {\bibfnamefont {R.-J.}\ \bibnamefont
  {Slager}}, \bibinfo {author} {\bibfnamefont {A.}~\bibnamefont {Mesaros}},
  \bibinfo {author} {\bibfnamefont {V.}~\bibnamefont {Juri{\v c}i{\'c}}}, \
  and\ \bibinfo {author} {\bibfnamefont {J.}~\bibnamefont {Zaanen}},\ }\href
  {http://dx.doi.org/10.1038/nphys2513} {\bibfield  {journal} {\bibinfo
  {journal} {Nature Physics}\ }\textbf {\bibinfo {volume} {9}},\ \bibinfo
  {pages} {98} (\bibinfo {year} {2012})}\BibitemShut {NoStop}%
\bibitem [{\citenamefont {Shiozaki}\ and\ \citenamefont
  {Sato}(2014)}]{Shiozaki14}%
  \BibitemOpen
  \bibfield  {author} {\bibinfo {author} {\bibfnamefont {K.}~\bibnamefont
  {Shiozaki}}\ and\ \bibinfo {author} {\bibfnamefont {M.}~\bibnamefont
  {Sato}},\ }\href {\doibase 10.1103/PhysRevB.90.165114} {\bibfield  {journal}
  {\bibinfo  {journal} {Phys. Rev. B}\ }\textbf {\bibinfo {volume} {90}},\
  \bibinfo {pages} {165114} (\bibinfo {year} {2014})}\BibitemShut {NoStop}%
\bibitem [{\citenamefont {Kruthoff}\ \emph {et~al.}(2017)\citenamefont
  {Kruthoff}, \citenamefont {de~Boer}, \citenamefont {van Wezel}, \citenamefont
  {Kane},\ and\ \citenamefont {Slager}}]{Kruthoff17}%
  \BibitemOpen
  \bibfield  {author} {\bibinfo {author} {\bibfnamefont {J.}~\bibnamefont
  {Kruthoff}}, \bibinfo {author} {\bibfnamefont {J.}~\bibnamefont {de~Boer}},
  \bibinfo {author} {\bibfnamefont {J.}~\bibnamefont {van Wezel}}, \bibinfo
  {author} {\bibfnamefont {C.~L.}\ \bibnamefont {Kane}}, \ and\ \bibinfo
  {author} {\bibfnamefont {R.-J.}\ \bibnamefont {Slager}},\ }\href {\doibase
  10.1103/PhysRevX.7.041069} {\bibfield  {journal} {\bibinfo  {journal} {Phys.
  Rev. X}\ }\textbf {\bibinfo {volume} {7}},\ \bibinfo {pages} {041069}
  (\bibinfo {year} {2017})}\BibitemShut {NoStop}%
\bibitem [{\citenamefont {Bradlyn}\ \emph {et~al.}(2017)\citenamefont
  {Bradlyn}, \citenamefont {Elcoro}, \citenamefont {Cano}, \citenamefont
  {Vergniory}, \citenamefont {Wang}, \citenamefont {Felser}, \citenamefont
  {Aroyo},\ and\ \citenamefont {Bernevig}}]{Bradlyn17}%
  \BibitemOpen
  \bibfield  {author} {\bibinfo {author} {\bibfnamefont {B.}~\bibnamefont
  {Bradlyn}}, \bibinfo {author} {\bibfnamefont {L.}~\bibnamefont {Elcoro}},
  \bibinfo {author} {\bibfnamefont {J.}~\bibnamefont {Cano}}, \bibinfo {author}
  {\bibfnamefont {M.~G.}\ \bibnamefont {Vergniory}}, \bibinfo {author}
  {\bibfnamefont {Z.}~\bibnamefont {Wang}}, \bibinfo {author} {\bibfnamefont
  {C.}~\bibnamefont {Felser}}, \bibinfo {author} {\bibfnamefont {M.~I.}\
  \bibnamefont {Aroyo}}, \ and\ \bibinfo {author} {\bibfnamefont {B.~A.}\
  \bibnamefont {Bernevig}},\ }\href {http://dx.doi.org/10.1038/nature23268}
  {\bibfield  {journal} {\bibinfo  {journal} {Nature}\ }\textbf {\bibinfo
  {volume} {547}},\ \bibinfo {pages} {298} (\bibinfo {year}
  {2017})}\BibitemShut {NoStop}%
\bibitem [{\citenamefont {Po}\ \emph {et~al.}(2017)\citenamefont {Po},
  \citenamefont {Vishwanath},\ and\ \citenamefont {Watanabe}}]{Po17}%
  \BibitemOpen
  \bibfield  {author} {\bibinfo {author} {\bibfnamefont {H.~C.}\ \bibnamefont
  {Po}}, \bibinfo {author} {\bibfnamefont {A.}~\bibnamefont {Vishwanath}}, \
  and\ \bibinfo {author} {\bibfnamefont {H.}~\bibnamefont {Watanabe}},\ }\href
  {\doibase 10.1038/s41467-017-00133-2} {\bibfield  {journal} {\bibinfo
  {journal} {Nature Communications}\ }\textbf {\bibinfo {volume} {8}},\
  \bibinfo {pages} {50} (\bibinfo {year} {2017})}\BibitemShut {NoStop}%
\bibitem [{\citenamefont {H\"oller}\ and\ \citenamefont
  {Alexandradinata}(2018)}]{Hoeller18}%
  \BibitemOpen
  \bibfield  {author} {\bibinfo {author} {\bibfnamefont {J.}~\bibnamefont
  {H\"oller}}\ and\ \bibinfo {author} {\bibfnamefont {A.}~\bibnamefont
  {Alexandradinata}},\ }\href {\doibase 10.1103/PhysRevB.98.024310} {\bibfield
  {journal} {\bibinfo  {journal} {Phys. Rev. B}\ }\textbf {\bibinfo {volume}
  {98}},\ \bibinfo {pages} {024310} (\bibinfo {year} {2018})}\BibitemShut
  {NoStop}%
\bibitem [{\citenamefont {Fang}\ and\ \citenamefont
  {Fu}(2015)}]{PhysRevB.91.161105}%
  \BibitemOpen
  \bibfield  {author} {\bibinfo {author} {\bibfnamefont {C.}~\bibnamefont
  {Fang}}\ and\ \bibinfo {author} {\bibfnamefont {L.}~\bibnamefont {Fu}},\
  }\href {\doibase 10.1103/PhysRevB.91.161105} {\bibfield  {journal} {\bibinfo
  {journal} {Phys. Rev. B}\ }\textbf {\bibinfo {volume} {91}},\ \bibinfo
  {pages} {161105} (\bibinfo {year} {2015})}\BibitemShut {NoStop}%
\bibitem [{\citenamefont {Lantagne-Hurtubise}\ and\ \citenamefont
  {Franz}(2019)}]{Hurtubise19}%
  \BibitemOpen
  \bibfield  {author} {\bibinfo {author} {\bibfnamefont {{\'E}.}~\bibnamefont
  {Lantagne-Hurtubise}}\ and\ \bibinfo {author} {\bibfnamefont
  {M.}~\bibnamefont {Franz}},\ }\href {\doibase 10.1038/s42254-019-0041-7}
  {\bibfield  {journal} {\bibinfo  {journal} {Nature Reviews Physics}\ }\textbf
  {\bibinfo {volume} {1}},\ \bibinfo {pages} {183} (\bibinfo {year}
  {2019})}\BibitemShut {NoStop}%
\bibitem [{\citenamefont {Oka}\ and\ \citenamefont {Aoki}(2009)}]{OkaAoki09}%
  \BibitemOpen
  \bibfield  {author} {\bibinfo {author} {\bibfnamefont {T.}~\bibnamefont
  {Oka}}\ and\ \bibinfo {author} {\bibfnamefont {H.}~\bibnamefont {Aoki}},\
  }\href {\doibase 10.1103/PhysRevB.79.081406} {\bibfield  {journal} {\bibinfo
  {journal} {Phys. Rev. B}\ }\textbf {\bibinfo {volume} {79}},\ \bibinfo
  {pages} {081406} (\bibinfo {year} {2009})}\BibitemShut {NoStop}%
\bibitem [{\citenamefont {Kitagawa}\ \emph
  {et~al.}(2010{\natexlab{a}})\citenamefont {Kitagawa}, \citenamefont {Rudner},
  \citenamefont {Berg},\ and\ \citenamefont {Demler}}]{Kitagawa10_walk}%
  \BibitemOpen
  \bibfield  {author} {\bibinfo {author} {\bibfnamefont {T.}~\bibnamefont
  {Kitagawa}}, \bibinfo {author} {\bibfnamefont {M.~S.}\ \bibnamefont
  {Rudner}}, \bibinfo {author} {\bibfnamefont {E.}~\bibnamefont {Berg}}, \ and\
  \bibinfo {author} {\bibfnamefont {E.}~\bibnamefont {Demler}},\ }\href
  {\doibase 10.1103/PhysRevA.82.033429} {\bibfield  {journal} {\bibinfo
  {journal} {Phys. Rev. A}\ }\textbf {\bibinfo {volume} {82}},\ \bibinfo
  {pages} {033429} (\bibinfo {year} {2010}{\natexlab{a}})}\BibitemShut
  {NoStop}%
\bibitem [{\citenamefont {Kitagawa}\ \emph
  {et~al.}(2010{\natexlab{b}})\citenamefont {Kitagawa}, \citenamefont {Berg},
  \citenamefont {Rudner},\ and\ \citenamefont {Demler}}]{Kitagawa10}%
  \BibitemOpen
  \bibfield  {author} {\bibinfo {author} {\bibfnamefont {T.}~\bibnamefont
  {Kitagawa}}, \bibinfo {author} {\bibfnamefont {E.}~\bibnamefont {Berg}},
  \bibinfo {author} {\bibfnamefont {M.}~\bibnamefont {Rudner}}, \ and\ \bibinfo
  {author} {\bibfnamefont {E.}~\bibnamefont {Demler}},\ }\href {\doibase
  10.1103/PhysRevB.82.235114} {\bibfield  {journal} {\bibinfo  {journal} {Phys.
  Rev. B}\ }\textbf {\bibinfo {volume} {82}},\ \bibinfo {pages} {235114}
  (\bibinfo {year} {2010}{\natexlab{b}})}\BibitemShut {NoStop}%
\bibitem [{\citenamefont {Kitagawa}\ \emph {et~al.}(2011)\citenamefont
  {Kitagawa}, \citenamefont {Oka}, \citenamefont {Brataas}, \citenamefont
  {Fu},\ and\ \citenamefont {Demler}}]{Kitagawa11}%
  \BibitemOpen
  \bibfield  {author} {\bibinfo {author} {\bibfnamefont {T.}~\bibnamefont
  {Kitagawa}}, \bibinfo {author} {\bibfnamefont {T.}~\bibnamefont {Oka}},
  \bibinfo {author} {\bibfnamefont {A.}~\bibnamefont {Brataas}}, \bibinfo
  {author} {\bibfnamefont {L.}~\bibnamefont {Fu}}, \ and\ \bibinfo {author}
  {\bibfnamefont {E.}~\bibnamefont {Demler}},\ }\href {\doibase
  10.1103/PhysRevB.84.235108} {\bibfield  {journal} {\bibinfo  {journal} {Phys.
  Rev. B}\ }\textbf {\bibinfo {volume} {84}},\ \bibinfo {pages} {235108}
  (\bibinfo {year} {2011})}\BibitemShut {NoStop}%
\bibitem [{\citenamefont {Lindner}\ \emph {et~al.}(2011)\citenamefont
  {Lindner}, \citenamefont {Refael},\ and\ \citenamefont
  {Galitski}}]{Lindner11}%
  \BibitemOpen
  \bibfield  {author} {\bibinfo {author} {\bibfnamefont {N.~H.}\ \bibnamefont
  {Lindner}}, \bibinfo {author} {\bibfnamefont {G.}~\bibnamefont {Refael}}, \
  and\ \bibinfo {author} {\bibfnamefont {V.}~\bibnamefont {Galitski}},\ }\href
  {https://www.nature.com/articles/nphys1926} {\bibfield  {journal} {\bibinfo
  {journal} {Nat. Phys.}\ }\textbf {\bibinfo {volume} {7}},\ \bibinfo {pages}
  {490} (\bibinfo {year} {2011})}\BibitemShut {NoStop}%
\bibitem [{\citenamefont {G\'omez-Le\'on}\ and\ \citenamefont
  {Platero}(2013)}]{GomezLeon13}%
  \BibitemOpen
  \bibfield  {author} {\bibinfo {author} {\bibfnamefont {A.}~\bibnamefont
  {G\'omez-Le\'on}}\ and\ \bibinfo {author} {\bibfnamefont {G.}~\bibnamefont
  {Platero}},\ }\href {\doibase 10.1103/PhysRevLett.110.200403} {\bibfield
  {journal} {\bibinfo  {journal} {Phys. Rev. Lett.}\ }\textbf {\bibinfo
  {volume} {110}},\ \bibinfo {pages} {200403} (\bibinfo {year}
  {2013})}\BibitemShut {NoStop}%
\bibitem [{\citenamefont {Kundu}\ \emph {et~al.}(2014)\citenamefont {Kundu},
  \citenamefont {Fertig},\ and\ \citenamefont {Seradjeh}}]{Kundu14}%
  \BibitemOpen
  \bibfield  {author} {\bibinfo {author} {\bibfnamefont {A.}~\bibnamefont
  {Kundu}}, \bibinfo {author} {\bibfnamefont {H.~A.}\ \bibnamefont {Fertig}}, \
  and\ \bibinfo {author} {\bibfnamefont {B.}~\bibnamefont {Seradjeh}},\ }\href
  {\doibase 10.1103/PhysRevLett.113.236803} {\bibfield  {journal} {\bibinfo
  {journal} {Phys. Rev. Lett.}\ }\textbf {\bibinfo {volume} {113}},\ \bibinfo
  {pages} {236803} (\bibinfo {year} {2014})}\BibitemShut {NoStop}%
\bibitem [{\citenamefont {Jiang}\ \emph {et~al.}(2011)\citenamefont {Jiang},
  \citenamefont {Kitagawa}, \citenamefont {Alicea}, \citenamefont {Akhmerov},
  \citenamefont {Pekker}, \citenamefont {Refael}, \citenamefont {Cirac},
  \citenamefont {Demler}, \citenamefont {Lukin},\ and\ \citenamefont
  {Zoller}}]{Jiang11}%
  \BibitemOpen
  \bibfield  {author} {\bibinfo {author} {\bibfnamefont {L.}~\bibnamefont
  {Jiang}}, \bibinfo {author} {\bibfnamefont {T.}~\bibnamefont {Kitagawa}},
  \bibinfo {author} {\bibfnamefont {J.}~\bibnamefont {Alicea}}, \bibinfo
  {author} {\bibfnamefont {A.~R.}\ \bibnamefont {Akhmerov}}, \bibinfo {author}
  {\bibfnamefont {D.}~\bibnamefont {Pekker}}, \bibinfo {author} {\bibfnamefont
  {G.}~\bibnamefont {Refael}}, \bibinfo {author} {\bibfnamefont {J.~I.}\
  \bibnamefont {Cirac}}, \bibinfo {author} {\bibfnamefont {E.}~\bibnamefont
  {Demler}}, \bibinfo {author} {\bibfnamefont {M.~D.}\ \bibnamefont {Lukin}}, \
  and\ \bibinfo {author} {\bibfnamefont {P.}~\bibnamefont {Zoller}},\ }\href
  {\doibase 10.1103/PhysRevLett.106.220402} {\bibfield  {journal} {\bibinfo
  {journal} {Phys. Rev. Lett.}\ }\textbf {\bibinfo {volume} {106}},\ \bibinfo
  {pages} {220402} (\bibinfo {year} {2011})}\BibitemShut {NoStop}%
\bibitem [{\citenamefont {Rudner}\ \emph {et~al.}(2013)\citenamefont {Rudner},
  \citenamefont {Lindner}, \citenamefont {Berg},\ and\ \citenamefont
  {Levin}}]{Rudner13}%
  \BibitemOpen
  \bibfield  {author} {\bibinfo {author} {\bibfnamefont {M.~S.}\ \bibnamefont
  {Rudner}}, \bibinfo {author} {\bibfnamefont {N.~H.}\ \bibnamefont {Lindner}},
  \bibinfo {author} {\bibfnamefont {E.}~\bibnamefont {Berg}}, \ and\ \bibinfo
  {author} {\bibfnamefont {M.}~\bibnamefont {Levin}},\ }\href {\doibase
  10.1103/PhysRevX.3.031005} {\bibfield  {journal} {\bibinfo  {journal} {Phys.
  Rev. X}\ }\textbf {\bibinfo {volume} {3}},\ \bibinfo {pages} {031005}
  (\bibinfo {year} {2013})}\BibitemShut {NoStop}%
\bibitem [{\citenamefont {Nathan}\ and\ \citenamefont
  {Rudner}(2015)}]{NathanRudner15}%
  \BibitemOpen
  \bibfield  {author} {\bibinfo {author} {\bibfnamefont {F.}~\bibnamefont
  {Nathan}}\ and\ \bibinfo {author} {\bibfnamefont {M.~S.}\ \bibnamefont
  {Rudner}},\ }\href {http://stacks.iop.org/1367-2630/17/i=12/a=125014}
  {\bibfield  {journal} {\bibinfo  {journal} {New Journal of Physics}\ }\textbf
  {\bibinfo {volume} {17}},\ \bibinfo {pages} {125014} (\bibinfo {year}
  {2015})}\BibitemShut {NoStop}%
\bibitem [{\citenamefont {Carpentier}\ \emph {et~al.}(2015)\citenamefont
  {Carpentier}, \citenamefont {Delplace}, \citenamefont {Fruchart},\ and\
  \citenamefont {Gaw\c{e}dzki}}]{Carpentier15}%
  \BibitemOpen
  \bibfield  {author} {\bibinfo {author} {\bibfnamefont {D.}~\bibnamefont
  {Carpentier}}, \bibinfo {author} {\bibfnamefont {P.}~\bibnamefont
  {Delplace}}, \bibinfo {author} {\bibfnamefont {M.}~\bibnamefont {Fruchart}},
  \ and\ \bibinfo {author} {\bibfnamefont {K.}~\bibnamefont {Gaw\c{e}dzki}},\
  }\href {\doibase 10.1103/PhysRevLett.114.106806} {\bibfield  {journal}
  {\bibinfo  {journal} {Phys. Rev. Lett.}\ }\textbf {\bibinfo {volume} {114}},\
  \bibinfo {pages} {106806} (\bibinfo {year} {2015})}\BibitemShut {NoStop}%
\bibitem [{\citenamefont {Roy}\ and\ \citenamefont
  {Harper}(2017{\natexlab{a}})}]{RoyHarper17}%
  \BibitemOpen
  \bibfield  {author} {\bibinfo {author} {\bibfnamefont {R.}~\bibnamefont
  {Roy}}\ and\ \bibinfo {author} {\bibfnamefont {F.}~\bibnamefont {Harper}},\
  }\href {\doibase 10.1103/PhysRevB.96.155118} {\bibfield  {journal} {\bibinfo
  {journal} {Phys. Rev. B}\ }\textbf {\bibinfo {volume} {96}},\ \bibinfo
  {pages} {155118} (\bibinfo {year} {2017}{\natexlab{a}})}\BibitemShut
  {NoStop}%
\bibitem [{\citenamefont {Morimoto}\ \emph {et~al.}(2017)\citenamefont
  {Morimoto}, \citenamefont {Po},\ and\ \citenamefont
  {Vishwanath}}]{MorimotoPoVishwanath}%
  \BibitemOpen
  \bibfield  {author} {\bibinfo {author} {\bibfnamefont {T.}~\bibnamefont
  {Morimoto}}, \bibinfo {author} {\bibfnamefont {H.~C.}\ \bibnamefont {Po}}, \
  and\ \bibinfo {author} {\bibfnamefont {A.}~\bibnamefont {Vishwanath}},\
  }\href {\doibase 10.1103/PhysRevB.95.195155} {\bibfield  {journal} {\bibinfo
  {journal} {Phys. Rev. B}\ }\textbf {\bibinfo {volume} {95}},\ \bibinfo
  {pages} {195155} (\bibinfo {year} {2017})}\BibitemShut {NoStop}%
\bibitem [{\citenamefont {Yao}\ \emph {et~al.}(2017)\citenamefont {Yao},
  \citenamefont {Yan},\ and\ \citenamefont {Wang}}]{Yao17}%
  \BibitemOpen
  \bibfield  {author} {\bibinfo {author} {\bibfnamefont {S.}~\bibnamefont
  {Yao}}, \bibinfo {author} {\bibfnamefont {Z.}~\bibnamefont {Yan}}, \ and\
  \bibinfo {author} {\bibfnamefont {Z.}~\bibnamefont {Wang}},\ }\href {\doibase
  10.1103/PhysRevB.96.195303} {\bibfield  {journal} {\bibinfo  {journal} {Phys.
  Rev. B}\ }\textbf {\bibinfo {volume} {96}},\ \bibinfo {pages} {195303}
  (\bibinfo {year} {2017})}\BibitemShut {NoStop}%
\bibitem [{\citenamefont {Budich}\ \emph {et~al.}(2017)\citenamefont {Budich},
  \citenamefont {Hu},\ and\ \citenamefont {Zoller}}]{Budich17}%
  \BibitemOpen
  \bibfield  {author} {\bibinfo {author} {\bibfnamefont {J.~C.}\ \bibnamefont
  {Budich}}, \bibinfo {author} {\bibfnamefont {Y.}~\bibnamefont {Hu}}, \ and\
  \bibinfo {author} {\bibfnamefont {P.}~\bibnamefont {Zoller}},\ }\href
  {\doibase 10.1103/PhysRevLett.118.105302} {\bibfield  {journal} {\bibinfo
  {journal} {Phys. Rev. Lett.}\ }\textbf {\bibinfo {volume} {118}},\ \bibinfo
  {pages} {105302} (\bibinfo {year} {2017})}\BibitemShut {NoStop}%
\bibitem [{\citenamefont {Higashikawa}\ \emph {et~al.}(2019)\citenamefont
  {Higashikawa}, \citenamefont {Nakagawa},\ and\ \citenamefont
  {Ueda}}]{Higashikawa18}%
  \BibitemOpen
  \bibfield  {author} {\bibinfo {author} {\bibfnamefont {S.}~\bibnamefont
  {Higashikawa}}, \bibinfo {author} {\bibfnamefont {M.}~\bibnamefont
  {Nakagawa}}, \ and\ \bibinfo {author} {\bibfnamefont {M.}~\bibnamefont
  {Ueda}},\ }\href {\doibase 10.1103/PhysRevLett.123.066403} {\bibfield
  {journal} {\bibinfo  {journal} {Phys. Rev. Lett.}\ }\textbf {\bibinfo
  {volume} {123}},\ \bibinfo {pages} {066403} (\bibinfo {year}
  {2019})}\BibitemShut {NoStop}%
\bibitem [{\citenamefont {Sun}\ \emph {et~al.}(2018)\citenamefont {Sun},
  \citenamefont {Xiao}, \citenamefont {Bzdu\ifmmode~\check{s}\else
  \v{s}\fi{}ek}, \citenamefont {Zhang},\ and\ \citenamefont {Fan}}]{Sun18}%
  \BibitemOpen
  \bibfield  {author} {\bibinfo {author} {\bibfnamefont {X.-Q.}\ \bibnamefont
  {Sun}}, \bibinfo {author} {\bibfnamefont {M.}~\bibnamefont {Xiao}}, \bibinfo
  {author} {\bibfnamefont {T.~c.~v.}\ \bibnamefont {Bzdu\ifmmode~\check{s}\else
  \v{s}\fi{}ek}}, \bibinfo {author} {\bibfnamefont {S.-C.}\ \bibnamefont
  {Zhang}}, \ and\ \bibinfo {author} {\bibfnamefont {S.}~\bibnamefont {Fan}},\
  }\href {\doibase 10.1103/PhysRevLett.121.196401} {\bibfield  {journal}
  {\bibinfo  {journal} {Phys. Rev. Lett.}\ }\textbf {\bibinfo {volume} {121}},\
  \bibinfo {pages} {196401} (\bibinfo {year} {2018})}\BibitemShut {NoStop}%
\bibitem [{\citenamefont {Eckardt}(2017)}]{Eckardt17}%
  \BibitemOpen
  \bibfield  {author} {\bibinfo {author} {\bibfnamefont {A.}~\bibnamefont
  {Eckardt}},\ }\href {\doibase 10.1103/RevModPhys.89.011004} {\bibfield
  {journal} {\bibinfo  {journal} {Rev. Mod. Phys.}\ }\textbf {\bibinfo {volume}
  {89}},\ \bibinfo {pages} {011004} (\bibinfo {year} {2017})}\BibitemShut
  {NoStop}%
\bibitem [{\citenamefont {Oka}\ and\ \citenamefont
  {Kitamura}(2019)}]{OkaKitamura19}%
  \BibitemOpen
  \bibfield  {author} {\bibinfo {author} {\bibfnamefont {T.}~\bibnamefont
  {Oka}}\ and\ \bibinfo {author} {\bibfnamefont {S.}~\bibnamefont {Kitamura}},\
  }\href {\doibase 10.1146/annurev-conmatphys-031218-013423} {\bibfield
  {journal} {\bibinfo  {journal} {Annual Review of Condensed Matter Physics}\
  }\textbf {\bibinfo {volume} {10}},\ \bibinfo {pages} {387} (\bibinfo {year}
  {2019})}\BibitemShut {NoStop}%
\bibitem [{\citenamefont {Wang}\ \emph {et~al.}(2013)\citenamefont {Wang},
  \citenamefont {Steinberg}, \citenamefont {Jarillo-Herrero},\ and\
  \citenamefont {Gedik}}]{Wang13}%
  \BibitemOpen
  \bibfield  {author} {\bibinfo {author} {\bibfnamefont {Y.~H.}\ \bibnamefont
  {Wang}}, \bibinfo {author} {\bibfnamefont {H.}~\bibnamefont {Steinberg}},
  \bibinfo {author} {\bibfnamefont {P.}~\bibnamefont {Jarillo-Herrero}}, \ and\
  \bibinfo {author} {\bibfnamefont {N.}~\bibnamefont {Gedik}},\ }\href
  {\doibase 10.1126/science.1239834} {\bibfield  {journal} {\bibinfo  {journal}
  {Science}\ }\textbf {\bibinfo {volume} {342}},\ \bibinfo {pages} {453}
  (\bibinfo {year} {2013})}\BibitemShut {NoStop}%
\bibitem [{\citenamefont {Mahmood}\ \emph {et~al.}(2016)\citenamefont
  {Mahmood}, \citenamefont {Chan}, \citenamefont {Alpichshev}, \citenamefont
  {Gardner}, \citenamefont {Lee}, \citenamefont {Lee},\ and\ \citenamefont
  {Gedik}}]{Mahmood16}%
  \BibitemOpen
  \bibfield  {author} {\bibinfo {author} {\bibfnamefont {F.}~\bibnamefont
  {Mahmood}}, \bibinfo {author} {\bibfnamefont {C.-K.}\ \bibnamefont {Chan}},
  \bibinfo {author} {\bibfnamefont {Z.}~\bibnamefont {Alpichshev}}, \bibinfo
  {author} {\bibfnamefont {D.}~\bibnamefont {Gardner}}, \bibinfo {author}
  {\bibfnamefont {Y.}~\bibnamefont {Lee}}, \bibinfo {author} {\bibfnamefont
  {P.~A.}\ \bibnamefont {Lee}}, \ and\ \bibinfo {author} {\bibfnamefont
  {N.}~\bibnamefont {Gedik}},\ }\href {\doibase
  https://doi.org/10.1038/nphys3609} {\bibfield  {journal} {\bibinfo  {journal}
  {Nat. Phys.}\ }\textbf {\bibinfo {volume} {12}},\ \bibinfo {pages} {306}
  (\bibinfo {year} {2016})}\BibitemShut {NoStop}%
\bibitem [{\citenamefont {McIver}\ \emph {et~al.}(2019)\citenamefont {McIver},
  \citenamefont {Schulte}, \citenamefont {Stein}, \citenamefont {Matsuyama},
  \citenamefont {Jotzu}, \citenamefont {Meier},\ and\ \citenamefont
  {Cavalleri}}]{McIver18}%
  \BibitemOpen
  \bibfield  {author} {\bibinfo {author} {\bibfnamefont {J.~W.}\ \bibnamefont
  {McIver}}, \bibinfo {author} {\bibfnamefont {B.}~\bibnamefont {Schulte}},
  \bibinfo {author} {\bibfnamefont {F.-U.}\ \bibnamefont {Stein}}, \bibinfo
  {author} {\bibfnamefont {T.}~\bibnamefont {Matsuyama}}, \bibinfo {author}
  {\bibfnamefont {G.}~\bibnamefont {Jotzu}}, \bibinfo {author} {\bibfnamefont
  {G.}~\bibnamefont {Meier}}, \ and\ \bibinfo {author} {\bibfnamefont
  {A.}~\bibnamefont {Cavalleri}},\ }\href {\doibase
  https://doi.org/10.1038/s41567-019-0698-y} {\bibfield  {journal} {\bibinfo
  {journal} {Nat. Phys.}\ } (\bibinfo {year} {2019}),\
  https://doi.org/10.1038/s41567-019-0698-y}\BibitemShut {NoStop}%
\bibitem [{\citenamefont {Kitagawa}\ \emph {et~al.}(2012)\citenamefont
  {Kitagawa}, \citenamefont {Broome}, \citenamefont {Fedrizzi}, \citenamefont
  {Rudner}, \citenamefont {Berg}, \citenamefont {Kassal}, \citenamefont
  {Aspuru-Guzik}, \citenamefont {Demler},\ and\ \citenamefont
  {White}}]{Kitagawa12}%
  \BibitemOpen
  \bibfield  {author} {\bibinfo {author} {\bibfnamefont {T.}~\bibnamefont
  {Kitagawa}}, \bibinfo {author} {\bibfnamefont {M.~A.}\ \bibnamefont
  {Broome}}, \bibinfo {author} {\bibfnamefont {A.}~\bibnamefont {Fedrizzi}},
  \bibinfo {author} {\bibfnamefont {M.~S.}\ \bibnamefont {Rudner}}, \bibinfo
  {author} {\bibfnamefont {E.}~\bibnamefont {Berg}}, \bibinfo {author}
  {\bibfnamefont {I.}~\bibnamefont {Kassal}}, \bibinfo {author} {\bibfnamefont
  {A.}~\bibnamefont {Aspuru-Guzik}}, \bibinfo {author} {\bibfnamefont
  {E.}~\bibnamefont {Demler}}, \ and\ \bibinfo {author} {\bibfnamefont {A.~G.}\
  \bibnamefont {White}},\ }\href {\doibase https://doi.org/10.1038/ncomms1872}
  {\bibfield  {journal} {\bibinfo  {journal} {Nat. Commun.}\ }\textbf {\bibinfo
  {volume} {3}},\ \bibinfo {pages} {882} (\bibinfo {year} {2012})}\BibitemShut
  {NoStop}%
\bibitem [{\citenamefont {Rechtsman}\ \emph {et~al.}(2013)\citenamefont
  {Rechtsman}, \citenamefont {Zeuner}, \citenamefont {Plotnik}, \citenamefont
  {Lumer}, \citenamefont {Podolsky}, \citenamefont {Dreisow}, \citenamefont
  {Nolte}, \citenamefont {Segev},\ and\ \citenamefont {Szameit}}]{Rechtsman13}%
  \BibitemOpen
  \bibfield  {author} {\bibinfo {author} {\bibfnamefont {M.~C.}\ \bibnamefont
  {Rechtsman}}, \bibinfo {author} {\bibfnamefont {J.~M.}\ \bibnamefont
  {Zeuner}}, \bibinfo {author} {\bibfnamefont {Y.}~\bibnamefont {Plotnik}},
  \bibinfo {author} {\bibfnamefont {Y.}~\bibnamefont {Lumer}}, \bibinfo
  {author} {\bibfnamefont {D.}~\bibnamefont {Podolsky}}, \bibinfo {author}
  {\bibfnamefont {F.}~\bibnamefont {Dreisow}}, \bibinfo {author} {\bibfnamefont
  {S.}~\bibnamefont {Nolte}}, \bibinfo {author} {\bibfnamefont
  {M.}~\bibnamefont {Segev}}, \ and\ \bibinfo {author} {\bibfnamefont
  {A.}~\bibnamefont {Szameit}},\ }\href {\doibase
  https://doi.org/10.1038/nature12066} {\bibfield  {journal} {\bibinfo
  {journal} {Nature}\ }\textbf {\bibinfo {volume} {496}},\ \bibinfo {pages}
  {196} (\bibinfo {year} {2013})}\BibitemShut {NoStop}%
\bibitem [{\citenamefont {Maczewsky}\ \emph {et~al.}(2017)\citenamefont
  {Maczewsky}, \citenamefont {Zeuner}, \citenamefont {Nolte},\ and\
  \citenamefont {Szameit}}]{Maczewsky17}%
  \BibitemOpen
  \bibfield  {author} {\bibinfo {author} {\bibfnamefont {L.~J.}\ \bibnamefont
  {Maczewsky}}, \bibinfo {author} {\bibfnamefont {J.~M.}\ \bibnamefont
  {Zeuner}}, \bibinfo {author} {\bibfnamefont {S.}~\bibnamefont {Nolte}}, \
  and\ \bibinfo {author} {\bibfnamefont {A.}~\bibnamefont {Szameit}},\ }\href
  {\doibase https://doi.org/10.1038/ncomms13756} {\bibfield  {journal}
  {\bibinfo  {journal} {Nat. Commun.}\ }\textbf {\bibinfo {volume} {8}},\
  \bibinfo {pages} {13756} (\bibinfo {year} {2017})}\BibitemShut {NoStop}%
\bibitem [{\citenamefont {Fleury}\ \emph {et~al.}(2016)\citenamefont {Fleury},
  \citenamefont {Khanikaev},\ and\ \citenamefont {Al\`{u}}}]{Fleury16}%
  \BibitemOpen
  \bibfield  {author} {\bibinfo {author} {\bibfnamefont {R.}~\bibnamefont
  {Fleury}}, \bibinfo {author} {\bibfnamefont {A.~B.}\ \bibnamefont
  {Khanikaev}}, \ and\ \bibinfo {author} {\bibfnamefont {A.}~\bibnamefont
  {Al\`{u}}},\ }\href {\doibase https://doi.org/10.1038/ncomms11744} {\bibfield
   {journal} {\bibinfo  {journal} {Nat. Commun.}\ }\textbf {\bibinfo {volume}
  {7}},\ \bibinfo {pages} {11744} (\bibinfo {year} {2016})}\BibitemShut
  {NoStop}%
\bibitem [{\citenamefont {Peng}\ \emph {et~al.}(2016)\citenamefont {Peng},
  \citenamefont {Qin}, \citenamefont {Zhao}, \citenamefont {Shen},
  \citenamefont {Xu}, \citenamefont {Bao}, \citenamefont {Jia},\ and\
  \citenamefont {Zhu}}]{Peng16}%
  \BibitemOpen
  \bibfield  {author} {\bibinfo {author} {\bibfnamefont {Y.-G.}\ \bibnamefont
  {Peng}}, \bibinfo {author} {\bibfnamefont {C.-Z.}\ \bibnamefont {Qin}},
  \bibinfo {author} {\bibfnamefont {D.-G.}\ \bibnamefont {Zhao}}, \bibinfo
  {author} {\bibfnamefont {Y.-X.}\ \bibnamefont {Shen}}, \bibinfo {author}
  {\bibfnamefont {X.-Y.}\ \bibnamefont {Xu}}, \bibinfo {author} {\bibfnamefont
  {M.}~\bibnamefont {Bao}}, \bibinfo {author} {\bibfnamefont {H.}~\bibnamefont
  {Jia}}, \ and\ \bibinfo {author} {\bibfnamefont {X.-F.}\ \bibnamefont
  {Zhu}},\ }\href {\doibase https://doi.org/10.1038/ncomms13368} {\bibfield
  {journal} {\bibinfo  {journal} {Nat. Commun.}\ }\textbf {\bibinfo {volume}
  {7}},\ \bibinfo {pages} {13368} (\bibinfo {year} {2016})}\BibitemShut
  {NoStop}%
\bibitem [{\citenamefont {Jotzu}\ \emph {et~al.}(2014)\citenamefont {Jotzu},
  \citenamefont {Messer}, \citenamefont {Desbuquois}, \citenamefont {Lebrat},
  \citenamefont {Uehlinger}, \citenamefont {Greif},\ and\ \citenamefont
  {Esslinger}}]{Jotzu14}%
  \BibitemOpen
  \bibfield  {author} {\bibinfo {author} {\bibfnamefont {G.}~\bibnamefont
  {Jotzu}}, \bibinfo {author} {\bibfnamefont {M.}~\bibnamefont {Messer}},
  \bibinfo {author} {\bibfnamefont {R.}~\bibnamefont {Desbuquois}}, \bibinfo
  {author} {\bibfnamefont {M.}~\bibnamefont {Lebrat}}, \bibinfo {author}
  {\bibfnamefont {T.}~\bibnamefont {Uehlinger}}, \bibinfo {author}
  {\bibfnamefont {D.}~\bibnamefont {Greif}}, \ and\ \bibinfo {author}
  {\bibfnamefont {T.}~\bibnamefont {Esslinger}},\ }\href {\doibase
  https://doi.org/10.1038/nature13915} {\bibfield  {journal} {\bibinfo
  {journal} {Nature}\ }\textbf {\bibinfo {volume} {515}},\ \bibinfo {pages}
  {237} (\bibinfo {year} {2014})}\BibitemShut {NoStop}%
\bibitem [{\citenamefont {Fl{\"a}schner}\ \emph {et~al.}(2016)\citenamefont
  {Fl{\"a}schner}, \citenamefont {Rem}, \citenamefont {Tarnowski},
  \citenamefont {Vogel}, \citenamefont {L{\"u}hmann}, \citenamefont
  {Sengstock},\ and\ \citenamefont {Weitenberg}}]{Flaschner16}%
  \BibitemOpen
  \bibfield  {author} {\bibinfo {author} {\bibfnamefont {N.}~\bibnamefont
  {Fl{\"a}schner}}, \bibinfo {author} {\bibfnamefont {B.~S.}\ \bibnamefont
  {Rem}}, \bibinfo {author} {\bibfnamefont {M.}~\bibnamefont {Tarnowski}},
  \bibinfo {author} {\bibfnamefont {D.}~\bibnamefont {Vogel}}, \bibinfo
  {author} {\bibfnamefont {D.-S.}\ \bibnamefont {L{\"u}hmann}}, \bibinfo
  {author} {\bibfnamefont {K.}~\bibnamefont {Sengstock}}, \ and\ \bibinfo
  {author} {\bibfnamefont {C.}~\bibnamefont {Weitenberg}},\ }\href {\doibase
  10.1126/science.aad4568} {\bibfield  {journal} {\bibinfo  {journal}
  {Science}\ }\textbf {\bibinfo {volume} {352}},\ \bibinfo {pages} {1091}
  (\bibinfo {year} {2016})}\BibitemShut {NoStop}%
\bibitem [{\citenamefont {Thonhauser}\ and\ \citenamefont
  {Vanderbilt}(2006)}]{Thonhauser06}%
  \BibitemOpen
  \bibfield  {author} {\bibinfo {author} {\bibfnamefont {T.}~\bibnamefont
  {Thonhauser}}\ and\ \bibinfo {author} {\bibfnamefont {D.}~\bibnamefont
  {Vanderbilt}},\ }\href {\doibase 10.1103/PhysRevB.74.235111} {\bibfield
  {journal} {\bibinfo  {journal} {Phys. Rev. B}\ }\textbf {\bibinfo {volume}
  {74}},\ \bibinfo {pages} {235111} (\bibinfo {year} {2006})}\BibitemShut
  {NoStop}%
\bibitem [{\citenamefont {Coh}\ and\ \citenamefont {Vanderbilt}(2009)}]{Coh09}%
  \BibitemOpen
  \bibfield  {author} {\bibinfo {author} {\bibfnamefont {S.}~\bibnamefont
  {Coh}}\ and\ \bibinfo {author} {\bibfnamefont {D.}~\bibnamefont
  {Vanderbilt}},\ }\href {\doibase 10.1103/PhysRevLett.102.107603} {\bibfield
  {journal} {\bibinfo  {journal} {Phys. Rev. Lett.}\ }\textbf {\bibinfo
  {volume} {102}},\ \bibinfo {pages} {107603} (\bibinfo {year}
  {2009})}\BibitemShut {NoStop}%
\bibitem [{\citenamefont {Soluyanov}\ and\ \citenamefont
  {Vanderbilt}(2011{\natexlab{a}})}]{Soluyanov11}%
  \BibitemOpen
  \bibfield  {author} {\bibinfo {author} {\bibfnamefont {A.~A.}\ \bibnamefont
  {Soluyanov}}\ and\ \bibinfo {author} {\bibfnamefont {D.}~\bibnamefont
  {Vanderbilt}},\ }\href {\doibase 10.1103/PhysRevB.83.035108} {\bibfield
  {journal} {\bibinfo  {journal} {Phys. Rev. B}\ }\textbf {\bibinfo {volume}
  {83}},\ \bibinfo {pages} {035108} (\bibinfo {year}
  {2011}{\natexlab{a}})}\BibitemShut {NoStop}%
\bibitem [{\citenamefont {Soluyanov}\ and\ \citenamefont
  {Vanderbilt}(2011{\natexlab{b}})}]{Soluyanov11_2}%
  \BibitemOpen
  \bibfield  {author} {\bibinfo {author} {\bibfnamefont {A.~A.}\ \bibnamefont
  {Soluyanov}}\ and\ \bibinfo {author} {\bibfnamefont {D.}~\bibnamefont
  {Vanderbilt}},\ }\href {\doibase 10.1103/PhysRevB.83.235401} {\bibfield
  {journal} {\bibinfo  {journal} {Phys. Rev. B}\ }\textbf {\bibinfo {volume}
  {83}},\ \bibinfo {pages} {235401} (\bibinfo {year}
  {2011}{\natexlab{b}})}\BibitemShut {NoStop}%
\bibitem [{\citenamefont {Yu}\ \emph {et~al.}(2011)\citenamefont {Yu},
  \citenamefont {Qi}, \citenamefont {Bernevig}, \citenamefont {Fang},\ and\
  \citenamefont {Dai}}]{Yu11}%
  \BibitemOpen
  \bibfield  {author} {\bibinfo {author} {\bibfnamefont {R.}~\bibnamefont
  {Yu}}, \bibinfo {author} {\bibfnamefont {X.~L.}\ \bibnamefont {Qi}}, \bibinfo
  {author} {\bibfnamefont {A.}~\bibnamefont {Bernevig}}, \bibinfo {author}
  {\bibfnamefont {Z.}~\bibnamefont {Fang}}, \ and\ \bibinfo {author}
  {\bibfnamefont {X.}~\bibnamefont {Dai}},\ }\href {\doibase
  10.1103/PhysRevB.84.075119} {\bibfield  {journal} {\bibinfo  {journal} {Phys.
  Rev. B}\ }\textbf {\bibinfo {volume} {84}},\ \bibinfo {pages} {075119}
  (\bibinfo {year} {2011})}\BibitemShut {NoStop}%
\bibitem [{\citenamefont {Read}(2017)}]{Read17}%
  \BibitemOpen
  \bibfield  {author} {\bibinfo {author} {\bibfnamefont {N.}~\bibnamefont
  {Read}},\ }\href {\doibase 10.1103/PhysRevB.95.115309} {\bibfield  {journal}
  {\bibinfo  {journal} {Phys. Rev. B}\ }\textbf {\bibinfo {volume} {95}},\
  \bibinfo {pages} {115309} (\bibinfo {year} {2017})}\BibitemShut {NoStop}%
\bibitem [{\citenamefont {Nathan}\ \emph {et~al.}(2017)\citenamefont {Nathan},
  \citenamefont {Rudner}, \citenamefont {Lindner}, \citenamefont {Berg},\ and\
  \citenamefont {Refael}}]{Nathan17}%
  \BibitemOpen
  \bibfield  {author} {\bibinfo {author} {\bibfnamefont {F.}~\bibnamefont
  {Nathan}}, \bibinfo {author} {\bibfnamefont {M.~S.}\ \bibnamefont {Rudner}},
  \bibinfo {author} {\bibfnamefont {N.~H.}\ \bibnamefont {Lindner}}, \bibinfo
  {author} {\bibfnamefont {E.}~\bibnamefont {Berg}}, \ and\ \bibinfo {author}
  {\bibfnamefont {G.}~\bibnamefont {Refael}},\ }\href {\doibase
  10.1103/PhysRevLett.119.186801} {\bibfield  {journal} {\bibinfo  {journal}
  {Phys. Rev. Lett.}\ }\textbf {\bibinfo {volume} {119}},\ \bibinfo {pages}
  {186801} (\bibinfo {year} {2017})}\BibitemShut {NoStop}%
\bibitem [{\citenamefont {Nathan}\ \emph {et~al.}(2019)\citenamefont {Nathan},
  \citenamefont {Abanin}, \citenamefont {Lindner}, \citenamefont {Berg},\ and\
  \citenamefont {Rudner}}]{Nathan19}%
  \BibitemOpen
  \bibfield  {author} {\bibinfo {author} {\bibfnamefont {F.}~\bibnamefont
  {Nathan}}, \bibinfo {author} {\bibfnamefont {D.~A.}\ \bibnamefont {Abanin}},
  \bibinfo {author} {\bibfnamefont {N.~H.}\ \bibnamefont {Lindner}}, \bibinfo
  {author} {\bibfnamefont {E.}~\bibnamefont {Berg}}, \ and\ \bibinfo {author}
  {\bibfnamefont {M.~S.}\ \bibnamefont {Rudner}},\ }\href
  {http://arxiv.org/abs/1907.12228} {\bibfield  {journal} {\bibinfo  {journal}
  {arXiv:1907.12228}\ } (\bibinfo {year} {2019})}\BibitemShut {NoStop}%
\bibitem [{\citenamefont {{Mondragon-Shem}}\ \emph {et~al.}(2018)\citenamefont
  {{Mondragon-Shem}}, \citenamefont {{Martin}}, \citenamefont
  {{Alexandradinata}},\ and\ \citenamefont {{Cheng}}}]{MondragonShem18}%
  \BibitemOpen
  \bibfield  {author} {\bibinfo {author} {\bibfnamefont {I.}~\bibnamefont
  {{Mondragon-Shem}}}, \bibinfo {author} {\bibfnamefont {I.}~\bibnamefont
  {{Martin}}}, \bibinfo {author} {\bibfnamefont {A.}~\bibnamefont
  {{Alexandradinata}}}, \ and\ \bibinfo {author} {\bibfnamefont
  {M.}~\bibnamefont {{Cheng}}},\ }\href {http://arxiv.org/abs/1811.10632}
  {\bibfield  {journal} {\bibinfo  {journal} {arXiv:1811.10632}\ } (\bibinfo
  {year} {2018})}\BibitemShut {NoStop}%
\bibitem [{\citenamefont {Shirley}(1965)}]{Shirley65}%
  \BibitemOpen
  \bibfield  {author} {\bibinfo {author} {\bibfnamefont {J.~H.}\ \bibnamefont
  {Shirley}},\ }\href {\doibase 10.1103/PhysRev.138.B979} {\bibfield  {journal}
  {\bibinfo  {journal} {Phys. Rev.}\ }\textbf {\bibinfo {volume} {138}},\
  \bibinfo {pages} {B979} (\bibinfo {year} {1965})}\BibitemShut {NoStop}%
\bibitem [{\citenamefont {Sambe}(1973)}]{Sambe73}%
  \BibitemOpen
  \bibfield  {author} {\bibinfo {author} {\bibfnamefont {H.}~\bibnamefont
  {Sambe}},\ }\href {\doibase 10.1103/PhysRevA.7.2203} {\bibfield  {journal}
  {\bibinfo  {journal} {Phys. Rev. A}\ }\textbf {\bibinfo {volume} {7}},\
  \bibinfo {pages} {2203} (\bibinfo {year} {1973})}\BibitemShut {NoStop}%
\bibitem [{\citenamefont {Marzari}\ \emph {et~al.}(2012)\citenamefont
  {Marzari}, \citenamefont {Mostofi}, \citenamefont {Yates}, \citenamefont
  {Souza},\ and\ \citenamefont {Vanderbilt}}]{Marzari12}%
  \BibitemOpen
  \bibfield  {author} {\bibinfo {author} {\bibfnamefont {N.}~\bibnamefont
  {Marzari}}, \bibinfo {author} {\bibfnamefont {A.~A.}\ \bibnamefont
  {Mostofi}}, \bibinfo {author} {\bibfnamefont {J.~R.}\ \bibnamefont {Yates}},
  \bibinfo {author} {\bibfnamefont {I.}~\bibnamefont {Souza}}, \ and\ \bibinfo
  {author} {\bibfnamefont {D.}~\bibnamefont {Vanderbilt}},\ }\href {\doibase
  10.1103/RevModPhys.84.1419} {\bibfield  {journal} {\bibinfo  {journal} {Rev.
  Mod. Phys.}\ }\textbf {\bibinfo {volume} {84}},\ \bibinfo {pages} {1419}
  (\bibinfo {year} {2012})}\BibitemShut {NoStop}%
\bibitem [{\citenamefont {Taherinejad}\ \emph {et~al.}(2014)\citenamefont
  {Taherinejad}, \citenamefont {Garrity},\ and\ \citenamefont
  {Vanderbilt}}]{Taherinejad14}%
  \BibitemOpen
  \bibfield  {author} {\bibinfo {author} {\bibfnamefont {M.}~\bibnamefont
  {Taherinejad}}, \bibinfo {author} {\bibfnamefont {K.~F.}\ \bibnamefont
  {Garrity}}, \ and\ \bibinfo {author} {\bibfnamefont {D.}~\bibnamefont
  {Vanderbilt}},\ }\href {\doibase 10.1103/PhysRevB.89.115102} {\bibfield
  {journal} {\bibinfo  {journal} {Phys. Rev. B}\ }\textbf {\bibinfo {volume}
  {89}},\ \bibinfo {pages} {115102} (\bibinfo {year} {2014})}\BibitemShut
  {NoStop}%
\bibitem [{\citenamefont {Taherinejad}\ and\ \citenamefont
  {Vanderbilt}(2015)}]{Taherinejad15}%
  \BibitemOpen
  \bibfield  {author} {\bibinfo {author} {\bibfnamefont {M.}~\bibnamefont
  {Taherinejad}}\ and\ \bibinfo {author} {\bibfnamefont {D.}~\bibnamefont
  {Vanderbilt}},\ }\href {\doibase 10.1103/PhysRevLett.114.096401} {\bibfield
  {journal} {\bibinfo  {journal} {Phys. Rev. Lett.}\ }\textbf {\bibinfo
  {volume} {114}},\ \bibinfo {pages} {096401} (\bibinfo {year}
  {2015})}\BibitemShut {NoStop}%
\bibitem [{\citenamefont {Gresch}\ \emph {et~al.}(2017)\citenamefont {Gresch},
  \citenamefont {Aut\`es}, \citenamefont {Yazyev}, \citenamefont {Troyer},
  \citenamefont {Vanderbilt}, \citenamefont {Bernevig},\ and\ \citenamefont
  {Soluyanov}}]{Gresch17}%
  \BibitemOpen
  \bibfield  {author} {\bibinfo {author} {\bibfnamefont {D.}~\bibnamefont
  {Gresch}}, \bibinfo {author} {\bibfnamefont {G.}~\bibnamefont {Aut\`es}},
  \bibinfo {author} {\bibfnamefont {O.~V.}\ \bibnamefont {Yazyev}}, \bibinfo
  {author} {\bibfnamefont {M.}~\bibnamefont {Troyer}}, \bibinfo {author}
  {\bibfnamefont {D.}~\bibnamefont {Vanderbilt}}, \bibinfo {author}
  {\bibfnamefont {B.~A.}\ \bibnamefont {Bernevig}}, \ and\ \bibinfo {author}
  {\bibfnamefont {A.~A.}\ \bibnamefont {Soluyanov}},\ }\href {\doibase
  10.1103/PhysRevB.95.075146} {\bibfield  {journal} {\bibinfo  {journal} {Phys.
  Rev. B}\ }\textbf {\bibinfo {volume} {95}},\ \bibinfo {pages} {075146}
  (\bibinfo {year} {2017})}\BibitemShut {NoStop}%
\bibitem [{\citenamefont {Vanderbilt}(2018)}]{Vanderbilt_book}%
  \BibitemOpen
  \bibfield  {author} {\bibinfo {author} {\bibfnamefont {D.}~\bibnamefont
  {Vanderbilt}},\ }\href@noop {} {\emph {\bibinfo {title} {Berry Phases in
  Electronic Structure Theory}}}\ (\bibinfo  {publisher} {Cambridge University
  Press},\ \bibinfo {address} {Cambridge},\ \bibinfo {year} {2018})\BibitemShut
  {NoStop}%
\bibitem [{\citenamefont {King-Smith}\ and\ \citenamefont
  {Vanderbilt}(1993)}]{KingSmith93}%
  \BibitemOpen
  \bibfield  {author} {\bibinfo {author} {\bibfnamefont {R.~D.}\ \bibnamefont
  {King-Smith}}\ and\ \bibinfo {author} {\bibfnamefont {D.}~\bibnamefont
  {Vanderbilt}},\ }\href {\doibase 10.1103/PhysRevB.47.1651} {\bibfield
  {journal} {\bibinfo  {journal} {Phys. Rev. B}\ }\textbf {\bibinfo {volume}
  {47}},\ \bibinfo {pages} {1651} (\bibinfo {year} {1993})}\BibitemShut
  {NoStop}%
\bibitem [{\citenamefont {Resta}(1994)}]{Resta94}%
  \BibitemOpen
  \bibfield  {author} {\bibinfo {author} {\bibfnamefont {R.}~\bibnamefont
  {Resta}},\ }\href {\doibase 10.1103/RevModPhys.66.899} {\bibfield  {journal}
  {\bibinfo  {journal} {Rev. Mod. Phys.}\ }\textbf {\bibinfo {volume} {66}},\
  \bibinfo {pages} {899} (\bibinfo {year} {1994})}\BibitemShut {NoStop}%
\bibitem [{\citenamefont {Berry}(1984)}]{Berry84}%
  \BibitemOpen
  \bibfield  {author} {\bibinfo {author} {\bibfnamefont {M.~V.}\ \bibnamefont
  {Berry}},\ }\href {\doibase 10.1098/rspa.1984.0023} {\bibfield  {journal}
  {\bibinfo  {journal} {Proceedings of the Royal Society of London. A.
  Mathematical and Physical Sciences}\ }\textbf {\bibinfo {volume} {392}},\
  \bibinfo {pages} {45} (\bibinfo {year} {1984})}\BibitemShut {NoStop}%
\bibitem [{\citenamefont {Zak}(1989)}]{Zak89}%
  \BibitemOpen
  \bibfield  {author} {\bibinfo {author} {\bibfnamefont {J.}~\bibnamefont
  {Zak}},\ }\href {\doibase 10.1103/PhysRevLett.62.2747} {\bibfield  {journal}
  {\bibinfo  {journal} {Phys. Rev. Lett.}\ }\textbf {\bibinfo {volume} {62}},\
  \bibinfo {pages} {2747} (\bibinfo {year} {1989})}\BibitemShut {NoStop}%
\bibitem [{\citenamefont {Wilczek}\ and\ \citenamefont
  {Zee}(1984)}]{WilczekZee84}%
  \BibitemOpen
  \bibfield  {author} {\bibinfo {author} {\bibfnamefont {F.}~\bibnamefont
  {Wilczek}}\ and\ \bibinfo {author} {\bibfnamefont {A.}~\bibnamefont {Zee}},\
  }\href {\doibase 10.1103/PhysRevLett.52.2111} {\bibfield  {journal} {\bibinfo
   {journal} {Phys. Rev. Lett.}\ }\textbf {\bibinfo {volume} {52}},\ \bibinfo
  {pages} {2111} (\bibinfo {year} {1984})}\BibitemShut {NoStop}%
\bibitem [{\citenamefont {Aharonov}\ and\ \citenamefont
  {Anandan}(1987)}]{Aharonov87}%
  \BibitemOpen
  \bibfield  {author} {\bibinfo {author} {\bibfnamefont {Y.}~\bibnamefont
  {Aharonov}}\ and\ \bibinfo {author} {\bibfnamefont {J.}~\bibnamefont
  {Anandan}},\ }\href {\doibase 10.1103/PhysRevLett.58.1593} {\bibfield
  {journal} {\bibinfo  {journal} {Phys. Rev. Lett.}\ }\textbf {\bibinfo
  {volume} {58}},\ \bibinfo {pages} {1593} (\bibinfo {year}
  {1987})}\BibitemShut {NoStop}%
\bibitem [{\citenamefont {Anandan}(1988)}]{Anandan88}%
  \BibitemOpen
  \bibfield  {author} {\bibinfo {author} {\bibfnamefont {J.}~\bibnamefont
  {Anandan}},\ }\href {\doibase https://doi.org/10.1016/0375-9601(88)91010-9}
  {\bibfield  {journal} {\bibinfo  {journal} {Physics Letters A}\ }\textbf
  {\bibinfo {volume} {133}},\ \bibinfo {pages} {171 } (\bibinfo {year}
  {1988})}\BibitemShut {NoStop}%
\bibitem [{\citenamefont {Ladovrechis}\ and\ \citenamefont
  {Fulga}(2019)}]{Ladovrechis18}%
  \BibitemOpen
  \bibfield  {author} {\bibinfo {author} {\bibfnamefont {K.}~\bibnamefont
  {Ladovrechis}}\ and\ \bibinfo {author} {\bibfnamefont {I.~C.}\ \bibnamefont
  {Fulga}},\ }\href {\doibase 10.1103/PhysRevB.99.195426} {\bibfield  {journal}
  {\bibinfo  {journal} {Phys. Rev. B}\ }\textbf {\bibinfo {volume} {99}},\
  \bibinfo {pages} {195426} (\bibinfo {year} {2019})}\BibitemShut {NoStop}%
\bibitem [{\citenamefont {Franca}\ \emph {et~al.}(2018)\citenamefont {Franca},
  \citenamefont {van~den Brink},\ and\ \citenamefont {Fulga}}]{Franca18}%
  \BibitemOpen
  \bibfield  {author} {\bibinfo {author} {\bibfnamefont {S.}~\bibnamefont
  {Franca}}, \bibinfo {author} {\bibfnamefont {J.}~\bibnamefont {van~den
  Brink}}, \ and\ \bibinfo {author} {\bibfnamefont {I.~C.}\ \bibnamefont
  {Fulga}},\ }\href {\doibase 10.1103/PhysRevB.98.201114} {\bibfield  {journal}
  {\bibinfo  {journal} {Phys. Rev. B}\ }\textbf {\bibinfo {volume} {98}},\
  \bibinfo {pages} {201114} (\bibinfo {year} {2018})}\BibitemShut {NoStop}%
\bibitem [{\citenamefont {Bomantara}\ \emph {et~al.}(2019)\citenamefont
  {Bomantara}, \citenamefont {Zhou}, \citenamefont {Pan},\ and\ \citenamefont
  {Gong}}]{Bomantara19}%
  \BibitemOpen
  \bibfield  {author} {\bibinfo {author} {\bibfnamefont {R.~W.}\ \bibnamefont
  {Bomantara}}, \bibinfo {author} {\bibfnamefont {L.}~\bibnamefont {Zhou}},
  \bibinfo {author} {\bibfnamefont {J.}~\bibnamefont {Pan}}, \ and\ \bibinfo
  {author} {\bibfnamefont {J.}~\bibnamefont {Gong}},\ }\href {\doibase
  10.1103/PhysRevB.99.045441} {\bibfield  {journal} {\bibinfo  {journal} {Phys.
  Rev. B}\ }\textbf {\bibinfo {volume} {99}},\ \bibinfo {pages} {045441}
  (\bibinfo {year} {2019})}\BibitemShut {NoStop}%
\bibitem [{\citenamefont {Rodriguez-Vega}\ \emph {et~al.}(2019)\citenamefont
  {Rodriguez-Vega}, \citenamefont {Kumar},\ and\ \citenamefont
  {Seradjeh}}]{RodriguezVega18}%
  \BibitemOpen
  \bibfield  {author} {\bibinfo {author} {\bibfnamefont {M.}~\bibnamefont
  {Rodriguez-Vega}}, \bibinfo {author} {\bibfnamefont {A.}~\bibnamefont
  {Kumar}}, \ and\ \bibinfo {author} {\bibfnamefont {B.}~\bibnamefont
  {Seradjeh}},\ }\href {\doibase 10.1103/PhysRevB.100.085138} {\bibfield
  {journal} {\bibinfo  {journal} {Phys. Rev. B}\ }\textbf {\bibinfo {volume}
  {100}},\ \bibinfo {pages} {085138} (\bibinfo {year} {2019})}\BibitemShut
  {NoStop}%
\bibitem [{\citenamefont {Huang}\ and\ \citenamefont {Liu}(2018)}]{Huang18}%
  \BibitemOpen
  \bibfield  {author} {\bibinfo {author} {\bibfnamefont {B.}~\bibnamefont
  {Huang}}\ and\ \bibinfo {author} {\bibfnamefont {W.~V.}\ \bibnamefont
  {Liu}},\ }\href {http://arxiv.org/abs/1811.00555} {\bibfield  {journal}
  {\bibinfo  {journal} {arXiv:1811.00555}\ } (\bibinfo {year}
  {2018})}\BibitemShut {NoStop}%
\bibitem [{\citenamefont {Peng}\ and\ \citenamefont {Refael}(2019)}]{Peng18}%
  \BibitemOpen
  \bibfield  {author} {\bibinfo {author} {\bibfnamefont {Y.}~\bibnamefont
  {Peng}}\ and\ \bibinfo {author} {\bibfnamefont {G.}~\bibnamefont {Refael}},\
  }\href {\doibase 10.1103/PhysRevLett.123.016806} {\bibfield  {journal}
  {\bibinfo  {journal} {Phys. Rev. Lett.}\ }\textbf {\bibinfo {volume} {123}},\
  \bibinfo {pages} {016806} (\bibinfo {year} {2019})}\BibitemShut {NoStop}%
\bibitem [{\citenamefont {Inoue}\ and\ \citenamefont {Tanaka}(2010)}]{Inoue10}%
  \BibitemOpen
  \bibfield  {author} {\bibinfo {author} {\bibfnamefont {J.-i.}\ \bibnamefont
  {Inoue}}\ and\ \bibinfo {author} {\bibfnamefont {A.}~\bibnamefont {Tanaka}},\
  }\href {\doibase 10.1103/PhysRevLett.105.017401} {\bibfield  {journal}
  {\bibinfo  {journal} {Phys. Rev. Lett.}\ }\textbf {\bibinfo {volume} {105}},\
  \bibinfo {pages} {017401} (\bibinfo {year} {2010})}\BibitemShut {NoStop}%
\bibitem [{\citenamefont {Perez-Piskunow}\ \emph {et~al.}(2014)\citenamefont
  {Perez-Piskunow}, \citenamefont {Usaj}, \citenamefont {Balseiro},\ and\
  \citenamefont {Torres}}]{PerezPiskunow14}%
  \BibitemOpen
  \bibfield  {author} {\bibinfo {author} {\bibfnamefont {P.~M.}\ \bibnamefont
  {Perez-Piskunow}}, \bibinfo {author} {\bibfnamefont {G.}~\bibnamefont
  {Usaj}}, \bibinfo {author} {\bibfnamefont {C.~A.}\ \bibnamefont {Balseiro}},
  \ and\ \bibinfo {author} {\bibfnamefont {L.~E. F.~F.}\ \bibnamefont
  {Torres}},\ }\href {\doibase 10.1103/PhysRevB.89.121401} {\bibfield
  {journal} {\bibinfo  {journal} {Phys. Rev. B}\ }\textbf {\bibinfo {volume}
  {89}},\ \bibinfo {pages} {121401} (\bibinfo {year} {2014})}\BibitemShut
  {NoStop}%
\bibitem [{\citenamefont {Usaj}\ \emph {et~al.}(2014)\citenamefont {Usaj},
  \citenamefont {Perez-Piskunow}, \citenamefont {Foa~Torres},\ and\
  \citenamefont {Balseiro}}]{Usaj14}%
  \BibitemOpen
  \bibfield  {author} {\bibinfo {author} {\bibfnamefont {G.}~\bibnamefont
  {Usaj}}, \bibinfo {author} {\bibfnamefont {P.~M.}\ \bibnamefont
  {Perez-Piskunow}}, \bibinfo {author} {\bibfnamefont {L.~E.~F.}\ \bibnamefont
  {Foa~Torres}}, \ and\ \bibinfo {author} {\bibfnamefont {C.~A.}\ \bibnamefont
  {Balseiro}},\ }\href {\doibase 10.1103/PhysRevB.90.115423} {\bibfield
  {journal} {\bibinfo  {journal} {Phys. Rev. B}\ }\textbf {\bibinfo {volume}
  {90}},\ \bibinfo {pages} {115423} (\bibinfo {year} {2014})}\BibitemShut
  {NoStop}%
\bibitem [{\citenamefont {Dehghani}\ \emph {et~al.}(2014)\citenamefont
  {Dehghani}, \citenamefont {Oka},\ and\ \citenamefont {Mitra}}]{Dehghani14}%
  \BibitemOpen
  \bibfield  {author} {\bibinfo {author} {\bibfnamefont {H.}~\bibnamefont
  {Dehghani}}, \bibinfo {author} {\bibfnamefont {T.}~\bibnamefont {Oka}}, \
  and\ \bibinfo {author} {\bibfnamefont {A.}~\bibnamefont {Mitra}},\ }\href
  {\doibase 10.1103/PhysRevB.90.195429} {\bibfield  {journal} {\bibinfo
  {journal} {Phys. Rev. B}\ }\textbf {\bibinfo {volume} {90}},\ \bibinfo
  {pages} {195429} (\bibinfo {year} {2014})}\BibitemShut {NoStop}%
\bibitem [{\citenamefont {Dehghani}\ \emph {et~al.}(2015)\citenamefont
  {Dehghani}, \citenamefont {Oka},\ and\ \citenamefont {Mitra}}]{Dehghani15_1}%
  \BibitemOpen
  \bibfield  {author} {\bibinfo {author} {\bibfnamefont {H.}~\bibnamefont
  {Dehghani}}, \bibinfo {author} {\bibfnamefont {T.}~\bibnamefont {Oka}}, \
  and\ \bibinfo {author} {\bibfnamefont {A.}~\bibnamefont {Mitra}},\ }\href
  {\doibase 10.1103/PhysRevB.91.155422} {\bibfield  {journal} {\bibinfo
  {journal} {Phys. Rev. B}\ }\textbf {\bibinfo {volume} {91}},\ \bibinfo
  {pages} {155422} (\bibinfo {year} {2015})}\BibitemShut {NoStop}%
\bibitem [{\citenamefont {Dehghani}\ and\ \citenamefont
  {Mitra}(2015)}]{Dehghani15_2}%
  \BibitemOpen
  \bibfield  {author} {\bibinfo {author} {\bibfnamefont {H.}~\bibnamefont
  {Dehghani}}\ and\ \bibinfo {author} {\bibfnamefont {A.}~\bibnamefont
  {Mitra}},\ }\href {\doibase 10.1103/PhysRevB.92.165111} {\bibfield  {journal}
  {\bibinfo  {journal} {Phys. Rev. B}\ }\textbf {\bibinfo {volume} {92}},\
  \bibinfo {pages} {165111} (\bibinfo {year} {2015})}\BibitemShut {NoStop}%
\bibitem [{\citenamefont {D'Alessio}\ and\ \citenamefont
  {Rigol}(2015)}]{DAlessio15}%
  \BibitemOpen
  \bibfield  {author} {\bibinfo {author} {\bibfnamefont {L.}~\bibnamefont
  {D'Alessio}}\ and\ \bibinfo {author} {\bibfnamefont {M.}~\bibnamefont
  {Rigol}},\ }\href {\doibase https://doi.org/10.1038/ncomms9336} {\bibfield
  {journal} {\bibinfo  {journal} {Nat. Commun.}\ }\textbf {\bibinfo {volume}
  {6}},\ \bibinfo {pages} {8336} (\bibinfo {year} {2015})}\BibitemShut
  {NoStop}%
\bibitem [{\citenamefont {Mikami}\ \emph {et~al.}(2016)\citenamefont {Mikami},
  \citenamefont {Kitamura}, \citenamefont {Yasuda}, \citenamefont {Tsuji},
  \citenamefont {Oka},\ and\ \citenamefont {Aoki}}]{Mikami16}%
  \BibitemOpen
  \bibfield  {author} {\bibinfo {author} {\bibfnamefont {T.}~\bibnamefont
  {Mikami}}, \bibinfo {author} {\bibfnamefont {S.}~\bibnamefont {Kitamura}},
  \bibinfo {author} {\bibfnamefont {K.}~\bibnamefont {Yasuda}}, \bibinfo
  {author} {\bibfnamefont {N.}~\bibnamefont {Tsuji}}, \bibinfo {author}
  {\bibfnamefont {T.}~\bibnamefont {Oka}}, \ and\ \bibinfo {author}
  {\bibfnamefont {H.}~\bibnamefont {Aoki}},\ }\href {\doibase
  10.1103/PhysRevB.93.144307} {\bibfield  {journal} {\bibinfo  {journal} {Phys.
  Rev. B}\ }\textbf {\bibinfo {volume} {93}},\ \bibinfo {pages} {144307}
  (\bibinfo {year} {2016})}\BibitemShut {NoStop}%
\bibitem [{\citenamefont {Yates}\ \emph {et~al.}(2016)\citenamefont {Yates},
  \citenamefont {Lemonik},\ and\ \citenamefont {Mitra}}]{Yates16}%
  \BibitemOpen
  \bibfield  {author} {\bibinfo {author} {\bibfnamefont {D.~J.}\ \bibnamefont
  {Yates}}, \bibinfo {author} {\bibfnamefont {Y.}~\bibnamefont {Lemonik}}, \
  and\ \bibinfo {author} {\bibfnamefont {A.}~\bibnamefont {Mitra}},\ }\href
  {\doibase 10.1103/PhysRevB.94.205422} {\bibfield  {journal} {\bibinfo
  {journal} {Phys. Rev. B}\ }\textbf {\bibinfo {volume} {94}},\ \bibinfo
  {pages} {205422} (\bibinfo {year} {2016})}\BibitemShut {NoStop}%
\bibitem [{\citenamefont {Alexandradinata}\ \emph {et~al.}(2014)\citenamefont
  {Alexandradinata}, \citenamefont {Dai},\ and\ \citenamefont
  {Bernevig}}]{Alexandradinata14}%
  \BibitemOpen
  \bibfield  {author} {\bibinfo {author} {\bibfnamefont {A.}~\bibnamefont
  {Alexandradinata}}, \bibinfo {author} {\bibfnamefont {X.}~\bibnamefont
  {Dai}}, \ and\ \bibinfo {author} {\bibfnamefont {B.~A.}\ \bibnamefont
  {Bernevig}},\ }\href {\doibase 10.1103/PhysRevB.89.155114} {\bibfield
  {journal} {\bibinfo  {journal} {Phys. Rev. B}\ }\textbf {\bibinfo {volume}
  {89}},\ \bibinfo {pages} {155114} (\bibinfo {year} {2014})}\BibitemShut
  {NoStop}%
\bibitem [{\citenamefont {Alexandradinata}\ and\ \citenamefont
  {Bernevig}(2016)}]{Alexandradinata16}%
  \BibitemOpen
  \bibfield  {author} {\bibinfo {author} {\bibfnamefont {A.}~\bibnamefont
  {Alexandradinata}}\ and\ \bibinfo {author} {\bibfnamefont {B.~A.}\
  \bibnamefont {Bernevig}},\ }\href {\doibase 10.1103/PhysRevB.93.205104}
  {\bibfield  {journal} {\bibinfo  {journal} {Phys. Rev. B}\ }\textbf {\bibinfo
  {volume} {93}},\ \bibinfo {pages} {205104} (\bibinfo {year}
  {2016})}\BibitemShut {NoStop}%
\bibitem [{\citenamefont {Bouhon}\ \emph {et~al.}(2019)\citenamefont {Bouhon},
  \citenamefont {Black-Schaffer},\ and\ \citenamefont {Slager}}]{Bouhon18}%
  \BibitemOpen
  \bibfield  {author} {\bibinfo {author} {\bibfnamefont {A.}~\bibnamefont
  {Bouhon}}, \bibinfo {author} {\bibfnamefont {A.~M.}\ \bibnamefont
  {Black-Schaffer}}, \ and\ \bibinfo {author} {\bibfnamefont {R.-J.}\
  \bibnamefont {Slager}},\ }\href {\doibase 10.1103/PhysRevB.100.195135}
  {\bibfield  {journal} {\bibinfo  {journal} {Phys. Rev. B}\ }\textbf {\bibinfo
  {volume} {100}},\ \bibinfo {pages} {195135} (\bibinfo {year}
  {2019})}\BibitemShut {NoStop}%
\bibitem [{\citenamefont {Lindner}\ \emph {et~al.}(2013)\citenamefont
  {Lindner}, \citenamefont {Bergman}, \citenamefont {Refael},\ and\
  \citenamefont {Galitski}}]{Lindner13}%
  \BibitemOpen
  \bibfield  {author} {\bibinfo {author} {\bibfnamefont {N.~H.}\ \bibnamefont
  {Lindner}}, \bibinfo {author} {\bibfnamefont {D.~L.}\ \bibnamefont
  {Bergman}}, \bibinfo {author} {\bibfnamefont {G.}~\bibnamefont {Refael}}, \
  and\ \bibinfo {author} {\bibfnamefont {V.}~\bibnamefont {Galitski}},\ }\href
  {\doibase 10.1103/PhysRevB.87.235131} {\bibfield  {journal} {\bibinfo
  {journal} {Phys. Rev. B}\ }\textbf {\bibinfo {volume} {87}},\ \bibinfo
  {pages} {235131} (\bibinfo {year} {2013})}\BibitemShut {NoStop}%
\bibitem [{\citenamefont {Nakagawa}\ and\ \citenamefont
  {Kawakami}(2014)}]{Nakagawa14}%
  \BibitemOpen
  \bibfield  {author} {\bibinfo {author} {\bibfnamefont {M.}~\bibnamefont
  {Nakagawa}}\ and\ \bibinfo {author} {\bibfnamefont {N.}~\bibnamefont
  {Kawakami}},\ }\href {\doibase 10.1103/PhysRevA.89.013627} {\bibfield
  {journal} {\bibinfo  {journal} {Phys. Rev. A}\ }\textbf {\bibinfo {volume}
  {89}},\ \bibinfo {pages} {013627} (\bibinfo {year} {2014})}\BibitemShut
  {NoStop}%
\bibitem [{\citenamefont {Takasan}\ \emph {et~al.}(2017)\citenamefont
  {Takasan}, \citenamefont {Nakagawa},\ and\ \citenamefont
  {Kawakami}}]{Takasan17}%
  \BibitemOpen
  \bibfield  {author} {\bibinfo {author} {\bibfnamefont {K.}~\bibnamefont
  {Takasan}}, \bibinfo {author} {\bibfnamefont {M.}~\bibnamefont {Nakagawa}}, \
  and\ \bibinfo {author} {\bibfnamefont {N.}~\bibnamefont {Kawakami}},\ }\href
  {\doibase 10.1103/PhysRevB.96.115120} {\bibfield  {journal} {\bibinfo
  {journal} {Phys. Rev. B}\ }\textbf {\bibinfo {volume} {96}},\ \bibinfo
  {pages} {115120} (\bibinfo {year} {2017})}\BibitemShut {NoStop}%
\bibitem [{\citenamefont {Roy}\ and\ \citenamefont
  {Harper}(2017{\natexlab{b}})}]{Roy17}%
  \BibitemOpen
  \bibfield  {author} {\bibinfo {author} {\bibfnamefont {R.}~\bibnamefont
  {Roy}}\ and\ \bibinfo {author} {\bibfnamefont {F.}~\bibnamefont {Harper}},\
  }\href {\doibase 10.1103/PhysRevB.95.195128} {\bibfield  {journal} {\bibinfo
  {journal} {Phys. Rev. B}\ }\textbf {\bibinfo {volume} {95}},\ \bibinfo
  {pages} {195128} (\bibinfo {year} {2017}{\natexlab{b}})}\BibitemShut
  {NoStop}%
\bibitem [{\citenamefont {Thouless}(1983)}]{Thouless83}%
  \BibitemOpen
  \bibfield  {author} {\bibinfo {author} {\bibfnamefont {D.~J.}\ \bibnamefont
  {Thouless}},\ }\href {\doibase 10.1103/PhysRevB.27.6083} {\bibfield
  {journal} {\bibinfo  {journal} {Phys. Rev. B}\ }\textbf {\bibinfo {volume}
  {27}},\ \bibinfo {pages} {6083} (\bibinfo {year} {1983})}\BibitemShut
  {NoStop}%
\bibitem [{\citenamefont {Mizuta}\ \emph {et~al.}(2018)\citenamefont {Mizuta},
  \citenamefont {Takasan}, \citenamefont {Nakagawa},\ and\ \citenamefont
  {Kawakami}}]{Mizuta18}%
  \BibitemOpen
  \bibfield  {author} {\bibinfo {author} {\bibfnamefont {K.}~\bibnamefont
  {Mizuta}}, \bibinfo {author} {\bibfnamefont {K.}~\bibnamefont {Takasan}},
  \bibinfo {author} {\bibfnamefont {M.}~\bibnamefont {Nakagawa}}, \ and\
  \bibinfo {author} {\bibfnamefont {N.}~\bibnamefont {Kawakami}},\ }\href
  {\doibase 10.1103/PhysRevLett.121.093001} {\bibfield  {journal} {\bibinfo
  {journal} {Phys. Rev. Lett.}\ }\textbf {\bibinfo {volume} {121}},\ \bibinfo
  {pages} {093001} (\bibinfo {year} {2018})}\BibitemShut {NoStop}%
\bibitem [{\citenamefont {Lindner}\ \emph {et~al.}(2017)\citenamefont
  {Lindner}, \citenamefont {Berg},\ and\ \citenamefont {Rudner}}]{Lindner17}%
  \BibitemOpen
  \bibfield  {author} {\bibinfo {author} {\bibfnamefont {N.~H.}\ \bibnamefont
  {Lindner}}, \bibinfo {author} {\bibfnamefont {E.}~\bibnamefont {Berg}}, \
  and\ \bibinfo {author} {\bibfnamefont {M.~S.}\ \bibnamefont {Rudner}},\
  }\href {\doibase 10.1103/PhysRevX.7.011018} {\bibfield  {journal} {\bibinfo
  {journal} {Phys. Rev. X}\ }\textbf {\bibinfo {volume} {7}},\ \bibinfo {pages}
  {011018} (\bibinfo {year} {2017})}\BibitemShut {NoStop}%
\bibitem [{\citenamefont {Privitera}\ \emph {et~al.}(2018)\citenamefont
  {Privitera}, \citenamefont {Russomanno}, \citenamefont {Citro},\ and\
  \citenamefont {Santoro}}]{Privitera18}%
  \BibitemOpen
  \bibfield  {author} {\bibinfo {author} {\bibfnamefont {L.}~\bibnamefont
  {Privitera}}, \bibinfo {author} {\bibfnamefont {A.}~\bibnamefont
  {Russomanno}}, \bibinfo {author} {\bibfnamefont {R.}~\bibnamefont {Citro}}, \
  and\ \bibinfo {author} {\bibfnamefont {G.~E.}\ \bibnamefont {Santoro}},\
  }\href {\doibase 10.1103/PhysRevLett.120.106601} {\bibfield  {journal}
  {\bibinfo  {journal} {Phys. Rev. Lett.}\ }\textbf {\bibinfo {volume} {120}},\
  \bibinfo {pages} {106601} (\bibinfo {year} {2018})}\BibitemShut {NoStop}%
\bibitem [{\citenamefont {Fu}\ and\ \citenamefont {Kane}(2006)}]{FuKane06}%
  \BibitemOpen
  \bibfield  {author} {\bibinfo {author} {\bibfnamefont {L.}~\bibnamefont
  {Fu}}\ and\ \bibinfo {author} {\bibfnamefont {C.~L.}\ \bibnamefont {Kane}},\
  }\href {\doibase 10.1103/PhysRevB.74.195312} {\bibfield  {journal} {\bibinfo
  {journal} {Phys. Rev. B}\ }\textbf {\bibinfo {volume} {74}},\ \bibinfo
  {pages} {195312} (\bibinfo {year} {2006})}\BibitemShut {NoStop}%
\bibitem [{Note1()}]{Note1}%
  \BibitemOpen
  \bibinfo {note} {This formula \protect \textup {\hbox {\mathsurround \z@
  \protect \normalfont (\ignorespaces \ref {eq_AIIgapless}\unskip \@@italiccorr
  )}} of the topological invariant has not been derived in earlier literature.
  Equation \protect \textup {\hbox {\mathsurround \z@ \protect \normalfont
  (\ignorespaces \ref {eq_AIIgapless}\unskip \@@italiccorr )}} is derived by
  combining the formula of a topological invariant of static $d = 1$ class DIII
  topological superconductors\cite {Qi10,Schnyder11} and the fact that the
  classification of class AII Floquet operators is performed by mapping a
  Floquet unitary to an artificial Hamiltonian of a static class DIII
  superconductor\cite {Higashikawa18}.}\BibitemShut {Stop}%
\bibitem [{\citenamefont {Shindou}(2005)}]{ShindouSpinPump}%
  \BibitemOpen
  \bibfield  {author} {\bibinfo {author} {\bibfnamefont {R.}~\bibnamefont
  {Shindou}},\ }\href {\doibase 10.1143/JPSJ.74.1214} {\bibfield  {journal}
  {\bibinfo  {journal} {Journal of the Physical Society of Japan}\ }\textbf
  {\bibinfo {volume} {74}},\ \bibinfo {pages} {1214} (\bibinfo {year}
  {2005})}\BibitemShut {NoStop}%
\bibitem [{\citenamefont {Nielsen}\ and\ \citenamefont
  {Ninomiya}(1981{\natexlab{a}})}]{NielsenNinomiya1}%
  \BibitemOpen
  \bibfield  {author} {\bibinfo {author} {\bibfnamefont {H.}~\bibnamefont
  {Nielsen}}\ and\ \bibinfo {author} {\bibfnamefont {M.}~\bibnamefont
  {Ninomiya}},\ }\href {\doibase
  http://dx.doi.org/10.1016/0550-3213(81)90361-8} {\bibfield  {journal}
  {\bibinfo  {journal} {Nuclear Physics B}\ }\textbf {\bibinfo {volume}
  {185}},\ \bibinfo {pages} {20 } (\bibinfo {year}
  {1981}{\natexlab{a}})}\BibitemShut {NoStop}%
\bibitem [{\citenamefont {Nielsen}\ and\ \citenamefont
  {Ninomiya}(1981{\natexlab{b}})}]{NielsenNinomiya2}%
  \BibitemOpen
  \bibfield  {author} {\bibinfo {author} {\bibfnamefont {H.}~\bibnamefont
  {Nielsen}}\ and\ \bibinfo {author} {\bibfnamefont {M.}~\bibnamefont
  {Ninomiya}},\ }\href {\doibase
  http://dx.doi.org/10.1016/0550-3213(81)90524-1} {\bibfield  {journal}
  {\bibinfo  {journal} {Nuclear Physics B}\ }\textbf {\bibinfo {volume}
  {193}},\ \bibinfo {pages} {173 } (\bibinfo {year}
  {1981}{\natexlab{b}})}\BibitemShut {NoStop}%
\bibitem [{\citenamefont {Qi}\ \emph {et~al.}(2008)\citenamefont {Qi},
  \citenamefont {Hughes},\ and\ \citenamefont {Zhang}}]{Qi08}%
  \BibitemOpen
  \bibfield  {author} {\bibinfo {author} {\bibfnamefont {X.-L.}\ \bibnamefont
  {Qi}}, \bibinfo {author} {\bibfnamefont {T.~L.}\ \bibnamefont {Hughes}}, \
  and\ \bibinfo {author} {\bibfnamefont {S.-C.}\ \bibnamefont {Zhang}},\ }\href
  {\doibase 10.1103/PhysRevB.78.195424} {\bibfield  {journal} {\bibinfo
  {journal} {Phys. Rev. B}\ }\textbf {\bibinfo {volume} {78}},\ \bibinfo
  {pages} {195424} (\bibinfo {year} {2008})}\BibitemShut {NoStop}%
\bibitem [{\citenamefont {Harper}\ \emph {et~al.}(2019)\citenamefont {Harper},
  \citenamefont {Roy}, \citenamefont {Rudner},\ and\ \citenamefont
  {Sondhi}}]{Harper19}%
  \BibitemOpen
  \bibfield  {author} {\bibinfo {author} {\bibfnamefont {F.}~\bibnamefont
  {Harper}}, \bibinfo {author} {\bibfnamefont {R.}~\bibnamefont {Roy}},
  \bibinfo {author} {\bibfnamefont {M.~S.}\ \bibnamefont {Rudner}}, \ and\
  \bibinfo {author} {\bibfnamefont {S.~L.}\ \bibnamefont {Sondhi}},\ }\href
  {http://arxiv.org/abs/1905.01317} {\bibfield  {journal} {\bibinfo  {journal}
  {arXiv:1905.01317}\ } (\bibinfo {year} {2019})}\BibitemShut {NoStop}%
\bibitem [{\citenamefont {Fruchart}(2016)}]{Fruchart16}%
  \BibitemOpen
  \bibfield  {author} {\bibinfo {author} {\bibfnamefont {M.}~\bibnamefont
  {Fruchart}},\ }\href {\doibase 10.1103/PhysRevB.93.115429} {\bibfield
  {journal} {\bibinfo  {journal} {Phys. Rev. B}\ }\textbf {\bibinfo {volume}
  {93}},\ \bibinfo {pages} {115429} (\bibinfo {year} {2016})}\BibitemShut
  {NoStop}%
\bibitem [{\citenamefont {Liu}\ \emph {et~al.}(2018)\citenamefont {Liu},
  \citenamefont {Harper},\ and\ \citenamefont {Roy}}]{Liu18}%
  \BibitemOpen
  \bibfield  {author} {\bibinfo {author} {\bibfnamefont {X.}~\bibnamefont
  {Liu}}, \bibinfo {author} {\bibfnamefont {F.}~\bibnamefont {Harper}}, \ and\
  \bibinfo {author} {\bibfnamefont {R.}~\bibnamefont {Roy}},\ }\href {\doibase
  10.1103/PhysRevB.98.165116} {\bibfield  {journal} {\bibinfo  {journal} {Phys.
  Rev. B}\ }\textbf {\bibinfo {volume} {98}},\ \bibinfo {pages} {165116}
  (\bibinfo {year} {2018})}\BibitemShut {NoStop}%
\bibitem [{Note2()}]{Note2}%
  \BibitemOpen
  \bibinfo {note} {A similar quantity has been discussed in a context of static
  chiral topological insulators in Refs.~\protect \rev@citealpnum {Shiozaki13,
  Daido19}.}\BibitemShut {Stop}%
\bibitem [{\citenamefont {Ryu}\ \emph {et~al.}(2010)\citenamefont {Ryu},
  \citenamefont {Schnyder}, \citenamefont {Furusaki},\ and\ \citenamefont
  {Ludwig}}]{Ryu10}%
  \BibitemOpen
  \bibfield  {author} {\bibinfo {author} {\bibfnamefont {S.}~\bibnamefont
  {Ryu}}, \bibinfo {author} {\bibfnamefont {A.~P.}\ \bibnamefont {Schnyder}},
  \bibinfo {author} {\bibfnamefont {A.}~\bibnamefont {Furusaki}}, \ and\
  \bibinfo {author} {\bibfnamefont {A.~W.~W.}\ \bibnamefont {Ludwig}},\ }\href
  {http://stacks.iop.org/1367-2630/12/i=6/a=065010} {\bibfield  {journal}
  {\bibinfo  {journal} {New Journal of Physics}\ }\textbf {\bibinfo {volume}
  {12}},\ \bibinfo {pages} {065010} (\bibinfo {year} {2010})}\BibitemShut
  {NoStop}%
\bibitem [{\citenamefont {Teo}\ and\ \citenamefont {Kane}(2010)}]{TeoKane}%
  \BibitemOpen
  \bibfield  {author} {\bibinfo {author} {\bibfnamefont {J.~C.~Y.}\
  \bibnamefont {Teo}}\ and\ \bibinfo {author} {\bibfnamefont {C.~L.}\
  \bibnamefont {Kane}},\ }\href {\doibase 10.1103/PhysRevB.82.115120}
  {\bibfield  {journal} {\bibinfo  {journal} {Phys. Rev. B}\ }\textbf {\bibinfo
  {volume} {82}},\ \bibinfo {pages} {115120} (\bibinfo {year}
  {2010})}\BibitemShut {NoStop}%
\bibitem [{\citenamefont {Po}\ \emph {et~al.}(2018)\citenamefont {Po},
  \citenamefont {Watanabe},\ and\ \citenamefont {Vishwanath}}]{Po18}%
  \BibitemOpen
  \bibfield  {author} {\bibinfo {author} {\bibfnamefont {H.~C.}\ \bibnamefont
  {Po}}, \bibinfo {author} {\bibfnamefont {H.}~\bibnamefont {Watanabe}}, \ and\
  \bibinfo {author} {\bibfnamefont {A.}~\bibnamefont {Vishwanath}},\ }\href
  {\doibase 10.1103/PhysRevLett.121.126402} {\bibfield  {journal} {\bibinfo
  {journal} {Phys. Rev. Lett.}\ }\textbf {\bibinfo {volume} {121}},\ \bibinfo
  {pages} {126402} (\bibinfo {year} {2018})}\BibitemShut {NoStop}%
\bibitem [{\citenamefont {Slager}\ \emph {et~al.}(2015)\citenamefont {Slager},
  \citenamefont {Rademaker}, \citenamefont {Zaanen},\ and\ \citenamefont
  {Balents}}]{Slager15}%
  \BibitemOpen
  \bibfield  {author} {\bibinfo {author} {\bibfnamefont {R.-J.}\ \bibnamefont
  {Slager}}, \bibinfo {author} {\bibfnamefont {L.}~\bibnamefont {Rademaker}},
  \bibinfo {author} {\bibfnamefont {J.}~\bibnamefont {Zaanen}}, \ and\ \bibinfo
  {author} {\bibfnamefont {L.}~\bibnamefont {Balents}},\ }\href {\doibase
  10.1103/PhysRevB.92.085126} {\bibfield  {journal} {\bibinfo  {journal} {Phys.
  Rev. B}\ }\textbf {\bibinfo {volume} {92}},\ \bibinfo {pages} {085126}
  (\bibinfo {year} {2015})}\BibitemShut {NoStop}%
\bibitem [{\citenamefont {Hatsugai}(1993)}]{Hatsugai93}%
  \BibitemOpen
  \bibfield  {author} {\bibinfo {author} {\bibfnamefont {Y.}~\bibnamefont
  {Hatsugai}},\ }\href {\doibase 10.1103/PhysRevLett.71.3697} {\bibfield
  {journal} {\bibinfo  {journal} {Phys. Rev. Lett.}\ }\textbf {\bibinfo
  {volume} {71}},\ \bibinfo {pages} {3697} (\bibinfo {year}
  {1993})}\BibitemShut {NoStop}%
\bibitem [{\citenamefont {Slager}\ \emph {et~al.}(2016)\citenamefont {Slager},
  \citenamefont {Juri\ifmmode \check{c}\else \v{c}\fi{}i\ifmmode~\acute{c}\else
  \'{c}\fi{}}, \citenamefont {Lahtinen},\ and\ \citenamefont
  {Zaanen}}]{Slager16}%
  \BibitemOpen
  \bibfield  {author} {\bibinfo {author} {\bibfnamefont {R.-J.}\ \bibnamefont
  {Slager}}, \bibinfo {author} {\bibfnamefont {V.}~\bibnamefont {Juri\ifmmode
  \check{c}\else \v{c}\fi{}i\ifmmode~\acute{c}\else \'{c}\fi{}}}, \bibinfo
  {author} {\bibfnamefont {V.}~\bibnamefont {Lahtinen}}, \ and\ \bibinfo
  {author} {\bibfnamefont {J.}~\bibnamefont {Zaanen}},\ }\href {\doibase
  10.1103/PhysRevB.93.245406} {\bibfield  {journal} {\bibinfo  {journal} {Phys.
  Rev. B}\ }\textbf {\bibinfo {volume} {93}},\ \bibinfo {pages} {245406}
  (\bibinfo {year} {2016})}\BibitemShut {NoStop}%
\bibitem [{\citenamefont {Essin}\ and\ \citenamefont
  {Gurarie}(2011)}]{Essin11}%
  \BibitemOpen
  \bibfield  {author} {\bibinfo {author} {\bibfnamefont {A.~M.}\ \bibnamefont
  {Essin}}\ and\ \bibinfo {author} {\bibfnamefont {V.}~\bibnamefont
  {Gurarie}},\ }\href {\doibase 10.1103/PhysRevB.84.125132} {\bibfield
  {journal} {\bibinfo  {journal} {Phys. Rev. B}\ }\textbf {\bibinfo {volume}
  {84}},\ \bibinfo {pages} {125132} (\bibinfo {year} {2011})}\BibitemShut
  {NoStop}%
\bibitem [{\citenamefont {Rhim}\ \emph {et~al.}(2018)\citenamefont {Rhim},
  \citenamefont {Bardarson},\ and\ \citenamefont {Slager}}]{Rhim18}%
  \BibitemOpen
  \bibfield  {author} {\bibinfo {author} {\bibfnamefont {J.-W.}\ \bibnamefont
  {Rhim}}, \bibinfo {author} {\bibfnamefont {J.~H.}\ \bibnamefont {Bardarson}},
  \ and\ \bibinfo {author} {\bibfnamefont {R.-J.}\ \bibnamefont {Slager}},\
  }\href {\doibase 10.1103/PhysRevB.97.115143} {\bibfield  {journal} {\bibinfo
  {journal} {Phys. Rev. B}\ }\textbf {\bibinfo {volume} {97}},\ \bibinfo
  {pages} {115143} (\bibinfo {year} {2018})}\BibitemShut {NoStop}%
\bibitem [{\citenamefont {Ran}\ \emph {et~al.}(2009)\citenamefont {Ran},
  \citenamefont {Zhang},\ and\ \citenamefont {Vishwanath}}]{Ran09}%
  \BibitemOpen
  \bibfield  {author} {\bibinfo {author} {\bibfnamefont {Y.}~\bibnamefont
  {Ran}}, \bibinfo {author} {\bibfnamefont {Y.}~\bibnamefont {Zhang}}, \ and\
  \bibinfo {author} {\bibfnamefont {A.}~\bibnamefont {Vishwanath}},\
  }\href@noop {} {\bibfield  {journal} {\bibinfo  {journal} {Nature Physics}\
  }\textbf {\bibinfo {volume} {5}},\ \bibinfo {pages} {298} (\bibinfo {year}
  {2009})}\BibitemShut {NoStop}%
\bibitem [{\citenamefont {Juri\ifmmode \check{c}\else
  \v{c}\fi{}i\ifmmode~\acute{c}\else \'{c}\fi{}}\ \emph
  {et~al.}(2012)\citenamefont {Juri\ifmmode \check{c}\else
  \v{c}\fi{}i\ifmmode~\acute{c}\else \'{c}\fi{}}, \citenamefont {Mesaros},
  \citenamefont {Slager},\ and\ \citenamefont {Zaanen}}]{Juricic12}%
  \BibitemOpen
  \bibfield  {author} {\bibinfo {author} {\bibfnamefont {V.}~\bibnamefont
  {Juri\ifmmode \check{c}\else \v{c}\fi{}i\ifmmode~\acute{c}\else \'{c}\fi{}}},
  \bibinfo {author} {\bibfnamefont {A.}~\bibnamefont {Mesaros}}, \bibinfo
  {author} {\bibfnamefont {R.-J.}\ \bibnamefont {Slager}}, \ and\ \bibinfo
  {author} {\bibfnamefont {J.}~\bibnamefont {Zaanen}},\ }\href {\doibase
  10.1103/PhysRevLett.108.106403} {\bibfield  {journal} {\bibinfo  {journal}
  {Phys. Rev. Lett.}\ }\textbf {\bibinfo {volume} {108}},\ \bibinfo {pages}
  {106403} (\bibinfo {year} {2012})}\BibitemShut {NoStop}%
\bibitem [{\citenamefont {Slager}\ \emph {et~al.}(2014)\citenamefont {Slager},
  \citenamefont {Mesaros}, \citenamefont {Juri\ifmmode \check{c}\else
  \v{c}\fi{}i\ifmmode~\acute{c}\else \'{c}\fi{}},\ and\ \citenamefont
  {Zaanen}}]{Slager14}%
  \BibitemOpen
  \bibfield  {author} {\bibinfo {author} {\bibfnamefont {R.-J.}\ \bibnamefont
  {Slager}}, \bibinfo {author} {\bibfnamefont {A.}~\bibnamefont {Mesaros}},
  \bibinfo {author} {\bibfnamefont {V.}~\bibnamefont {Juri\ifmmode
  \check{c}\else \v{c}\fi{}i\ifmmode~\acute{c}\else \'{c}\fi{}}}, \ and\
  \bibinfo {author} {\bibfnamefont {J.}~\bibnamefont {Zaanen}},\ }\href
  {\doibase 10.1103/PhysRevB.90.241403} {\bibfield  {journal} {\bibinfo
  {journal} {Phys. Rev. B}\ }\textbf {\bibinfo {volume} {90}},\ \bibinfo
  {pages} {241403} (\bibinfo {year} {2014})}\BibitemShut {NoStop}%
\bibitem [{\citenamefont {Slager}(2019)}]{SlagerJPC}%
  \BibitemOpen
  \bibfield  {author} {\bibinfo {author} {\bibfnamefont {R.-J.}\ \bibnamefont
  {Slager}},\ }\href {\doibase https://doi.org/10.1016/j.jpcs.2018.01.023}
  {\bibfield  {journal} {\bibinfo  {journal} {Journal of Physics and Chemistry
  of Solids}\ }\textbf {\bibinfo {volume} {128}},\ \bibinfo {pages} {24 }
  (\bibinfo {year} {2019})},\ \bibinfo {note} {spin-Orbit Coupled
  Materials}\BibitemShut {NoStop}%
\bibitem [{\citenamefont {Slager}\ \emph {et~al.}(2017)\citenamefont {Slager},
  \citenamefont {Juri\ifmmode \check{c}\else \v{c}\fi{}i\ifmmode~\acute{c}\else
  \'{c}\fi{}},\ and\ \citenamefont {Roy}}]{bbcweyl}%
  \BibitemOpen
  \bibfield  {author} {\bibinfo {author} {\bibfnamefont {R.-J.}\ \bibnamefont
  {Slager}}, \bibinfo {author} {\bibfnamefont {V.}~\bibnamefont {Juri\ifmmode
  \check{c}\else \v{c}\fi{}i\ifmmode~\acute{c}\else \'{c}\fi{}}}, \ and\
  \bibinfo {author} {\bibfnamefont {B.}~\bibnamefont {Roy}},\ }\href {\doibase
  10.1103/PhysRevB.96.201401} {\bibfield  {journal} {\bibinfo  {journal} {Phys.
  Rev. B}\ }\textbf {\bibinfo {volume} {96}},\ \bibinfo {pages} {201401}
  (\bibinfo {year} {2017})}\BibitemShut {NoStop}%
\bibitem [{\citenamefont {Mesaros}\ \emph {et~al.}(2013)\citenamefont
  {Mesaros}, \citenamefont {Slager}, \citenamefont {Zaanen},\ and\
  \citenamefont {Juri{\v{c}}i{\'c}}}]{Mesaros13}%
  \BibitemOpen
  \bibfield  {author} {\bibinfo {author} {\bibfnamefont {A.}~\bibnamefont
  {Mesaros}}, \bibinfo {author} {\bibfnamefont {R.-J.}\ \bibnamefont {Slager}},
  \bibinfo {author} {\bibfnamefont {J.}~\bibnamefont {Zaanen}}, \ and\ \bibinfo
  {author} {\bibfnamefont {V.}~\bibnamefont {Juri{\v{c}}i{\'c}}},\ }\href@noop
  {} {\bibfield  {journal} {\bibinfo  {journal} {Nuclear Physics B}\ }\textbf
  {\bibinfo {volume} {867}},\ \bibinfo {pages} {977} (\bibinfo {year}
  {2013})}\BibitemShut {NoStop}%
\bibitem [{\citenamefont {Imura}\ \emph {et~al.}(2011)\citenamefont {Imura},
  \citenamefont {Takane},\ and\ \citenamefont {Tanaka}}]{Imura11}%
  \BibitemOpen
  \bibfield  {author} {\bibinfo {author} {\bibfnamefont {K.-I.}\ \bibnamefont
  {Imura}}, \bibinfo {author} {\bibfnamefont {Y.}~\bibnamefont {Takane}}, \
  and\ \bibinfo {author} {\bibfnamefont {A.}~\bibnamefont {Tanaka}},\
  }\href@noop {} {\bibfield  {journal} {\bibinfo  {journal} {Phys. Rev. B}\
  }\textbf {\bibinfo {volume} {84}},\ \bibinfo {pages} {035443} (\bibinfo
  {year} {2011})}\BibitemShut {NoStop}%
\bibitem [{\citenamefont {Benalcazar}\ \emph {et~al.}(2017)\citenamefont
  {Benalcazar}, \citenamefont {Bernevig},\ and\ \citenamefont
  {Hughes}}]{Benalcazar17}%
  \BibitemOpen
  \bibfield  {author} {\bibinfo {author} {\bibfnamefont {W.~A.}\ \bibnamefont
  {Benalcazar}}, \bibinfo {author} {\bibfnamefont {B.~A.}\ \bibnamefont
  {Bernevig}}, \ and\ \bibinfo {author} {\bibfnamefont {T.~L.}\ \bibnamefont
  {Hughes}},\ }\href@noop {} {\bibfield  {journal} {\bibinfo  {journal}
  {Science}\ }\textbf {\bibinfo {volume} {357}},\ \bibinfo {pages} {61}
  (\bibinfo {year} {2017})}\BibitemShut {NoStop}%
\bibitem [{\citenamefont {Khalaf}\ \emph {et~al.}(2018)\citenamefont {Khalaf},
  \citenamefont {Po}, \citenamefont {Vishwanath},\ and\ \citenamefont
  {Watanabe}}]{Khalaf18}%
  \BibitemOpen
  \bibfield  {author} {\bibinfo {author} {\bibfnamefont {E.}~\bibnamefont
  {Khalaf}}, \bibinfo {author} {\bibfnamefont {H.~C.}\ \bibnamefont {Po}},
  \bibinfo {author} {\bibfnamefont {A.}~\bibnamefont {Vishwanath}}, \ and\
  \bibinfo {author} {\bibfnamefont {H.}~\bibnamefont {Watanabe}},\ }\href@noop
  {} {\bibfield  {journal} {\bibinfo  {journal} {Physical Review X}\ }\textbf
  {\bibinfo {volume} {8}},\ \bibinfo {pages} {031070} (\bibinfo {year}
  {2018})}\BibitemShut {NoStop}%
\bibitem [{\citenamefont {Qi}\ \emph {et~al.}(2010)\citenamefont {Qi},
  \citenamefont {Hughes},\ and\ \citenamefont {Zhang}}]{Qi10}%
  \BibitemOpen
  \bibfield  {author} {\bibinfo {author} {\bibfnamefont {X.-L.}\ \bibnamefont
  {Qi}}, \bibinfo {author} {\bibfnamefont {T.~L.}\ \bibnamefont {Hughes}}, \
  and\ \bibinfo {author} {\bibfnamefont {S.-C.}\ \bibnamefont {Zhang}},\ }\href
  {\doibase 10.1103/PhysRevB.81.134508} {\bibfield  {journal} {\bibinfo
  {journal} {Phys. Rev. B}\ }\textbf {\bibinfo {volume} {81}},\ \bibinfo
  {pages} {134508} (\bibinfo {year} {2010})}\BibitemShut {NoStop}%
\bibitem [{\citenamefont {Schnyder}\ and\ \citenamefont
  {Ryu}(2011)}]{Schnyder11}%
  \BibitemOpen
  \bibfield  {author} {\bibinfo {author} {\bibfnamefont {A.~P.}\ \bibnamefont
  {Schnyder}}\ and\ \bibinfo {author} {\bibfnamefont {S.}~\bibnamefont {Ryu}},\
  }\href {\doibase 10.1103/PhysRevB.84.060504} {\bibfield  {journal} {\bibinfo
  {journal} {Phys. Rev. B}\ }\textbf {\bibinfo {volume} {84}},\ \bibinfo
  {pages} {060504} (\bibinfo {year} {2011})}\BibitemShut {NoStop}%
\bibitem [{\citenamefont {Shiozaki}\ and\ \citenamefont
  {Fujimoto}(2013)}]{Shiozaki13}%
  \BibitemOpen
  \bibfield  {author} {\bibinfo {author} {\bibfnamefont {K.}~\bibnamefont
  {Shiozaki}}\ and\ \bibinfo {author} {\bibfnamefont {S.}~\bibnamefont
  {Fujimoto}},\ }\href {\doibase 10.1103/PhysRevLett.110.076804} {\bibfield
  {journal} {\bibinfo  {journal} {Phys. Rev. Lett.}\ }\textbf {\bibinfo
  {volume} {110}},\ \bibinfo {pages} {076804} (\bibinfo {year}
  {2013})}\BibitemShut {NoStop}%
\bibitem [{\citenamefont {Daido}\ and\ \citenamefont {Yanase}(2019)}]{Daido19}%
  \BibitemOpen
  \bibfield  {author} {\bibinfo {author} {\bibfnamefont {A.}~\bibnamefont
  {Daido}}\ and\ \bibinfo {author} {\bibfnamefont {Y.}~\bibnamefont {Yanase}},\
  }\href {\doibase 10.1103/PhysRevB.100.174512} {\bibfield  {journal} {\bibinfo
   {journal} {Phys. Rev. B}\ }\textbf {\bibinfo {volume} {100}},\ \bibinfo
  {pages} {174512} (\bibinfo {year} {2019})}\BibitemShut {NoStop}%
\end{thebibliography}%

\end{document}